%% file: main.tex
\let\mypdfximage\pdfximage
\def\pdfximage{\immediate\mypdfximage}

\documentclass[letterpaper, 11pt]{article}

\usepackage{algorithm}
\usepackage{algorithmic}
\usepackage{amsmath}
\usepackage{amssymb}
\usepackage{amsthm}
\usepackage[font=small]{caption} \usepackage{epsfig}
\usepackage{geometry}
\usepackage{graphicx}
\usepackage[colorlinks=true, citecolor=blue, filecolor=black, linkcolor=red, urlcolor=blue]{hyperref}
\usepackage{subfigure}
\usepackage[textsize=tiny]{todonotes}
\usepackage{soul} \setstcolor{red}
\setulcolor{blue}

\usepackage[capitalise,nameinlink,noabbrev]{cleveref} 
\include{newcommands_all}

\title{Compressive Sensing with Cross-Validation and
  Stop-Sampling\\for Sparse Polynomial Chaos Expansions}

\author{Xun Huan\footnote{Corresponding author:
    \href{mailto:xhuan@sandia.gov}{xhuan@sandia.gov}, Sandia National
    Laboratories, Livermore, CA 94550, USA.}, Cosmin
  Safta\footnote{Sandia National Laboratories, Livermore, CA 94550,
    USA.}, Khachik Sargsyan\footnotemark[2], Zachary
  P. Vane\footnotemark[2],\\Guilhem Lacaze\footnotemark[2], Joseph
  C. Oefelein\footnotemark[2], and Habib N. Najm\footnotemark[2]}

\begin {document}
\maketitle

\begin{abstract}
  Compressive sensing is a powerful technique for recovering sparse
  solutions of underdetermined linear systems, which is often
  encountered in uncertainty quantification analysis of expensive and
  high-dimensional physical models.  We perform numerical
  investigations employing several compressive sensing solvers that
  target the unconstrained LASSO formulation, with a focus on linear
  systems that arise in the construction of polynomial chaos
  expansions. With core solvers of \lls{}, \sparsa{}, \cgist{},
  \fpcas{}, and \admm{}, we develop techniques to mitigate overfitting
  through an automated selection of regularization constant based on
  cross-validation, and a heuristic strategy to guide the
  stop-sampling decision. Practical recommendations on parameter
  settings for these techniques are provided and discussed.  The
  overall method is applied to a series of numerical examples of
  increasing complexity, including large eddy simulations of
  supersonic turbulent jet-in-crossflow involving a 24-dimensional
  input.  Through empirical phase-transition diagrams and convergence
  plots, we illustrate sparse recovery performance under structures
  induced by polynomial chaos, accuracy and computational tradeoffs
  between polynomial bases of different degrees, and practicability of
  conducting compressive sensing for a realistic, high-dimensional
  physical application.  Across test cases studied in this paper, we
  find \admm{} to have demonstrated empirical advantages through
  consistent lower errors and faster computational times.
\end{abstract}

\input{sections/introduction}

\input{sections/methodology}

\input{sections/pce}

\input{sections/results}

\input{sections/conclusions}

\input{sections/acknowledgments}
\input{sections/appendix}

\clearpage
\bibliography{local}
\bibliographystyle{siamplain}

\end{document}

%% file: newcommands_all.tex
\newcommand{\lls}{l1\underline{\hspace{0.5em}}ls}
\newcommand{\sparsa}{SpaRSA}
\newcommand{\cgist}{CGIST}
\newcommand{\fpcas}{FPC\underline{\hspace{0.5em}}AS}
\newcommand{\admm}{ADMM}

\renewcommand{\[}{\left[}
\renewcommand{\]}{\right]}
\renewcommand{\(}{\left(}
\renewcommand{\)}{\right)}

\newcommand{\norm}[2]{\left\|\, #1 \,\right\|_{#2}}

\newcommand{\vvvert}{|\kern-1pt|\kern-1pt|}

\newcommand{\EE}{\mathbb{E}}

\newcommand{\NN}{\mathbb{N}}

\newcommand{\RR}{\mathbb{R}}

\newcommand{\CJ}{\mathcal{J}}

\newcommand{\CN}{\mathcal{N}}

\newcommand{\CU}{\mathcal{U}}

\newcommand{\argmin}{\operatornamewithlimits{argmin}}

\crefname{section}{section}{sections}
\crefname{subsection}{subsection}{subsections}

\Crefname{figure}{Figure}{Figures}

\crefformat{equation}{\textup{#2(#1)#3}}
\crefrangeformat{equation}{\textup{#3(#1)#4--#5(#2)#6}}
\crefmultiformat{equation}{\textup{#2(#1)#3}}{ and \textup{#2(#1)#3}}
{, \textup{#2(#1)#3}}{, and \textup{#2(#1)#3}}
\crefrangemultiformat{equation}{\textup{#3(#1)#4--#5(#2)#6}}%
{ and \textup{#3(#1)#4--#5(#2)#6}}{, \textup{#3(#1)#4--#5(#2)#6}}{, and \textup{#3(#1)#4--#5(#2)#6}}

\Crefformat{equation}{#2Equation~\textup{(#1)}#3}
\Crefrangeformat{equation}{Equations~\textup{#3(#1)#4--#5(#2)#6}}
\Crefmultiformat{equation}{Equations~\textup{#2(#1)#3}}{ and \textup{#2(#1)#3}}
{, \textup{#2(#1)#3}}{, and \textup{#2(#1)#3}}
\Crefrangemultiformat{equation}{Equations~\textup{#3(#1)#4--#5(#2)#6}}%
{ and \textup{#3(#1)#4--#5(#2)#6}}{, \textup{#3(#1)#4--#5(#2)#6}}{, and \textup{#3(#1)#4--#5(#2)#6}}

\crefdefaultlabelformat{#2\textup{#1}#3}

%% file: sections/introduction.tex
\section{Introduction}
\label{sec:intro}

Compressive sensing (CS) started as a breakthrough technique in signal
processing around a decade ago~\cite{Candes2006a, Donoho2006a}. It has
since been broadly employed to recover sparse solutions of
underdetermined linear systems where the number of unknowns exceeds
the number of data measurements. Such a system has an infinite number
of solutions, and CS seeks the sparsest solution---that is, solution
with the fewest number of non-zero components, minimizing its
$\ell_0$-norm. This $\ell_0$-minimization is an NP-hard problem,
however, and a simpler convex relaxation minimizing the $\ell_1$-norm
is often used as an approximation, which can be shown to reach the
$\ell_0$ solution under certain conditions~\cite{Donoho2006}.  Several
variants of $\ell_1$-sparse recovery are thus frequently studied:
\begin{align}
  & \textnormal{(BP)} &  \min_x \norm{x}{1} && \textnormal{subject
    to} && Ax = y,\label{e:CS_BP} \\
  & \textnormal{(BPDN)} & \min_x \norm{x}{1} &&
  \textnormal{subject to} && \norm{Ax -
    y}{2}^2 < \epsilon,\label{e:CS_BPDN} \\
  &\textnormal{(LASSO)}&  \min_x \norm{Ax - y}{2}^2 && \textnormal{subject to} &&
  \norm{x}{1} < \tau,\label{e:CS_LASSO} \\
  &\textnormal{(uLASSO)} &  \min_x \norm{Ax - y}{2}^2 + \lambda
  \norm{x}{1}, &&\label{e:CS_uLASSO}
\end{align}
where $A\in \RR^{m\times n}$ ($m$ is the number of data samples, and
$n$ the number of basis functions), $x\in\RR^{n}$, $y\in\RR^m$,
$\epsilon>0$ and $\tau>0$ are scalar tolerances, and $\lambda\geq 0$
is a scalar regularization parameter. \Cref{e:CS_BP} (BP) is known as
the \textit{basis pursuit} problem; \cref{e:CS_BPDN} (BPDN) is the
\textit{basis pursuit denoising}; \cref{e:CS_LASSO} (LASSO) is the
\textit{(original) least absolute shrinkage and selection operator};
and \cref{e:CS_uLASSO} (uLASSO) is the \textit{unconstrained LASSO}
(also known as the Lagrangian BPDN, or simply the $\ell_1$-regularized
least squares). These variants are closely related to each other, and
in fact \cref{e:CS_BPDN,e:CS_LASSO,e:CS_uLASSO} can be shown to arrive
at the same solution for appropriate choices of $\epsilon$, $\tau$,
and $\lambda$, although their relationships are difficult to determine
a priori~\cite{VandenBerg2009a}.

The connections among these optimization statements do not imply they
are equally easy or difficult to solve, and different algorithms and
solvers have been developed to specifically target each form. For
example, least angle regression (LARS)~\cite{Efron2004} accommodates
\cref{e:CS_LASSO} by iteratively expanding a basis following an
equiangular direction of the residual; primal-dual interior point
methods (e.g.,~\cite{Chen2001a}) can be used to solve
\cref{e:CS_BP,e:CS_BPDN} by reformulating them as ``perturbed'' linear
programs; $\ell_1$-\textit{magic}~\cite{Candes2006a} recasts
\cref{e:CS_BPDN} as a second-order cone program
and leverages efficient log-barrier optimization; and many more.
Greedy algorithms such as orthogonal matching pursuit
(OMP)~\cite{Davis1997} rely on heuristics, do not target any
particular optimization formulation, and have been demonstrated to
work very well in many situations.

We focus on the uLASSO problem \cref{e:CS_uLASSO}.  This unconstrained
setup is of significant interest, being related to convex quadratic
programming, for which many algorithms and solvers have already been
developed~\cite{VandenBerg2009a}. The uLASSO problem also shares a
close connection with Bayesian statistics, where the minimizer can be
interpreted to be the posterior mode corresponding to a likelihood
encompassing additive Gaussian noise on a linear model and a log-prior
with the $\ell_1$ form. Bayesian compressive sensing (BCS)
methods~\cite{Ji2008,Babacan2009} take advantage of this perspective,
and leverage Bayesian inference methods to explore the posterior
distribution that can be useful in assessing solution robustness
around the mode.  We do not investigate BCS in this paper, and focus
only on non-Bayesian approaches for the time being.  Various
algorithms have been developed to directly target \cref{e:CS_uLASSO}
relying on different mathematical principles, and they may perform
differently in practice depending on the problem structure. It is thus
valuable to investigate and compare them under problems and scenarios
we are interested in.  In this study, we perform numerical experiments
using several off-the-shelf solvers: \lls{}~\cite{Kim2007, Koh2008},
sparse reconstruction by separable approximation
(\sparsa{})~\cite{Wright2009, Wright2009a}, conjugate gradient
iterative shrinkage/thresholding (\cgist{})~\cite{Goldstein2010,
  Goldstein2011}, fixed point continuation with active set
(\fpcas{})~\cite{Wen2010, Yin2010}, and alternating direction method
of multipliers (\admm{})~\cite{Boyd2010, Boyd2011a}.

The \lls{} algorithm
transforms \cref{e:CS_uLASSO} to a convex quadratic problem with
linear inequality constraints. The resulting formulation is solved
using a primal interior-point method with logarithmic barrier
functions while invoking iterations of conjugate gradient (CG) or
truncated Newton.
\sparsa{}
takes advantage of a sequence of related optimization subproblems with
quadratic terms and diagonal Hessians, where their special structure
leads to rapid solutions and convergence.
\cgist{} is
based on forward-backward splitting with adaptive step size and
parallel tangent (partan) CG acceleration. The eventual
  stabilization of the active set from the splitting updates renders
  the problem quadratic, at which point partan CG produces
  convergence in a finite number of steps.
\fpcas{} alternates between two stages: establishing a working index set
using iterative shrinkage and line search, and solving a smooth
subproblem defined by the (usually lower dimensional) working index
set through a second-order method such as CG.
Finally, \admm{} combines techniques from dual descent and the method
of multipliers, and naturally decomposes the $\ell_1$ term from
$\ell_2$, requiring only iterations of small local subproblems where
analytical updates can be obtained from ridge regression and soft
thresholding formulas.

We further target the use of CS for polynomial chaos expansion (PCE)
construction. PCE is a spectral expansion for random variables, and
offers an inexpensive surrogate modeling alternative for representing
probabilistic input-output relationships. It is a valuable tool for
enabling computationally feasible uncertainty quantification (UQ)
analysis of expensive engineering and science applications
(e.g.,~\cite{Ghanem1991, Najm2009, Xiu2009, LeMaitre2010}). The number
of PCE basis terms grows drastically with the parameter dimension and
polynomial order, while the number of available model evaluations are
often few and limited due to high simulation costs. Consequently,
constructing \textit{sparse} PCEs is both desirable and necessary
especially for high-dimensional problems, and research efforts are
growing across several fronts to tackle this challenge.

In general, sampling efficiency is a major topic of interest, mainly
due to the potential computational expense of each isolated model
evaluation. Rauhut and Ward~\cite{Rauhut2012} demonstrated advantages
of Chebyshev sampling for Legendre PCEs while Hampton and
Doostan~\cite{Hampton2015} proposed coherence-optimal sampling for a
general orthonormal basis using Markov chain Monte Carlo.  Recently,
Jakeman, Narayan, and Zhou~\cite{Jakeman2017} illustrated the
advantages of sampling with respect to a weighted equilibrium measure
followed by solving a preconditioned $\ell_1$-minimization problem.
Fajraoui, Marelli, and Sudret~\cite{Fajraoui2017} also adopted linear
optimal experimental design practice to iteratively select samples
maximizing metrics based on the Fisher information matrix.  Another
strategy involves tailoring the objective function directly, such as
using weighted $\ell_1$-minimization~\cite{Peng2014} for more targeted
recovery of PCE coefficients that are often observed to decay in
physical phenomena.  Eldred et al.~\cite{Eldred2015} described a
different perspective that combined CS in a multifidelity framework,
which can achieve high overall computation efficiency by trading off
between accuracy and cost across different models.  All these
approaches, however, maintain a static set of basis functions (i.e.,
regressors or features). A promising avenue of advancement involves
adapting the basis dictionary iteratively based on specified
objectives. For example, works by Sargsyan et al.~\cite{Sargsyan2014}
and Jakeman, Eldred, and Sargsyan et al.~\cite{Jakeman2015}
incorporated the concept of ``front-tracking'', where the basis set is
iteratively pruned and enriched according to criteria reflecting the
importance of each term. In our study, we investigate different
numerical aspects of several CS solvers under standard sampling
strategies and fixed basis sets.

By promoting sparsity, CS is designed to reduce overfitting.  An
overfit solution is observed when the error on the training set (i.e.,
data used to define the underdetermined linear system) is very
different (much smaller) than error on a separate validation set, and
the use of a different training set could lead to entirely different
results. Such a solution has poor predictive capability and is thus
unreliable. However, CS is not always successful in preventing
overfitting, such as when emphases of fitting the data and
regularization are not properly balanced, or if there is simply too
little data.  In the context of sparse PCE fitting, Blatman and
Sudret~\cite{Blatman:2011} developed a LARS-based algorithm combined
with a model selection score utilizing leave-one-out cross-validation,
which helped \emph{both} to avoid overfitting \emph{and} to inform the
sampling strategy.  In this paper, we also explore approaches for
addressing these challenges while focusing on solvers aimed at the
uLASSO problem. Specifically, we use techniques to help improve the
mitigation of overfitting on two levels.  First, \textit{for a given
  set of data points}, we employ cross-validation (CV) error to
reflect the degree of overfitting of solutions obtained under
different $\lambda$, and choose the solution that minimizes CV
error. However, when sample size is too small, then the solutions
could be overfit no matter what $\lambda$ is used. A
minimally-informative sample size is problem-dependent, difficult to
determine a priori, and may be challenging to even define and detect
in real applications where noise and modeling errors are large.  We
provide a practical procedure to use information from existing data to
help guide decisions as to whether additional samples would be
worthwhile to obtain---i.e., \textit{a stop-sampling strategy}. While
previous work focused on rules based on the stabilization of solution
versus sample size~\cite{Malioutov2010}, we take a goal-oriented
approach to target overfitting, and devise a strategy using heuristics
based on CV error levels and their rates of improvement.

The main objectives and contributions of this paper are as follows.
\begin{itemize}
\item We conduct numerical investigations to compare several CS
  solvers that target the uLASSO problem \cref{e:CS_uLASSO}. The scope
  of study involves using solver implementations from the algorithm
  authors, and focusing on assessments of linear systems that emerge
  from PCE constructions. The solvers employ their default parameter
  settings.
\item We develop techniques to help mitigate overfitting through
  \begin{itemize}
  \item an automated selection of regularization constant $\lambda$
    based on CV error and
  \item a heuristic strategy to guide the stop-sampling decision.
  \end{itemize}

\item We demonstrate the overall methodology in a realistic
  engineering application, where a 24-dimensional PCE is constructed
  with expensive large eddy simulations of supersonic turbulent
  jet-in-crossflow (JXF). We examine performance in terms of recovery
  error and timing.
\end{itemize}

This paper is outlined as follows. \Cref{s:methodology} describes the
numerical methodology used to solve the overall CS
problem. \Cref{s:pce} provides a brief introduction to PCE, which is
the main form of linear systems we focus on. Numerical results on
different cases of increasing complexity are presented in
\cref{s:results}. The paper then ends with conclusions and future work
in \cref{s:conclusions}.

%% file: sections/methodology.tex
\section{Methodology}
\label{s:methodology}

We target the uLASSO problem \cref{e:CS_uLASSO} based on several
existing solvers, as outlined in \cref{sec:intro}.  Additionally, we
aim to reduce overfitting through two techniques: (1) selecting the
regularization parameter $\lambda$ via CV error, and (2) stop-sampling
strategy for efficient sparse recovery.

\subsection{Selecting $\lambda$ via cross-validation}

Consider \cref{e:CS_uLASSO} for a \textit{fixed} system $A$ and $b$
(and thus $m$ and $n$). The non-negative constant $\lambda$ is the
relative weight between the $\ell_2$ and $\ell_1$ terms, with the
first term reflecting how well training data is fit, and the latter
imposing sparsity via regularization. A large $\lambda$ heavily
penalizes nonzero terms of the solution vector, forcing them toward
zero (underfitting); a small $\lambda$ emphasizes fitting the training
data, and may lead to solutions that are not sparse and that
\textit{only} fit the training data but otherwise do not predict well
(overfitting). A useful solution thus requires an intricate choice of
$\lambda$, which is a problem-dependent and nontrivial task.

The selection of $\lambda$ can be viewed as a model selection problem,
where different models are parameterized by $\lambda$. For example,
when a Bayesian perspective is adopted, the Bayes factor
(e.g.,~\cite{Kass1995, Wasserman2000}) is a rigorous criterion for
model selection, but it generally does not have closed forms for
non-Gaussian (e.g., $\ell_1$ form) priors on these linear
systems. Quantities simplified from the Bayes factor, such as the
Akaike information criterion (AIC)~\cite{Akaike1974} and the Bayesian
information criterion (BIC)~\cite{Schwarz1978}, further reduce to
formulas involving the maximum likelihood, parameter dimension, and
data sample size. In fact, a fully Bayesian approach would assimilate
the model-selection problem into the inference procedure altogether,
and treat the parameter $\lambda$ (equivalent to the ratio between
prior and likelihood ``standard deviations'') as a hyper-parameter
that would be inferred from data as well.

In this study, we utilize a criterion that more directly reflects and
addresses our concern of overfitting: the CV error
(e.g., see Chapter 7 in~\cite{Hastie2009}), in particular
the $K$-fold CV error. The procedure involves first partitioning the
full set of $m$ training points into $K$ equal (or approximately
equal) subsets.  For each of the subsets, a reduced version of the
original CS problem is solved:
\begin{align}
  x_{[\sim k]}(\lambda) = \argmin_x \norm{A_{[\sim k]}x - y_{[\sim
        k]}}{2}^2 + \lambda \norm{x}{1}, \label{e:CS_uLASSO_reduced}
\end{align}
where $A_{[\sim k]}$ denotes $A$ but with rows corresponding the $k$th
subset removed, $y_{[\sim k]}$ is $y$ with elements corresponding to
the $k$th subset removed, and $x_{[\sim k]}(\lambda)$ is the solution
vector from solving this reduced CS problem. The $\ell_2$ residual
from validation using the $k$th subset that was left out is therefore
\begin{align}
  R_{[k]}(\lambda) \equiv \norm{A_{[k]} x_{[\sim
        k]}(\lambda)-y_{[k]}}{2},
  \label{e:CV_residual}
\end{align}
where $A_{[k]}$ denotes $A$ that only contains rows corresponding to
the $k$th subset, and $y_{[k]}$ is $y$ containing only elements
corresponding to the $k$th subset.  Combining the residuals from all
$k$ subsets, we arrive at the (normalized) $K$-fold CV error:
\begin{align}
  E_{\textnormal{CV}}(\lambda) \equiv
    \frac{\sqrt{\sum_{k=1}^{K}\[R_{[k]}(\lambda)\]^2}}{\norm{y}{2}}.
  \label{e:CV_kfold}
\end{align}
The CV error thus provides an estimate of the validation error using
only the training data set at hand and without needing additional
validation points, and reflects the predictive capability of solutions
generated by a given solver.  The CS problem with $\lambda$ selection
through CV error is thus
\begin{align}
  &\min_x \norm{Ax - y}{2}^2 + \lambda^{\ast}
  \norm{x}{1}, \label{e:CS_uLASSO_opt}\\
  &\textnormal{where}\hspace{0.5em} \lambda^{\ast} = \argmin_{\lambda\ge 0}
  E_{\textnormal{CV}}(\lambda)
      \label{e:CV_opt}
\end{align}
Note that solving \cref{e:CV_opt} does not require the solution from
the full CS system, only the $K$ reduced systems.

In practice, solutions to \cref{e:CS_uLASSO_reduced,e:CS_uLASSO_opt}
are evaluated numerically and approximately, and so, not only do they
depend on the problem (i.e., the linear system, size of $m$ and $n$,
etc.) but also on the numerical solver.  Hence, $x_{[\sim k]}(\lambda,
S)$ should also depend on the solver $S$ (this encompasses the
algorithm, implementation, solver parameters and tolerances, etc.),
and subsequently $\lambda^{\ast}(S)$ and $E_{\textnormal{CV}}(\lambda,
S)$. The numerical results presented later on will involve numerical
experiments comparing several different CS solvers.

For simplicity, we solve the minimization problem \cref{e:CV_opt} for
$\lambda^{\ast}$ using a simple grid-search across a discretized
log-$\lambda$ space.  More sophisticated optimization methods, such as
bisection, linear-programming, and gradient-based approaches, are
certainly possible. One might be tempted to take advantage of the fact
that the solution to \cref{e:CS_uLASSO} is piecewise linear in
$\lambda$~\cite{Efron2004}. However, the CV error, especially when
numerical solvers are involved, generally no longer enjoys such
guarantees.  The gain in optimization efficiency would also be small
for this one-dimensional optimization problem, and given the much more
expensive application simulations and other computational
components. Therefore, we do not pursue more sophisticated search
techniques.

\subsection{Stop-sampling strategy for sparse recovery}
\label{ss:stop_sampling}

When sample size is small, the solutions could be overfit no matter
what $\lambda$ is used. Indeed, past work in precise undersampling
theorems~\cite{Donoho2010}, phase-transition
diagrams~\cite{Donoho2009}, and various numerical experiments suggest
drastic improvement in the probability of successful sparse
reconstruction when a critical sample size is reached.  However, this
critical quantity is dependent on the true solution sparsity,
correlation structure of matrix $A$, and numerical methods being used
(e.g., CS solver), and therefore prediction a priori is difficult,
especially in the presence of noise and modeling error.  Instead, we
introduce a heuristic procedure using CV error trends from currently
available data to reflect the rate of improvement, and to guide
decisions of whether additional samples would be worthwhile to obtain.

\begin{table}[htb]
\caption{Summary of parameters relevant for stop-sampling described in
  this section.}
\label{t:my_alg_params}
\begin{center}
\begin{tabular}{cl}
\hline
Parameter &  Description \\
\hline
$m_0$ & Initial sample size \\
$\Delta m$ & Sample increment size \\
$m$ & Total sample size thus far \\
$m_p$ & Available parallel batch size for data gathering\\
$n$ & Number of basis functions (regressors) \\
$q$ & Moving window size for estimating log-CV error slope\\
$\eta$ & Tolerance for log-CV error slope rebound fraction\\
$r$ & Slope criterion activation tolerance\\
$a$ & Tolerance for absolute CV error\\
$K$ & Number of folds in the $K$-fold CV\\
$J$ & Number of grid points in discretizing $\lambda$\\
$\lambda_j$ & Discretized grid points of $\lambda$\\
$\lambda^{\ast}$ &  $\lambda$ value that produces the lowest CV
error\\
$S$ & Variable reflecting the choice of CS solver\\
\hline
\end{tabular}
\end{center}
\end{table}

Before we describe our procedure, we first summarize a list of
parameters in \cref{t:my_alg_params} that are relevant for
stop-sampling and the eventual algorithm we propose in this section.
We start with some initial small sample size $m_0$, and a decision is
made whether to obtain an additional batch of $\Delta m$ new samples
or to stop sampling. In the spirit of using only currently available
data, we base the decision criteria on the $\lambda$-optimal CV error,
$E_{\textnormal{CV}, \lambda^{\ast}}(m) \equiv
E_{\textnormal{CV}}(\lambda^{\ast})$ for sample size $m$. More
specifically, multiple stop-sampling criteria are evaluated.  Our
primary criterion is the slope of $\log E_{\textnormal{CV},
  \lambda^{\ast}}(m)$ with respect to $m$, as an effective error decay
rate. A simple approximation involves using a moving window of the
past $q$ values of $\log E_{\textnormal{CV}, \lambda^{\ast}}(m)$ and
estimating its slope through ordinary least squares. We stop sampling
when the current slope estimate rebounds to a certain fraction $\eta$
of the steepest slope estimate encountered so far, thus indicates
crossing the critical sample size of sharp performance improvement.
Since the samples are iteratively appended, its nestedness helps
produce smoother $E_{\textnormal{CV}, \lambda^{\ast}}(m)$ than if
entirely new sample sets are generated at each $m$. Nonetheless,
oscillation may still be present from numerical computations, and
larger choices of $\Delta m$ and $q$ can further help with smoothing,
but at the cost of lower resolution on $m$ and increased influence of
non-local behavior. To guard against premature stop-sampling due to
oscillations, we activate the slope criterion only after
$E_{\textnormal{CV}, \lambda^{\ast}}(m)$ drops below a threshold of
$r$.  Additionally, we also stop sampling if the value of
$E_{\textnormal{CV}, \lambda^{\ast}}(m)$ drops below some absolute
tolerance $a$. This is useful for cases where the drop occurs for very
small $m$ and is not captured from the starting $m_0$.

\subsection{Overall method and implementation}

The pseudocode for the overall method is presented in
\cref{a:CS_alg_overall}, and parameter descriptions can be found in
\cref{t:my_alg_params}. We provide some heuristic guidelines below for
setting its parameters. These are also the settings used for numerical
examples in this paper.
\begin{itemize}
\item For the $K$ in $K$-fold CV error, a small $K$ tends toward low
  variance and high bias, while a large $K$ tends toward low bias and
  high variance as well as higher computational costs since it needs
  to solve more instances of the reduced problem~\cite{Hastie2009}.
  For problem sizes encountered in this paper, $K$ between 20-50
  appears to work well. We revert to leave-one-out CV when $K\ge m$.

\item A reasonable choice of $m_0$ is 5\% of $n$, and together with
  $\Delta m=m_0$, corresponds to a growing uniform grid of 20 nodes
  over $m\in[m_0,n]$. In practice, sample acquisition is expected to
  be much more expensive than CS solves. We then recommend adopting
  the finest resolution $m_0=\Delta m = m_p$ where $m_p$ is the
  available parallel batch size for data gathering (e.g., $m_p=1$ for
  serial setups).

\item Stop-sampling parameters $q=4$, $\eta=0.1$, $r=0.5$, $a=10^{-4}$
  are a good starting point. However, $\eta=0.1$ and $r=0.5$ are a
  somewhat conservative combination (i.e., less likely to stop
  sampling prematurely, and more likely to arrive at larger sample
  sizes).  More relaxed values (larger $\eta$ and smaller $r$) may be
  used especially for more difficult problems where noise and modeling
  error are present, and where the CV error may not exhibit a sharp
  dropoff versus $m$.

\item The selection of $J$ and $\lambda_j$ should cover a good range
  in the logarithmic scale, such as from $10^{-4}$ to $10^4$ through
  15 log-spaced points. The upper bound can also be set to the
  theoretical $\lambda_{max} = 2\norm{A^{T}y}{\infty}$ that starts to
  produce the zero solution.  Experience also indicates cases with
  small values of $\lambda$ take substantially longer for the CS
  solvers to return a solution. We thus set $\lambda$ ranges in our
  numerical experiments based on some initial trial-and-error. This
  could be improved with grid adaptation techniques so that the
  $\lambda$-grid can extend beyond initial bounds when not wide enough
  to capture $\lambda^{\ast}$, and also to refine its resolution to
  zoom in on $\lambda^{\ast}$.

\end{itemize}

\begin{algorithm}[htb]
  \caption{Pseudocode for our method using (1) selecting the
    regularization parameter $\lambda$ via CV error, and (2)
    stop-sampling strategy for sparse recovery. Parameter descriptions
    can be found in \cref{t:my_alg_params}.}
  \label{a:CS_alg_overall}
  \begin{algorithmic}[1]
    \STATE{\textbf{Set parameters:} choose CS solver $S$ and set its
      corresponding algorithm parameters; select CV fold size $K$;
      initial sample size $m_0$; sample increment size $\Delta m$;
      stop-sampling parameters $q$, $\eta$, $r$, and $a$; and $J$ grid
      points $\left\{\lambda_j\right\}$}
    \STATE{$m = m_0$}
    \WHILE{stop-sampling conditions not met (see text in
      \cref{ss:stop_sampling})}
    \FOR{$j=1,\ldots,J$}
    \FOR{$k=1,\ldots,K$}
    \STATE{Solve reduced CS problem \cref{e:CS_uLASSO_reduced} using
      solver $S$ to obtain $x_{[\sim k]}(\lambda_j, S)$}
    \ENDFOR
    \STATE{Evaluate $K$-fold CV error \cref{e:CV_kfold} to
      obtain $E_{\textnormal{CV}}(\lambda_j, S)$}
    \ENDFOR
    \STATE{Select $\lambda^{\ast}(S)$ from $\left\{\lambda_j\right\}$ that
      produced the smallest $E_{\textnormal{CV}}$, denote it as
      $E_{\textnormal{CV}}^{\ast}(m, S)$}
    \STATE{Evaluate stop-sampling conditions using
      $E_{\textnormal{CV}}^{\ast}(m, S)$}
    \IF{stop-sampling conditions met}
    \STATE{Solve the full CS problem \cref{e:CS_uLASSO_opt} using
       $\lambda^{\ast}(S)$; this is the final solution}
    \ELSE
    \STATE{$m = m + \Delta m$}
    \ENDIF
  \ENDWHILE
  \end{algorithmic}
\end{algorithm}

%% file: sections/pce.tex
\section{Polynomial chaos expansion (PCE)}
\label{s:pce}

We now introduce the CS problem emerging from constructing sparse
PCEs.
PCE offers an inexpensive surrogate modeling alternative for
representing probabilistic input and output relationships, and is a
valuable tool for enabling computationally-feasible UQ analysis of
expensive engineering and science applications.  As we shall see
below, the number of columns, $n$, in these PCE-induced linear systems
becomes very large under high-dimensional and nonlinear settings.
Therefore, it is important to find \textit{sparse} PCEs, and CS
provides a useful framework under which this is possible.
We present a brief description of the PCE construction below, and
refer readers to several references for further detailed
discussions~\cite{Ghanem1991, Najm2009, Xiu2009, LeMaitre2010}.

A PCE for a real-valued, finite-variance random vector $\theta$ can be
expressed in the form~\cite{Ernst2012}
\begin{align}
  \theta = \sum_{\norm{\beta}{1}=0}^{\infty} \theta_{\beta}
  \Psi_{\beta}(\xi_1, \ldots, \xi_{n_s}),
  \label{e:theta_PCE_multivariate}
\end{align}
where $\theta_{\beta}$ are the expansion coefficients,
$\beta=\(\beta_1,\ldots,\beta_{n_s}\),\,\forall \beta_j\in\NN_0$, is a
multi-index, $n_s$ is the stochastic dimension (often convenient to be
set equal to the number of uncertain model inputs), $\xi_j$ are a
chosen set of independent random variables, and
$\Psi_{\beta}(\xi_1,\ldots,\xi_{n_s})$ are multivariate polynomials of
the product form
\begin{align}
  \Psi_{\beta}(\xi_1,\ldots, \xi_{n_s}) = \prod_{j=1}^{n_s}
  \psi_{\beta_j}(\xi_j),
\end{align}
with $\psi_{\beta_j}$ being degree-$\beta_j$ polynomials orthonormal
with respect to the probability density function of $\xi_j$ (i.e.,
$p\(\xi_j\)$):
\begin{align}
  \EE\[\psi_k(\xi_j)\psi_n(\xi_j)\] = \int_{\Xi_j}
  \psi_k\(\xi_j\)\psi_n\(\xi_j\)p\(\xi_j\)\,d\xi_j = \delta_{k,n}.
  \label{e:orthogonality}
\end{align}
Different choices of $\xi_j$ and $\psi_{\beta_j}$ are available under
the generalized Askey family~\cite{Xiu2002}. Two most commonly used
PCE forms are the Legendre PCE with uniform $\xi_j\sim \CU(-1,1)$, and
Hermite PCE with Gaussian $\xi_j\sim\CN(0,1)$.
For computational purposes, the infinite sum in the expansion
\cref{e:theta_PCE_multivariate} must be truncated:
\begin{align}
  \theta \approx \sum_{\beta \in \CJ} \theta_{\beta}
  \Psi_{\beta}(\xi_1,\ldots,\xi_{n_s}),
  \label{e:theta_PCE_truncate}
\end{align}
where $\CJ$ is some finite index set.  For simplicity, we focus only
on ``total-order'' expansion of degree $p$ in this study, where
$\CJ=\{\beta : \norm{\beta}{1} \leq p \}$, containing a total of
$\frac{\(n_s+p\)!}{n_s!p!}$ basis terms.

A major element of UQ involves the forward propagation of uncertainty
through a model, typically involving characterizing some uncertain
model output quantity of interest (QoI) $f(\theta)$ given an uncertain
model input $\theta$. PCE offers a convenient forum to accomplish
this, and the QoI can be approximated by a truncated PCE
\begin{align}
  f \approx \sum_{\beta \in \CJ} c_{\beta}
  \Psi_{\beta}(\xi_1,\ldots,\xi_{n_s}).
  \label{e:QoI_PCE}
\end{align}
Methods for computing the coefficients $c_{\beta}$ are broadly divided
into two groups---intrusive and non-intrusive. The former involves
substituting the expansions into the governing equations of the model,
resulting in a new, usually larger, system that needs to be solved
only once. The latter encompasses finding an approximation in the
subspace spanned by the basis functions, which typically requires
evaluating the original model many times under different input
values. Since we often encounter models that are highly complicated
and only available as a black-box, we focus on the non-intrusive
route.

One such non-intrusive method relies on Galerkin projection of the
solution, known as the non-intrusive spectral projection (NISP)
method:
\begin{align}
  c_{\beta} = \EE\[f(\theta)
    \Psi_{\beta}\] = \int_{\Xi}
    f\(\theta(\xi)\) \Psi_{\beta}(\xi) p(\xi)
    \,d\xi.
    \label{e:NISP}
\end{align}
Generally, the integral must be estimated numerically and
approximately via, for example, sparse
quadrature~\cite{Barthelmann2000, Gerstner1998, Gerstner2003}. When
the dimension of $\xi$ is high, the model is expensive, and only few
evaluations are available, however, even sparse quadrature becomes
impractical. In such situations, regression is a more effective
method. It involves solving the following regression linear system $Ax
= y$:
\begin{align}
  \underbrace{\begin{bmatrix} \Psi_{\beta^1}(\xi^{(1)}) & \cdots &
  \Psi_{\beta^n}(\xi^{(1)}) \\ \vdots & & \vdots \\
  \Psi_{\beta^1}(\xi^{(m)}) & \cdots &
  \Psi_{\beta^n}(\xi^{(m)}) \end{bmatrix}}_{A}
  \underbrace{\begin{bmatrix} c_{\beta^1} \\  \vdots
  \\ c_{\beta^n} \end{bmatrix}}_{x} =
  \underbrace{\begin{bmatrix} f(\theta(\xi^{(1)})) \\  \vdots
  \\ f(\theta(\xi^{(m)})) \end{bmatrix}}_{y},\label{e:sparse_regression}
\end{align}
where the notation $\Psi_{\beta^n}$ refers to the $n$th basis
function, $c_{\beta^n}$ is the coefficient corresponding to that
basis, and $\xi^{(m)}$ is the $m$th regression point. Common practice
for constructing PCEs from regression involves eliminating the mean
(constant) term from the $A$ matrix and correspondingly centering the
$y$ vector.  $A$ is thus the regression matrix where each column
corresponds to a basis (except the constant basis term) and each row
corresponds to a regression point. The number of columns $n$ can
easily become quite large in high-dimensional settings; for example, a
total-order expansion of degree 3 in 24 dimensions contains
$n=\frac{(3+24)!}{3!24!} - 1= 2924$ terms. When each sample is an
expensive physical model simulation, the number of runs $m$ that can
be afforded may be much smaller than $n$. At the same time, PCEs
describing physical phenomena are often observed to be sparse where
responses are dominated by only a subset of inputs.  CS thus provides
a natural means to discover the sparse structure in PCE by finding a
sparse solution for the underdetermined system in
\cref{e:sparse_regression}.

%% file: sections/results.tex
\section{Numerical examples}
\label{s:results}

We perform numerical investigations on several test cases of
increasing complexity. The following MATLAB implementations of CS
solvers are used within \cref{a:CS_alg_overall}:
\lls{}~\cite{Koh2008}, \sparsa{}~\cite{Wright2009a},
\cgist{}~\cite{Goldstein2011}, \fpcas{}~\cite{Yin2010}, and
\admm{}~\cite{Boyd2011a}.  We do not tune the algorithm parameters for
practical reasons, since we expect to encounter millions of CS solves
across a wide range of problem sizes and sparsity, and the optimal
setting certainly would vary. Instead, we adopt default algorithm
settings provided by the solver authors. A summary of relevant
algorithm parameters can be found in \cref{app:default_params}.
Several error quantities are used in our results when available:
\begin{align}
  & \textnormal{Training error} &
   \frac{\norm{Ax-y}{2}}{\norm{y}{2}}\\
  & \textnormal{Cross-validation error} &
   \textnormal{from \cref{e:CV_kfold}}\\
  & \textnormal{Validation error} &
   \frac{\norm{A_{V}x-y_{V}}{2}}{\norm{y_V}{2}} \\
  & \textnormal{Solution error} &
   \frac{\norm{x-x^{\ast}}{2}}{\norm{x^{\ast}}{2}}.
\end{align}
All of the above are normalized quantities.  For the validation error,
$A_V$ is formed from a separate (external) validation data set, and
$y_V$ is its corresponding data vector generated from the same
measurement process. For expensive applications, only the training and
CV errors are available in practice (validation is possible in
principle but likely too expensive, and true solution would not be
known). However, validation and solution errors are available for our
synthetic test cases when the simulation is inexpensive and when the
true solution is known. One may interpret the solution and validation
errors as the true errors for assessing recovery and prediction
performance, respectively, the training error as a data-fitting
indicator, and CV as an approximation to the validation error using
only available training data. All of the following numerical results
employ $K=50$ folds for CV, except for the phase-transition diagrams
which use a smaller $K=20$ for faster computations.

\subsection{Case 1: Gaussian random matrix}

The first example is a Gaussian random matrix; it is not a PCE. This
example is often used as a benchmark in CS studies to verify
theoretical analysis.  In this system, the elements in $A$ are drawn
from independent and identically distributed (i.i.d.) standard normal
$\CN(0,1)$, and a true solution $x^{\ast}$ is constructed to have $s$
randomly selected nonzero elements with values drawn from i.i.d.
uniform $\CU(-1,1)$.

\begin{figure}[htb]
  \centering
      \includegraphics[width=0.32\textwidth, trim={1em 0 4.5em 2em}, clip]{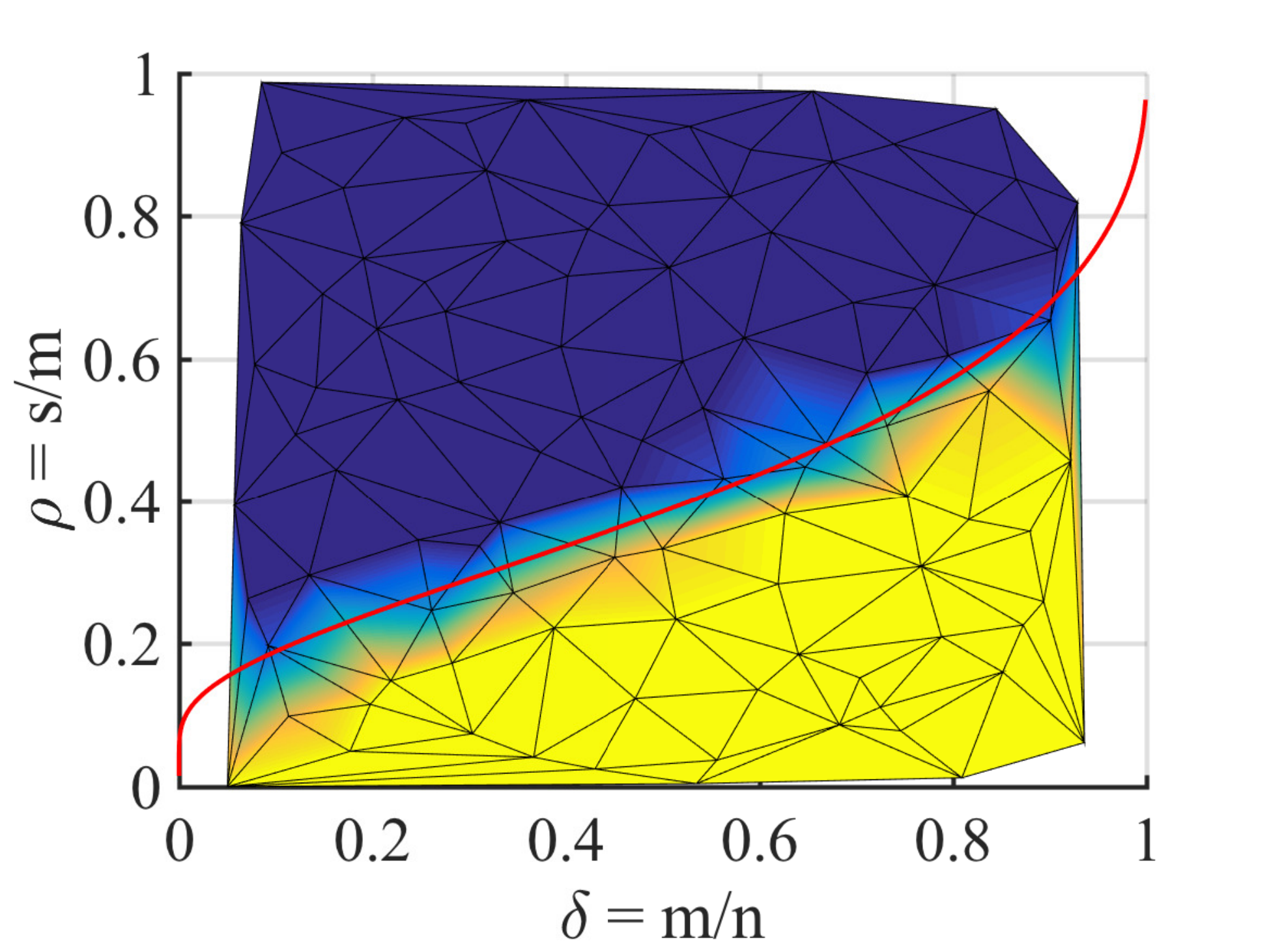}
  \includegraphics[width=0.32\textwidth, trim={1em 0 4.5em 2em}, clip]{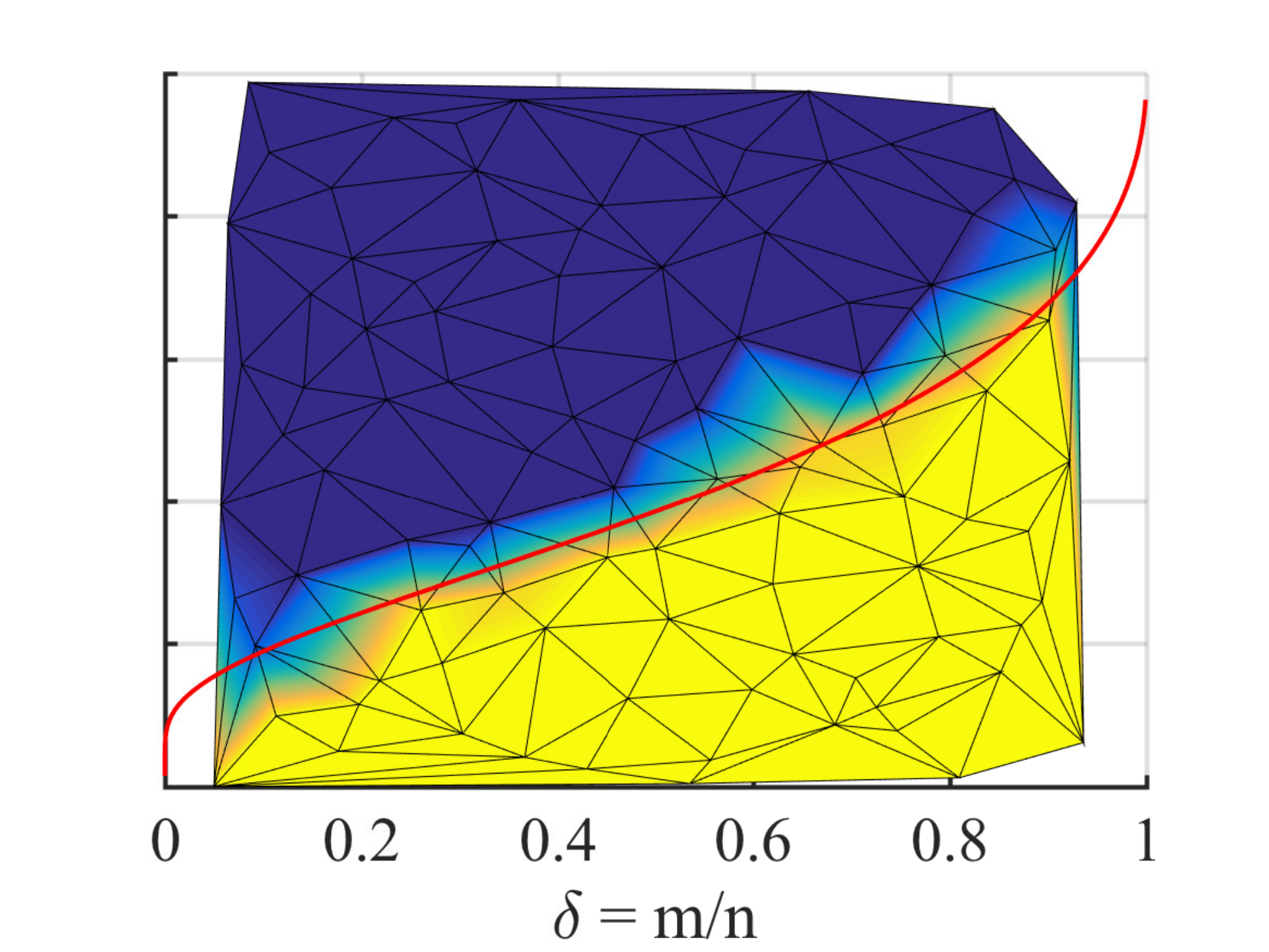}
  \includegraphics[width=0.32\textwidth, trim={1em 0 4.5em 2em}, clip]{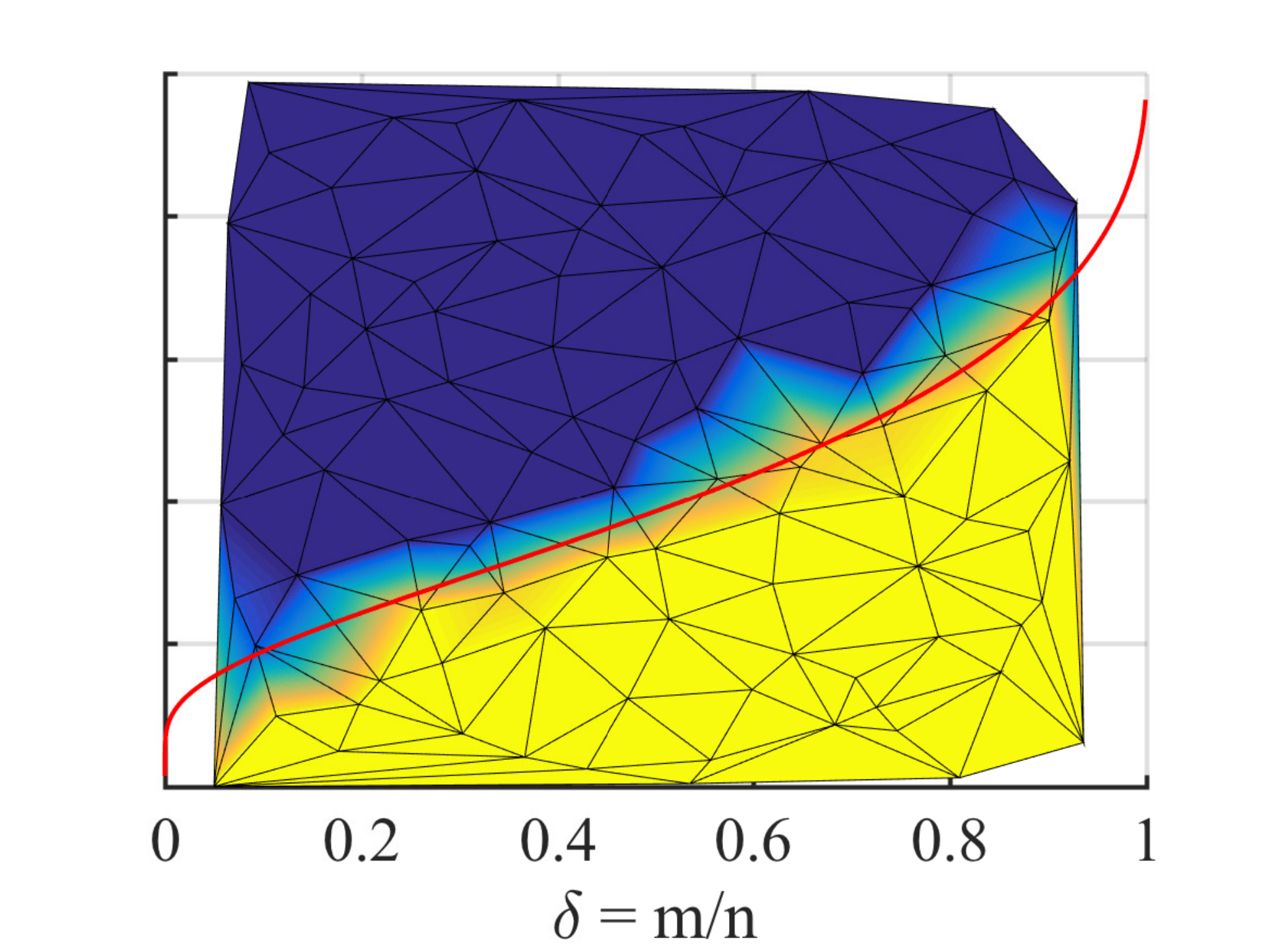}\\
  \includegraphics[width=0.95\textwidth, trim={3em 0 0 43em}, clip]{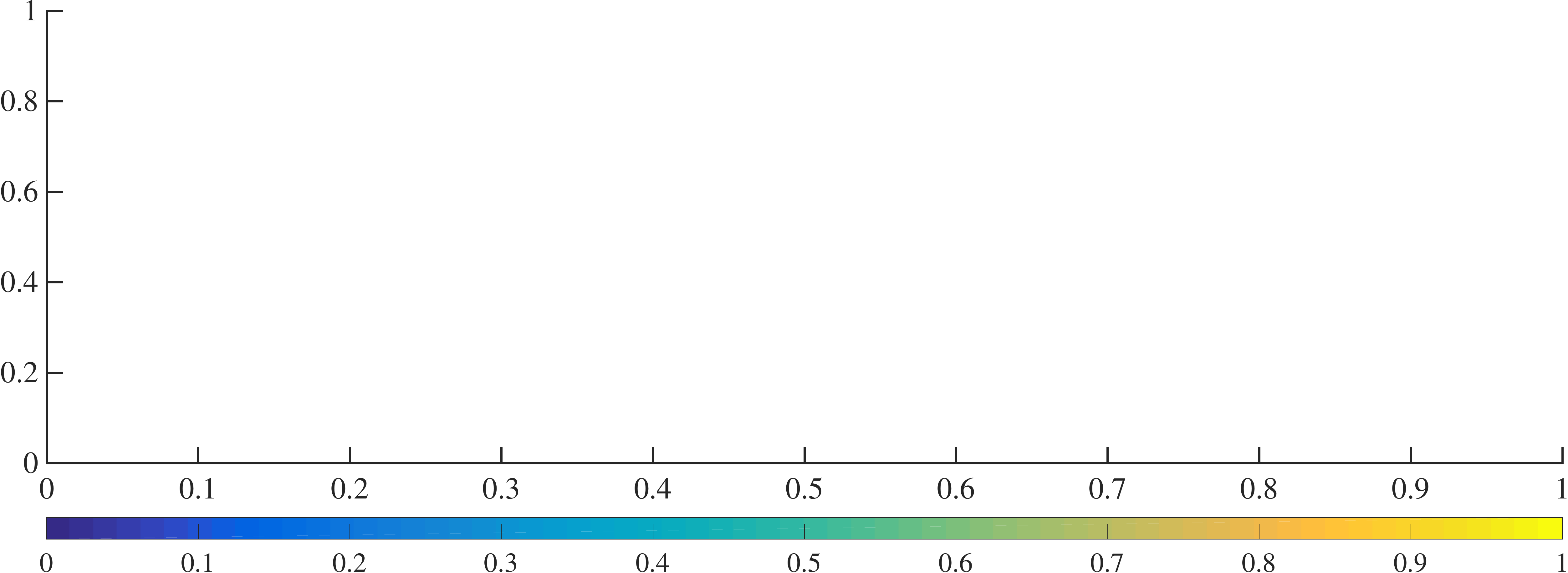}
  \caption{Case 1 Gaussian random matrix: phase-transition diagrams
    plotted from CV (left), validation (middle), and solution (right)
    errors where a success recovery is defined by when the respective
    error quantity is less than $0.1$, using \admm{}. Plots using
    \lls{}, \sparsa{}, \cgist{}, and \fpcas{} are very similar to the
    \admm{} results, and thus omitted to avoid repetition.}
              \label{f:case1_PTDiagrams}
\end{figure}

We first perform a sanity check, and ensure our setup is consistent
with observations from previous research studies.  To do this, we
construct phase-transition diagrams, solving the CS problem with
different undersampling ratios $\delta = m/n$ and sparsity ratios
$\rho = s/m$ (recall that $m$ is the number of rows of matrix $A$, $n$
is the number of columns, and $s$ is the number of nonzero elements in
the true solution vector). For a given combination of $(\delta,
\rho)$, this exercise is repeated $b$ times where each trial has a
newly generated $A$ and $x^{\ast}$. Donoho and
Tanner~\cite{Donoho2009} described the phase-transition behavior with
a geometric interpretation, where the likelihood of a sparse recovery
rapidly drops when a threshold is crossed in the $\delta$-$\rho$
space. This observation is also shown to exhibit universality for
various ensembles of $A$, and was formalized further through the
precise undersampling theorems~\cite{Donoho2010}.
We build these diagrams from numerical experiments. The number of
columns of $A$ is fixed at $n=500$, and $b=10$ repeats are performed
at each $(\delta,\rho)$.  Instead of a traditional structured
discretization of $\delta$ and $\rho$, we employ a quasi Monte Carlo
(QMC) sampling technique using Halton sequences~\cite{Halton1960} and
produce an unstructured grid that can characterize the transition
cliff with fewer points. \Cref{f:case1_PTDiagrams} shows the
phase-transition diagrams using \admm{},
and plotted based on assessing CV (left), validation (middle), and
solution (right) errors. We define a successful reconstruction if the
error quantity of interest from a run is less than $0.1$, and the
diagram plots the empirical success rate from the $b=10$ repeated
trials at each $(\delta,\rho)$ node.  The choice of error threshold
value depends on the solver and its algorithm settings. We selected a
value, based on trial-and-error, that produced reasonably well defined
transition boundaries. Values too high or too low can lead to diagrams
that have success or failure over the entire $\delta$-$\rho$ space.
The diagrams are overlayed with the theoretical transition curve using
tabulated data from Tanner's website~\cite{Tanner2012}, and excellent
agreement is observed. Overall, the CV error results are
representative of those from the validation and solution errors. Plots
using \lls{}, \sparsa{}, \cgist{}, and \fpcas{} are very similar to
the \admm{} results, and are thus omitted to avoid repetition.

\begin{figure}[htb]
  \centering
  \includegraphics[width=0.32\textwidth]{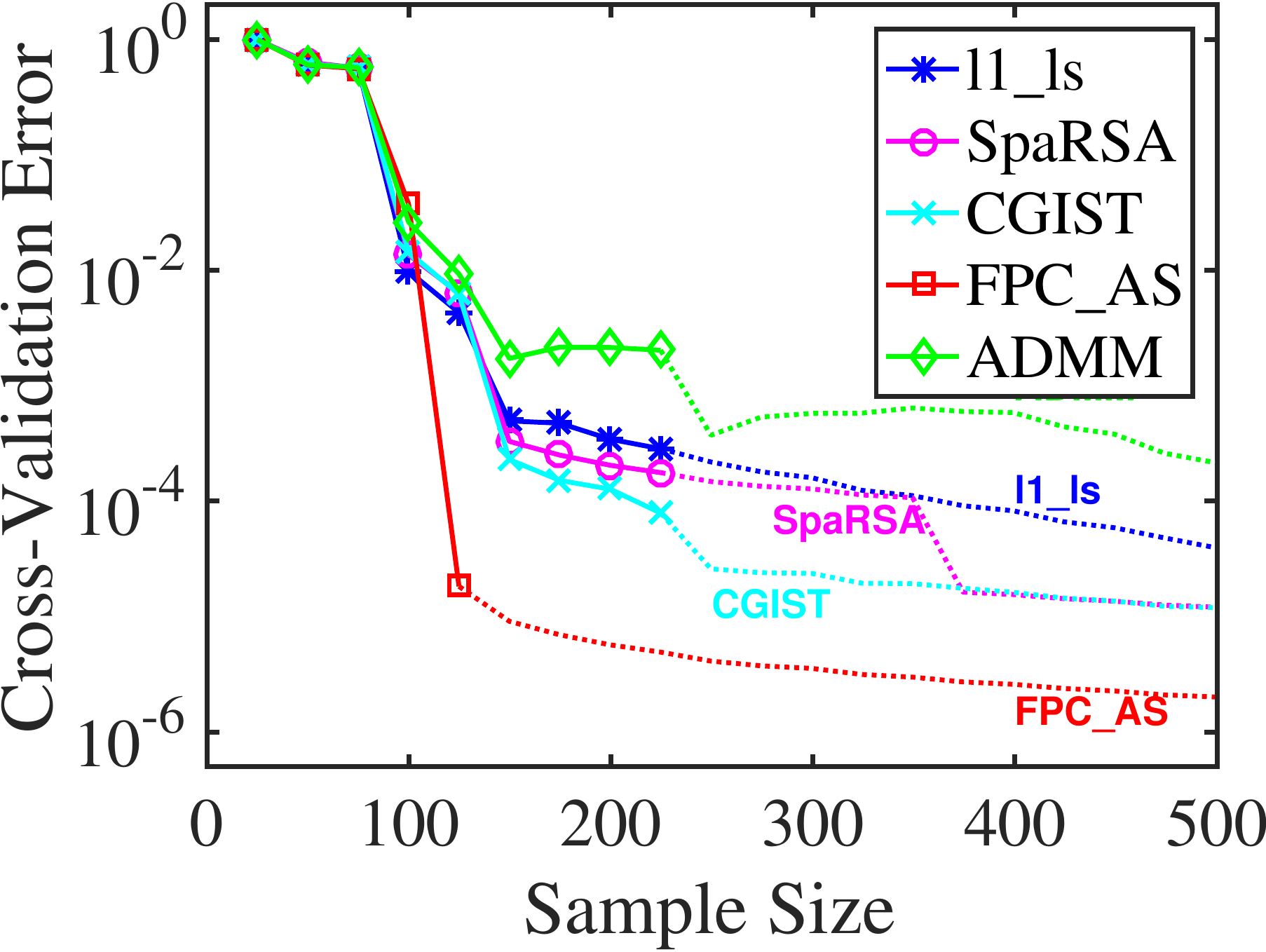}
  \includegraphics[width=0.32\textwidth]{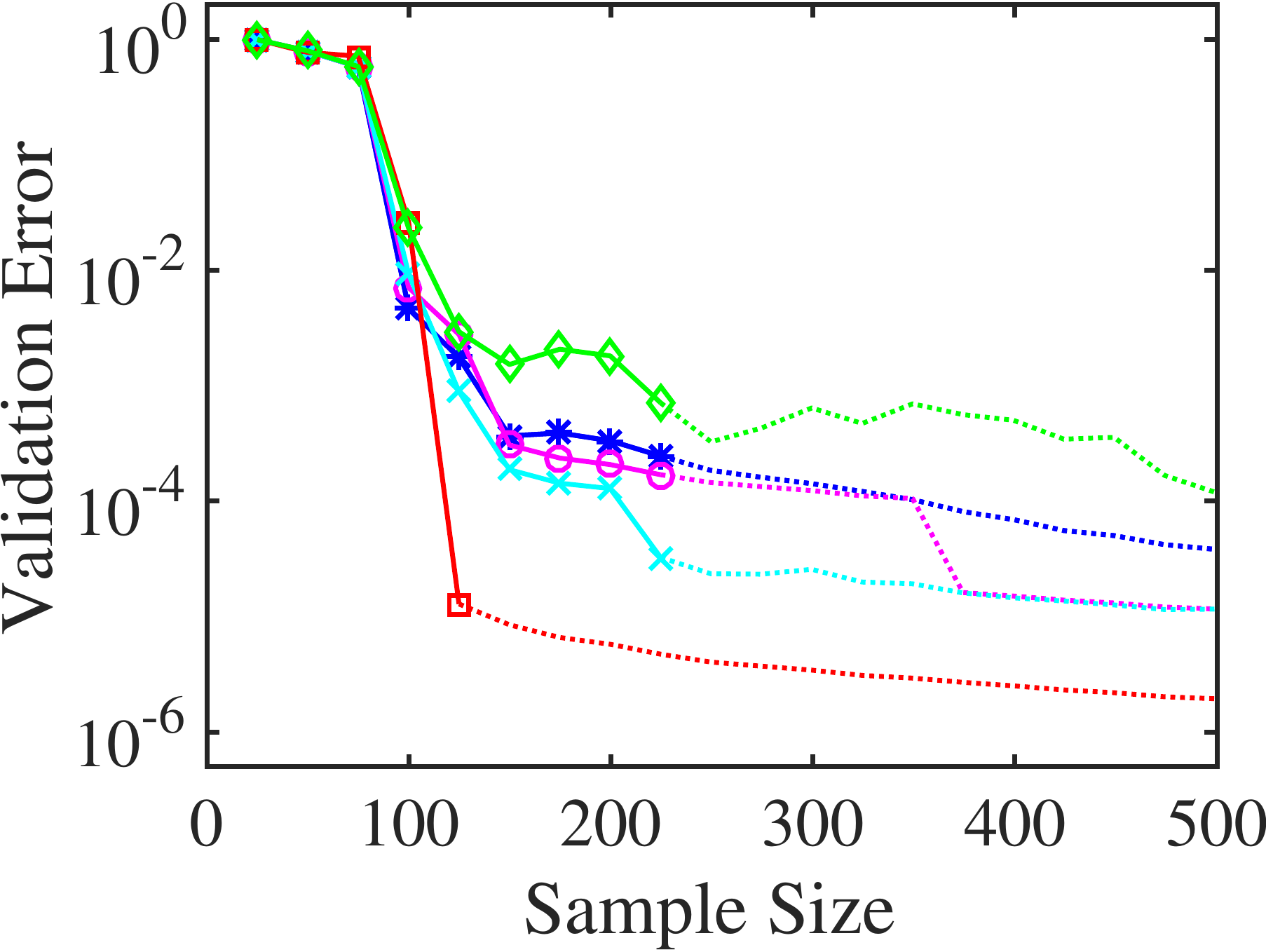}
  \includegraphics[width=0.32\textwidth]{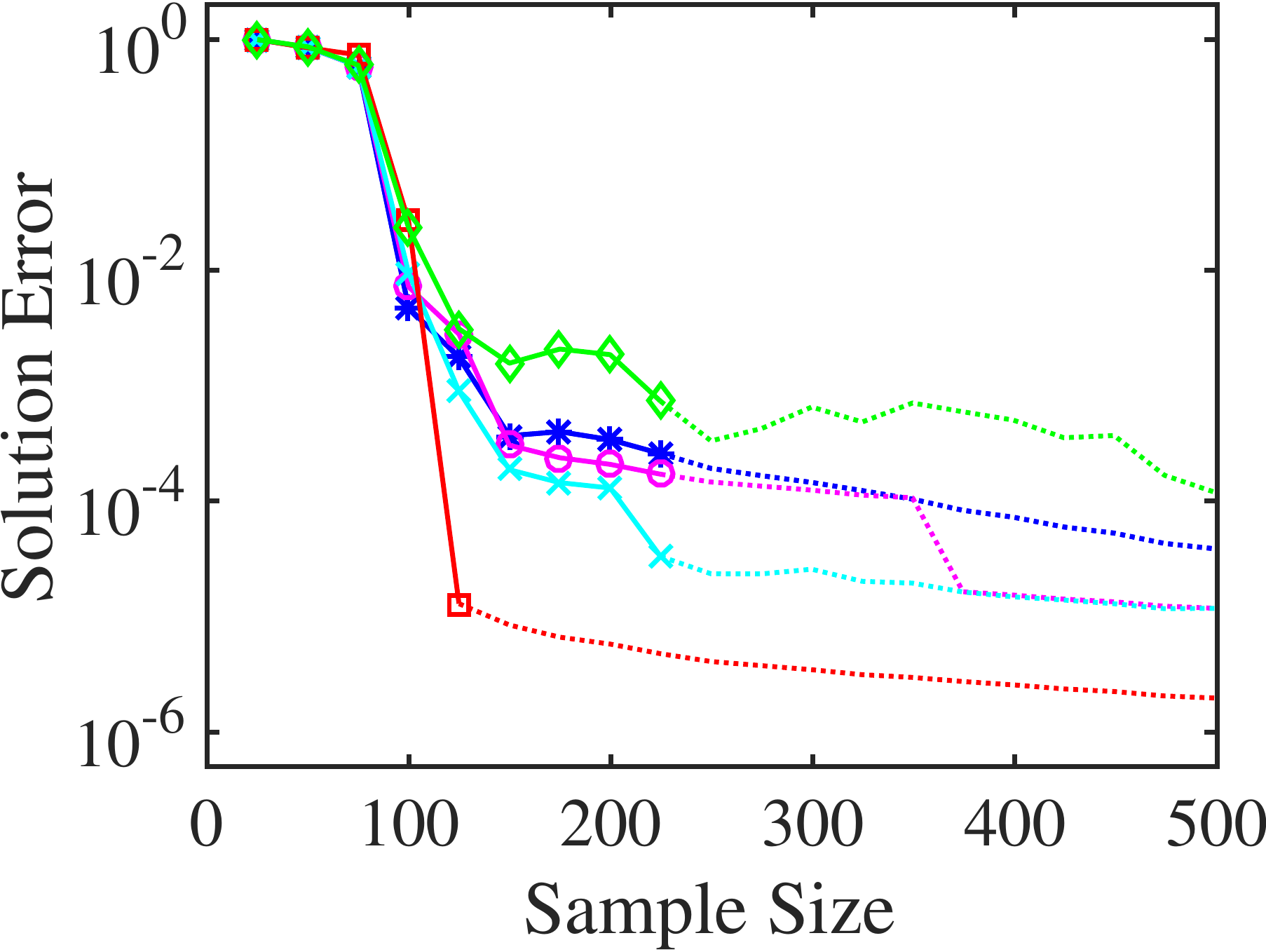}
  \caption{Case 1 Gaussian random matrix: CV (left), validation
    (middle), and solution (right) errors for fixed problem instance
    with $n=500$ and $s=25$. Stop-sampling is activated when the plot
    line turns from solid (with symbols) to dotted (without symbols).}
  \label{f:case1_single}
\end{figure}

We now exercise \cref{a:CS_alg_overall} to a fixed problem instance
that has a sparse solution with $s=25$ nonzero entries.
\Cref{f:case1_single} plots the CV, validation, and solution errors
versus $m$.  Both \cref{f:case1_single} and a horizontal slice of
\cref{f:case1_PTDiagrams} display the sharp performance change when
the critical sample size is crossed, but there are some differences
between the two: the former directly plots errors while the latter
plots success rates, and the data sets are nested for different $m$ in
the former while independently regenerated for each trial in the
latter.  In \cref{f:case1_single}, the error lines for all solvers
share similar behavior, with \fpcas{} yielding lower errors compared
to others.  All solvers display a drastic drop in errors at around 100
sample points, which corresponds to the ``cliff'' in the
phase-transition diagrams---this would be the sample size we want to
detect and use for efficient sparse recovery.  The change of plot
lines from solid (with symbol) to dotted (with no symbol) indicates
the stop-sampling point when our algorithm is applied. Overall, they
work fairly well and stop sampling in regions around just past the
bottom of the sharp drop.  Finally, we note that CV errors agree very
well with the solution and validation error trends.

\begin{figure}[htb]
  \centering
  \includegraphics[width=0.32\textwidth, trim={0 6em -0.5em 0}, clip]{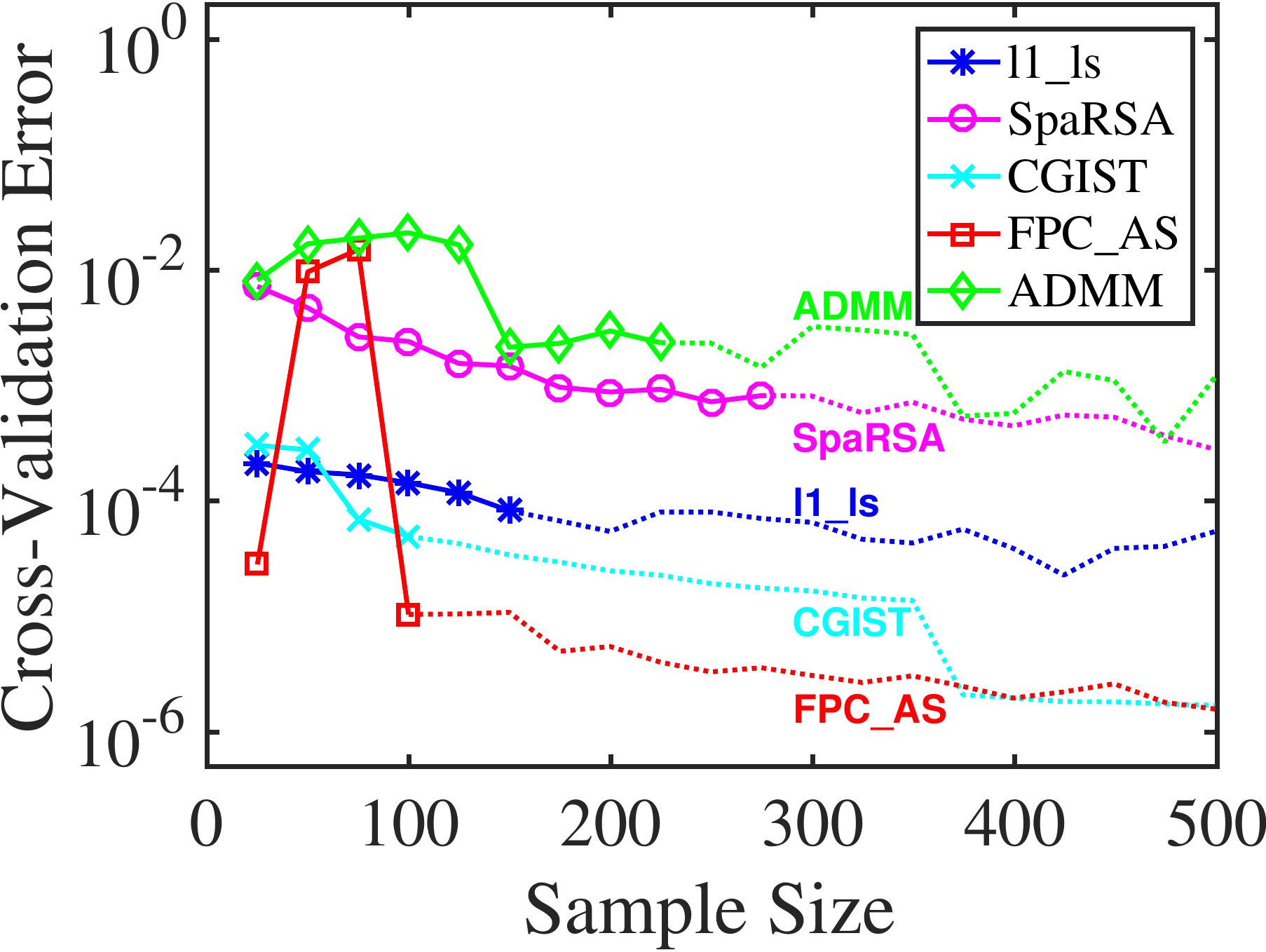}
  \includegraphics[width=0.32\textwidth, trim={0 6em -0.5em 0}, clip]{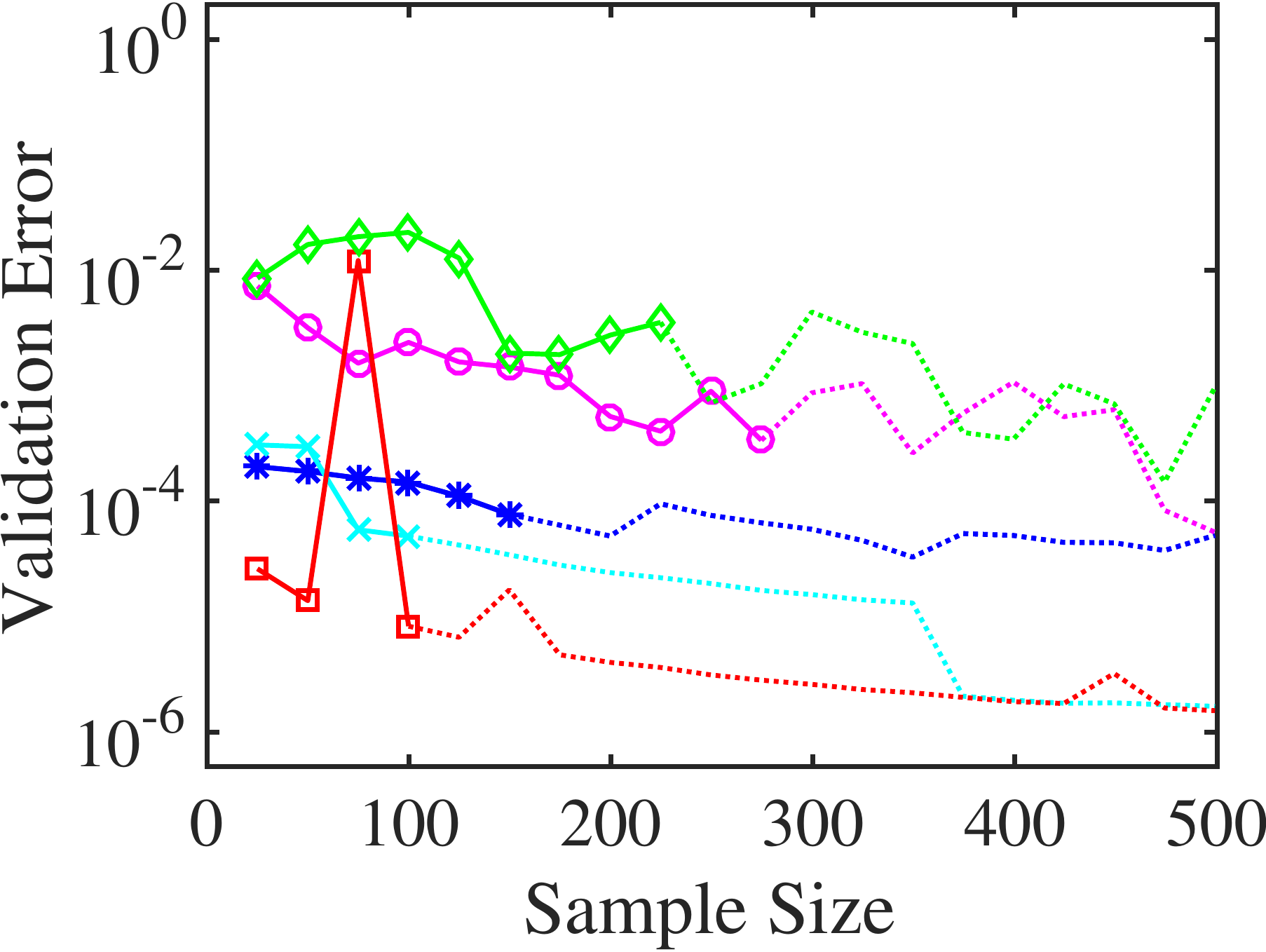}
  \includegraphics[width=0.32\textwidth, trim={0 6em -0.5em 0}, clip]{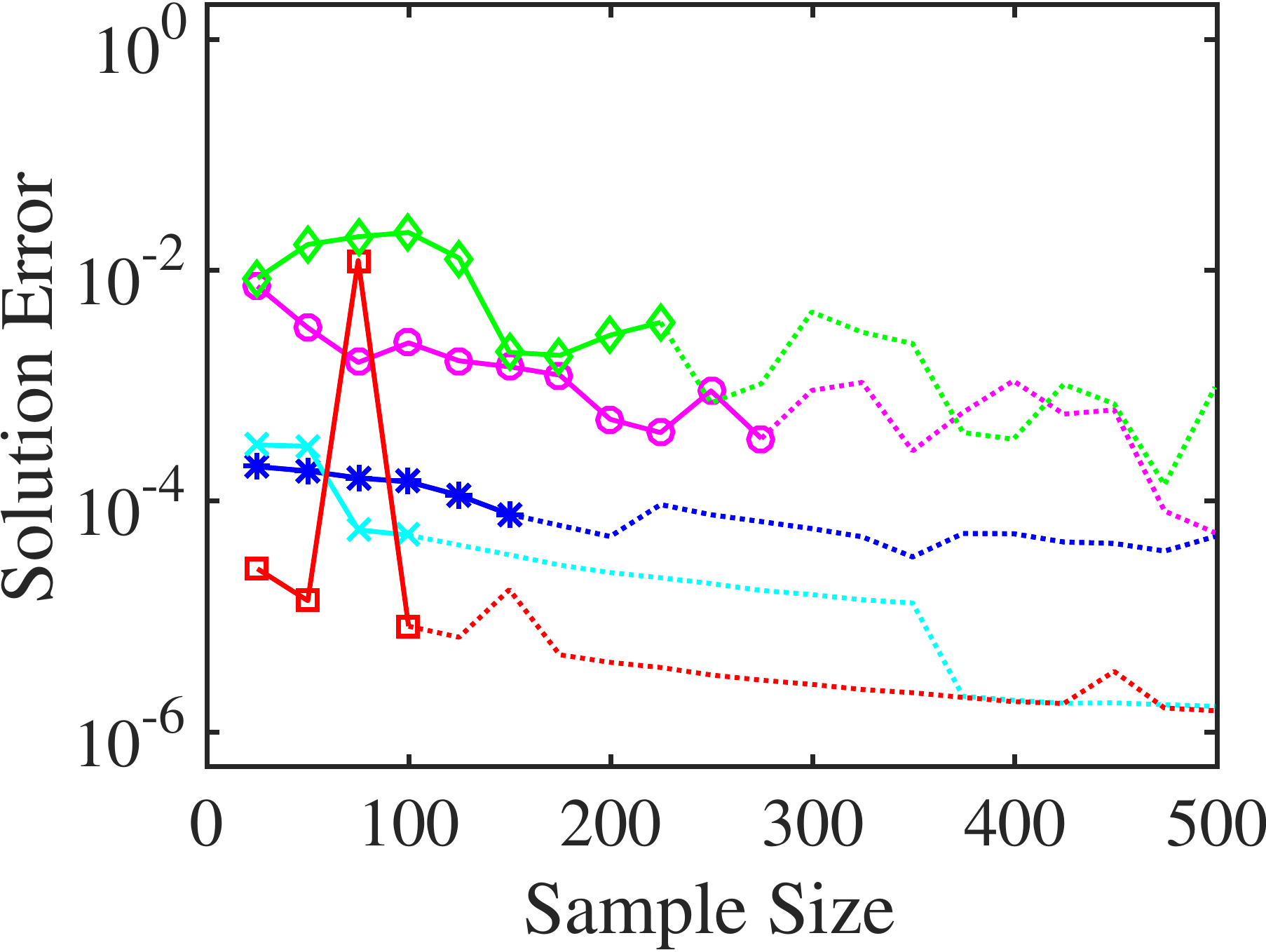}\\[0.5em]
  \includegraphics[width=0.32\textwidth]{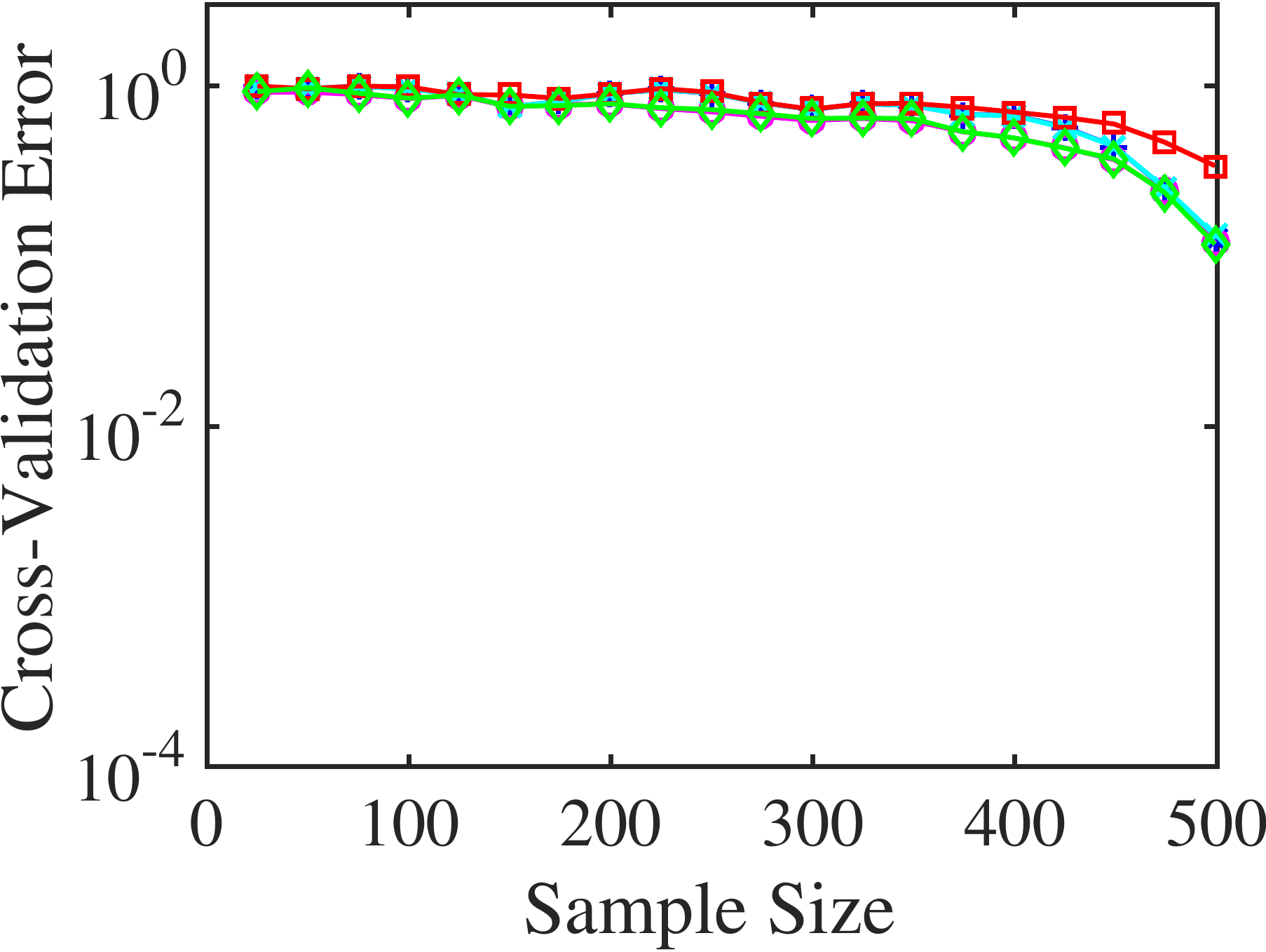}
  \includegraphics[width=0.32\textwidth]{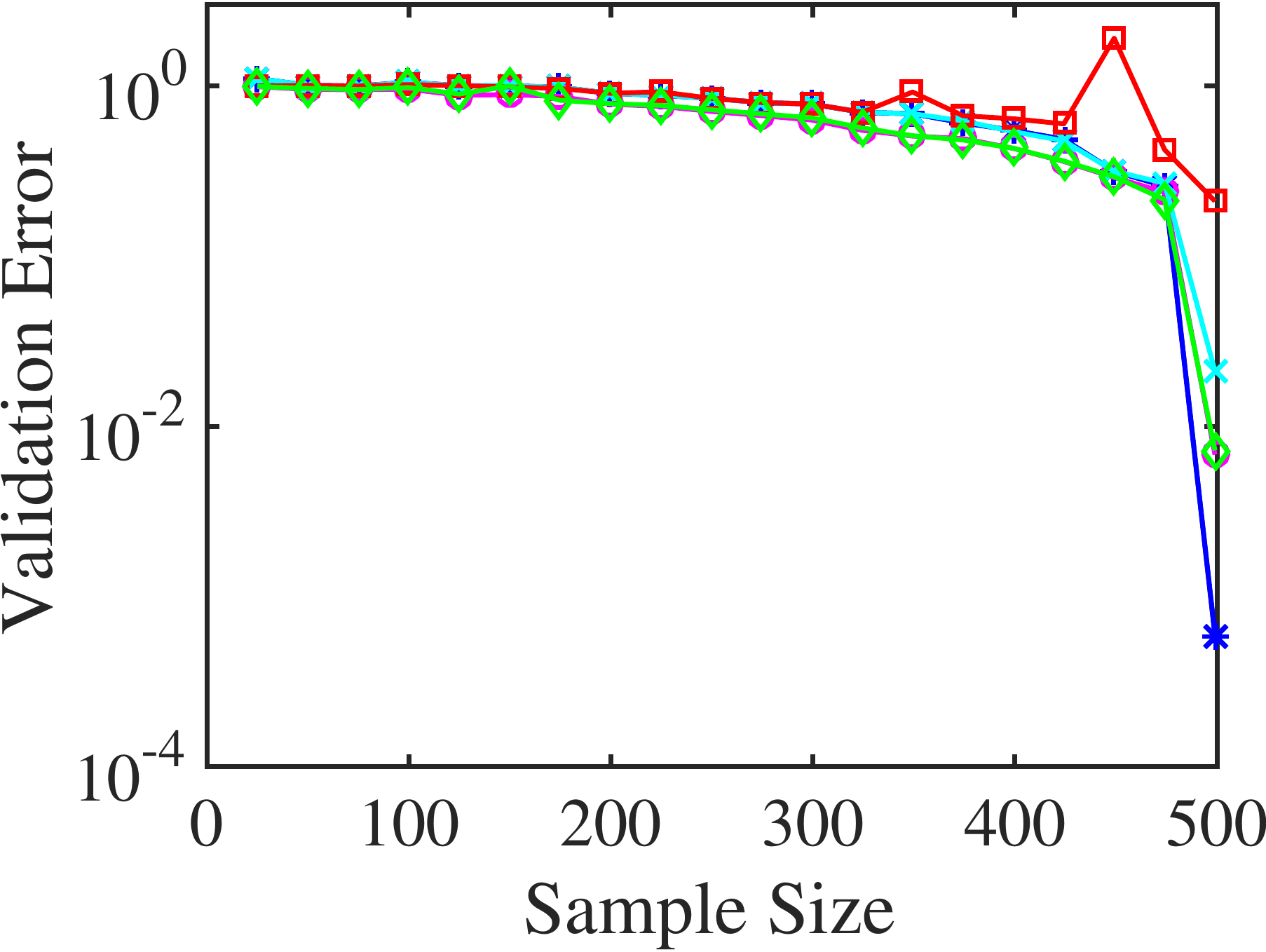}
  \includegraphics[width=0.32\textwidth]{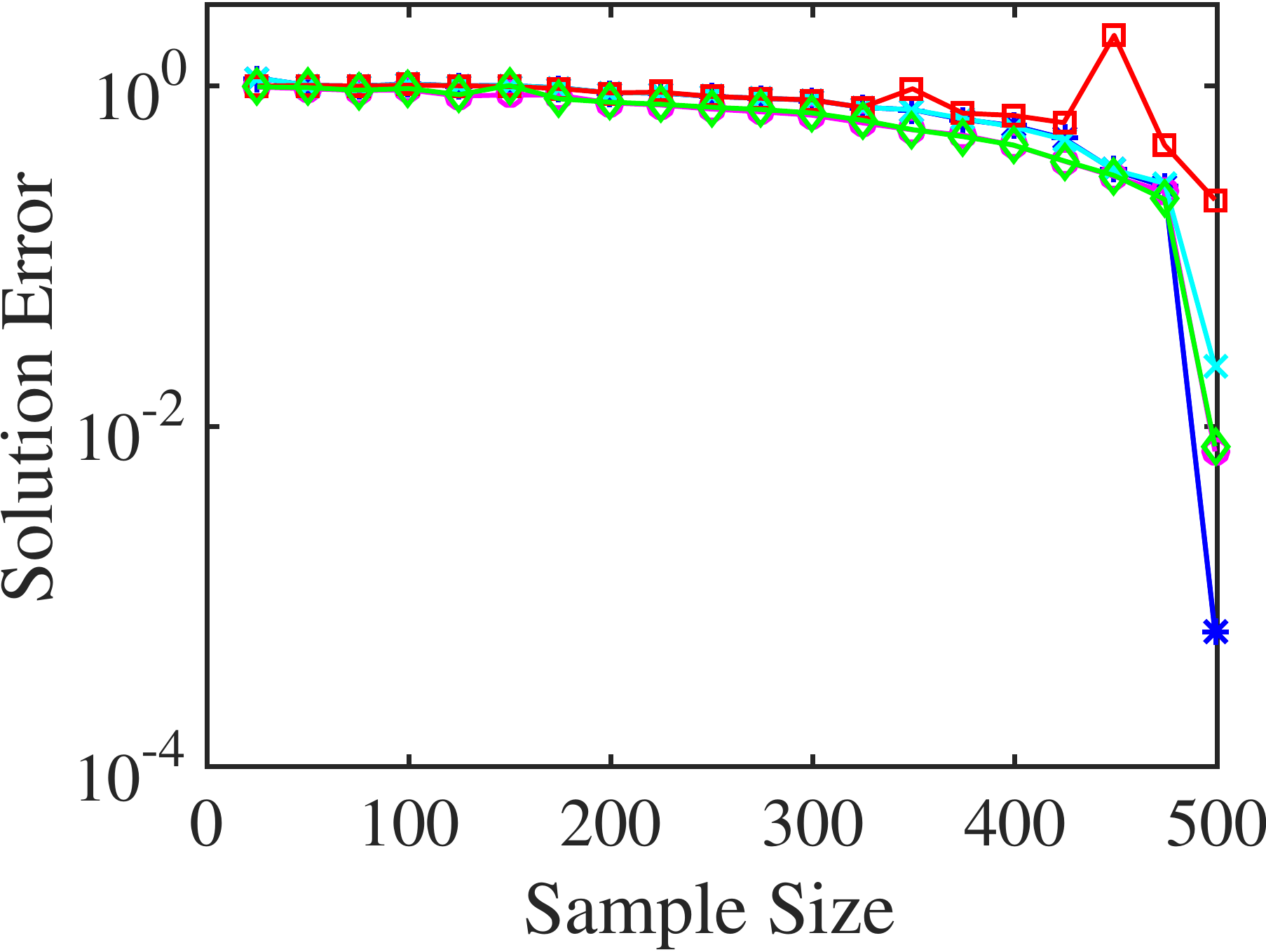}
  \caption{Case 1 Gaussian random matrix: CV (left), validation
    (middle), and solution (right) errors for fixed problem instance
    with $n=500$, on a very sparse example $s=1$ (top) and a very
    dense example $s=400$ (bottom). Stop-sampling is activated when
    the plot line turns from solid (with symbols) to dotted (without
    symbols).}
  \label{f:case1_single_very_sparse_dense}
\end{figure}

\Cref{f:case1_single_very_sparse_dense} shows the errors from two
systems with very sparse (top, $s=1$) and very dense (bottom, $s=400$)
solutions, respectively. In both cases, the sharp drop is not
immediately noticeable, especially if $\Delta m$ is coarse. For the
sparse case, all the errors are already very low even with a very
small $m$; for the dense case, the drop occurs only when a large $m$
is achieved (around $500$). Our algorithm is still able to offer
reasonable stop-sampling points in both situations. \fpcas{} has
noticeable spikes in these plots. Additional inspections indicate the
spikes are not caused by the grid resolution of $\lambda$, and are
likely resulted from within the \fpcas{} algorithm.

\begin{figure}[htb]
  \centering
  \includegraphics[width=0.47\textwidth, trim={0 2em 0 0}, clip]{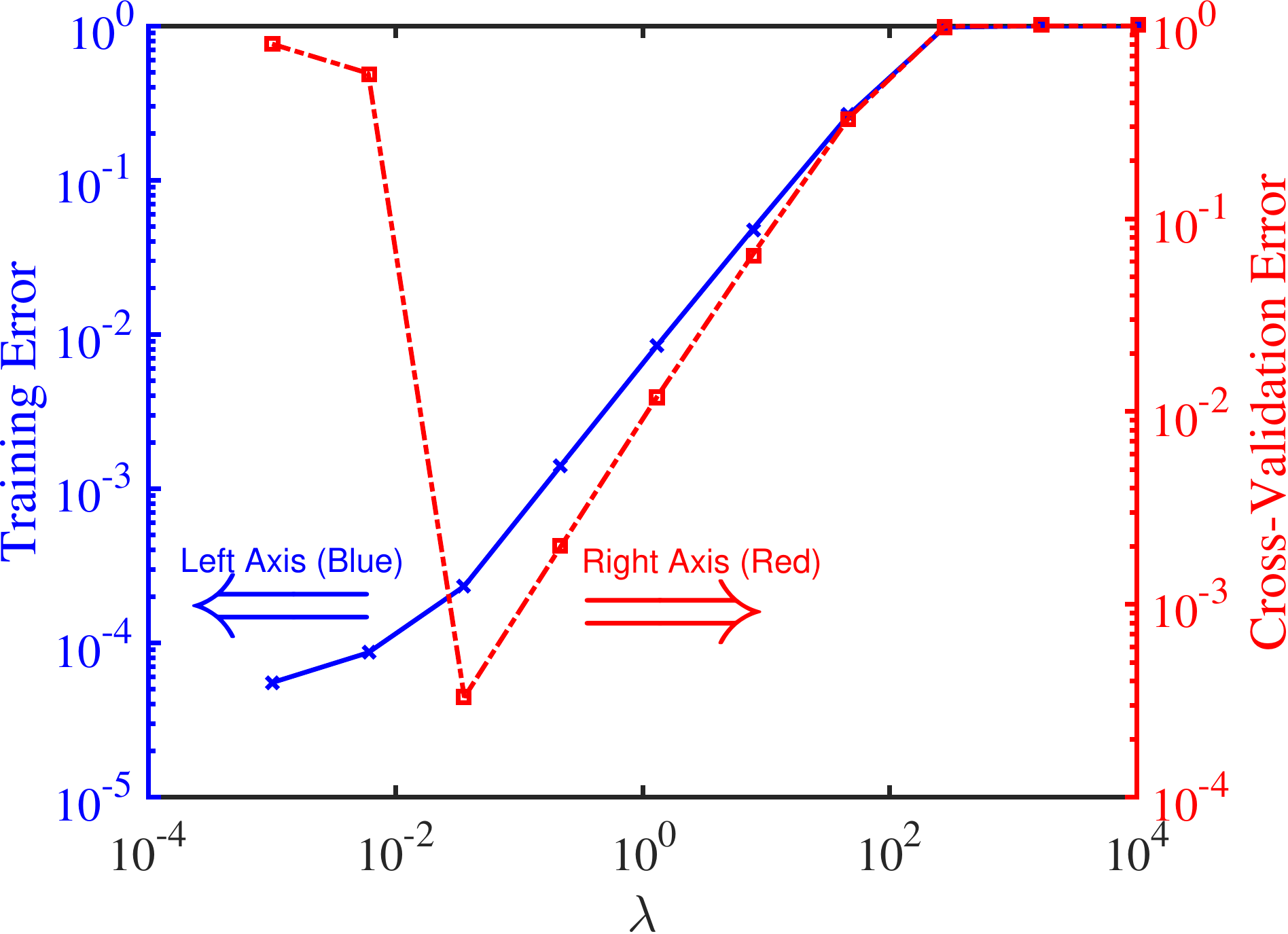}
  \includegraphics[width=0.45\textwidth, trim={0 2em 0 0}, clip]{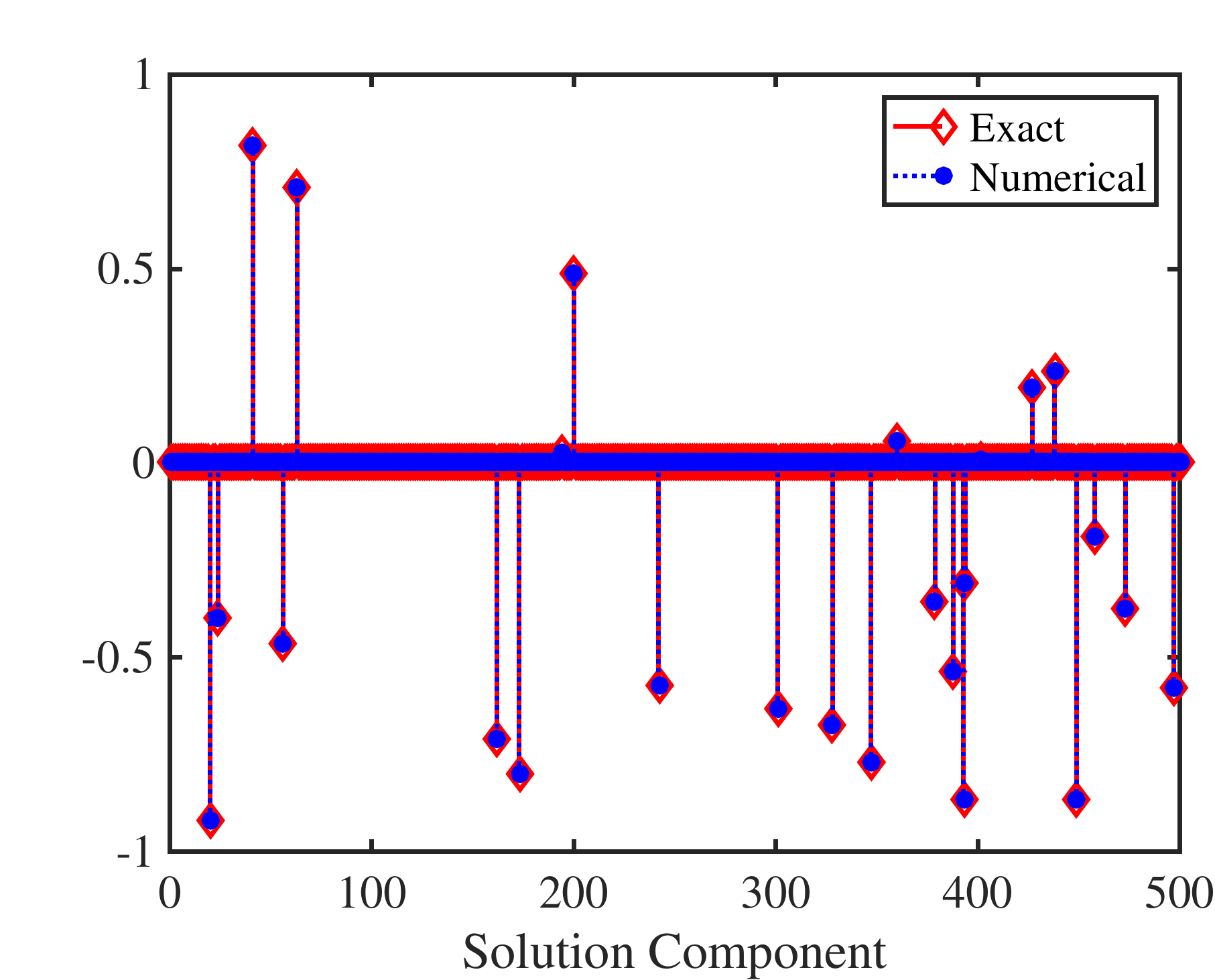}\\
  \includegraphics[width=0.47\textwidth]{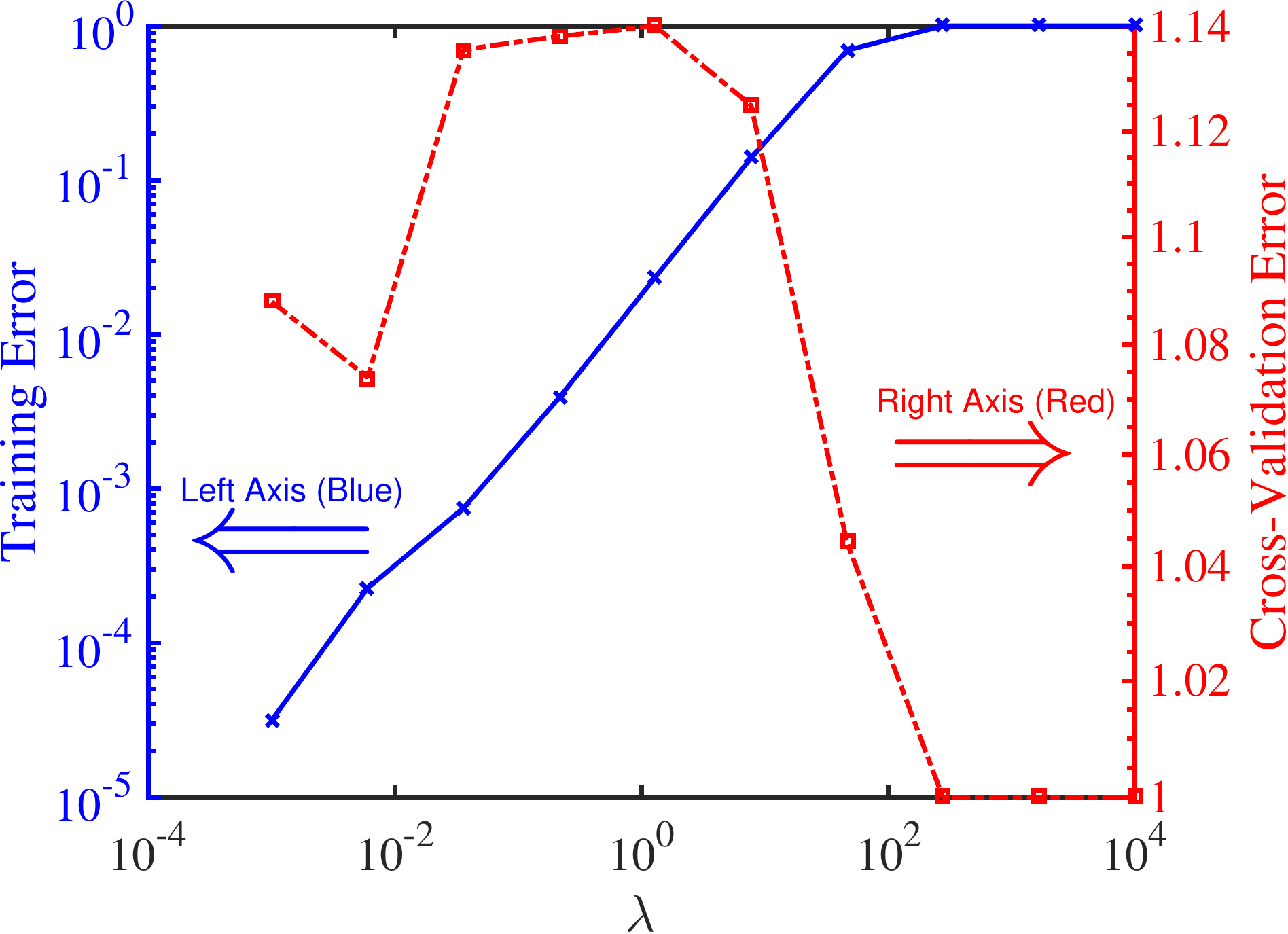}
  \includegraphics[width=0.45\textwidth]{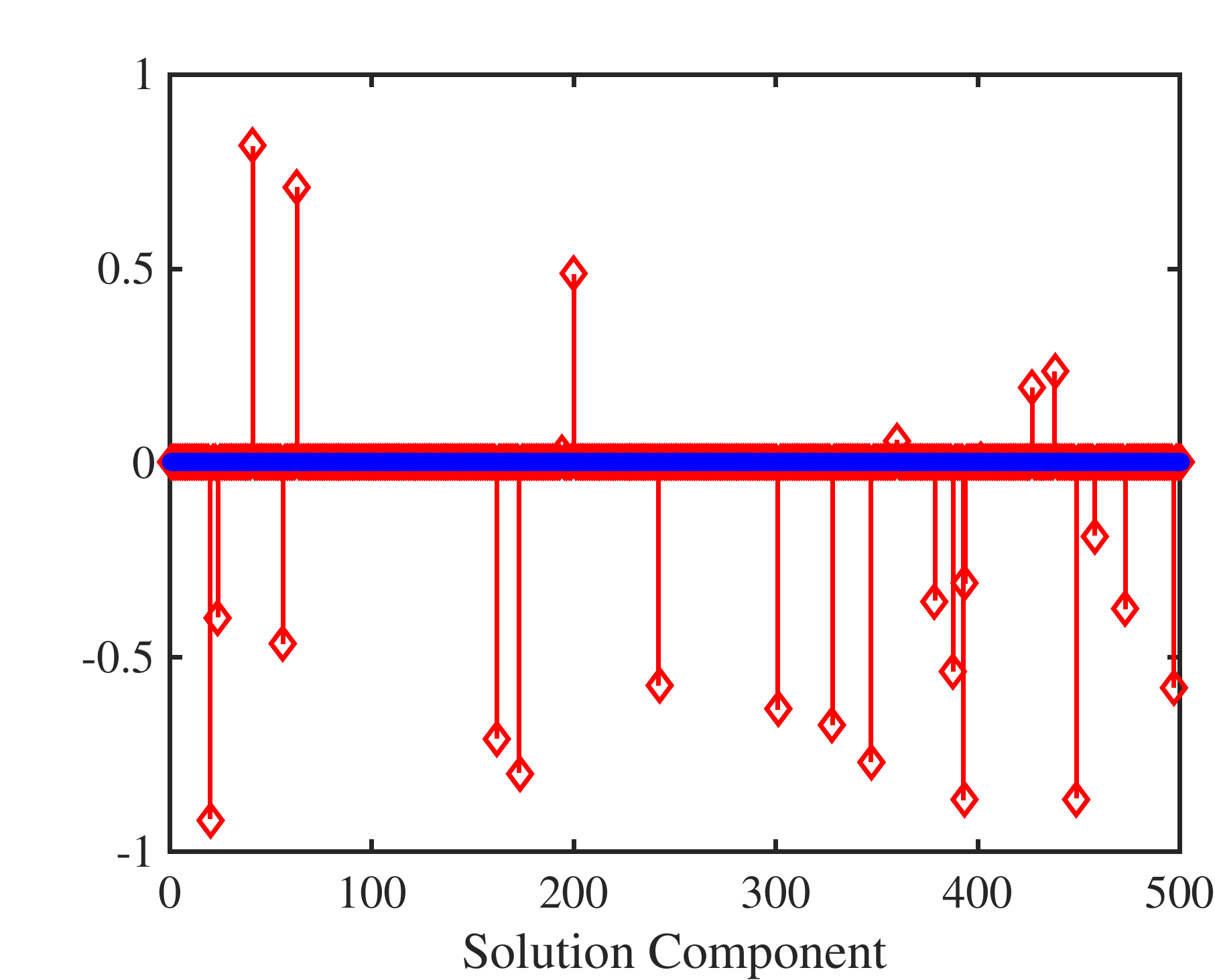}
  \caption{Case 1 Gaussian random matrix: training and CV errors
    (left; training error in solid blue using left $y$-axis, and CV
    error in dotted red using right $y$-axis) and corresponding
    lowest-CV-error solution stem plots (right) for fixed problem
    instance with $n=500$ and $s=25$ (i.e., encountered in
    \cref{f:case1_single}), showing an example with good solution
    (top, using \sparsa{} with $m=150$), and an example with bad
    solution (bottom, using \sparsa{} with $m=25$).}
  \label{f:case1_single_select}
\end{figure}

Lastly, we present in \cref{f:case1_single_select} detailed results of
two cases from the fixed problem instance of \cref{f:case1_single},
with an example of good solution using \sparsa{} and $m=150$ (top),
and an example of bad solution using \sparsa{} and $m=25$
(bottom). The training and CV errors at different $\lambda$ are shown
in the left column, and the corresponding lowest-CV-error solution
stem plots are in the right column. As expected, the training error
always decreases as $\lambda$ decreases.  For the good solution, the
CV error initially decreases as $\lambda$ decreases, and then
increases, reflecting overfitting when $\lambda$ becomes too small.
The optimal solution is chosen to be at $\lambda^{\ast}$ corresponding
to minimal CV error. The stem plot of that solution on the top-right
panel shows excellent agreement with the exact solution.  For the
example of bad solution, the CV error has a plateau of high magnitude
in comparison, and the lowest-CV-error solution is in fact at high
$\lambda$. The stem plot verifies it to be the zero vector. This
outcome implies that when the sample size is too small, CV is still
able to reflect and prevent overfitting by reverting to recommend the
zero solution (in a sense, a ``none'' solution is better than the bad
solution).

\subsection{Case 2: random polynomial chaos expansion}

We now move on to PCE examples. Consider a Gauss-Hermite PCE of
stochastic dimension $n_s=5$ and total-order polynomial basis of
degree $p=5$, which translates to $n=251$. We use a simple sampling
strategy, sampling $\xi_j$ corresponding to the PCE germ distribution,
i.e., i.i.d. standard normal $\CN(0,1)$ for Gauss-Hermite.

\begin{figure}[htb]
  \centering
      \includegraphics[width=0.32\textwidth, trim={1em 5.5em 4.5em 2em}, clip]{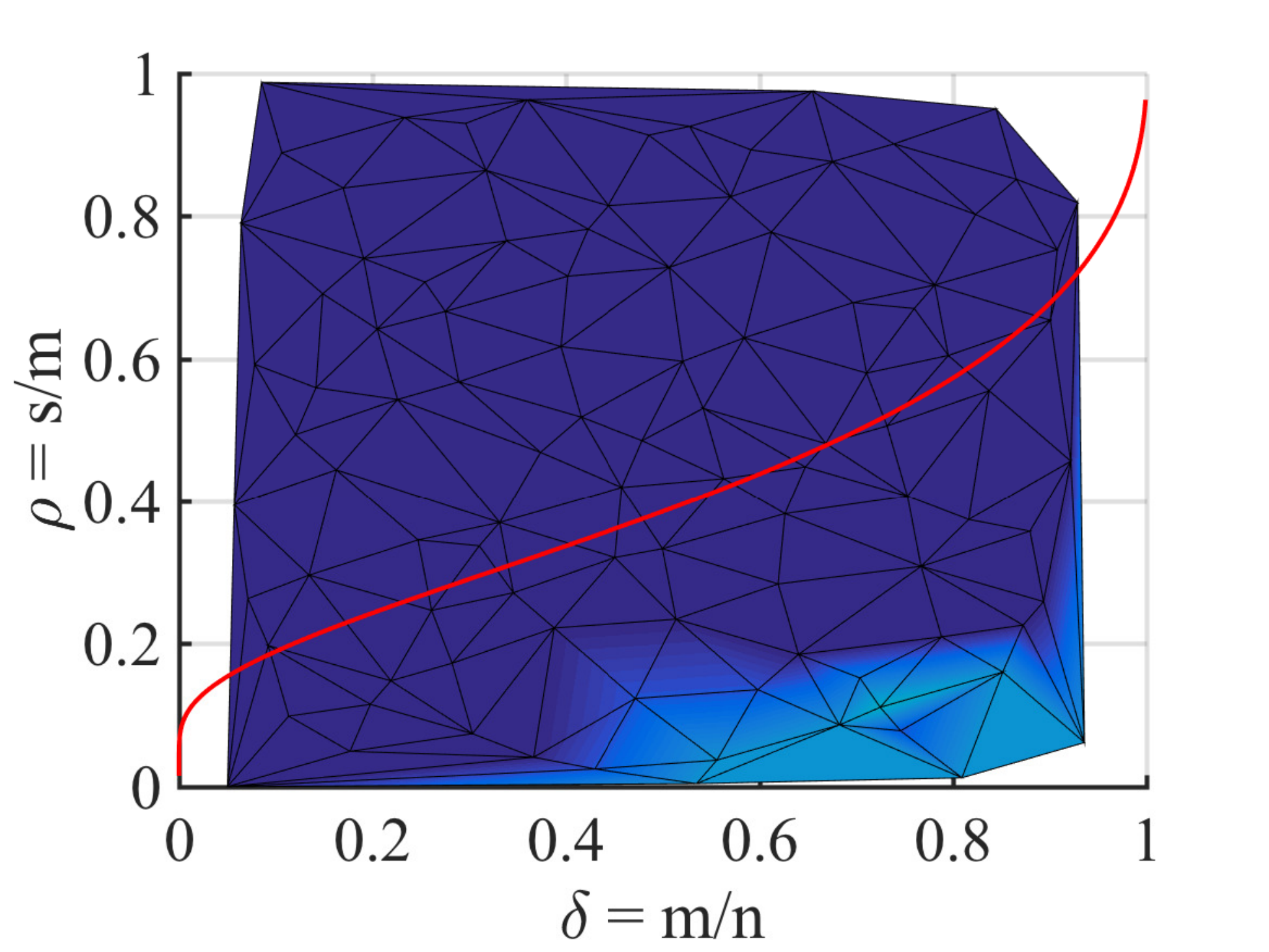}
  \includegraphics[width=0.32\textwidth, trim={1em 5.5em 4.5em 2em}, clip]{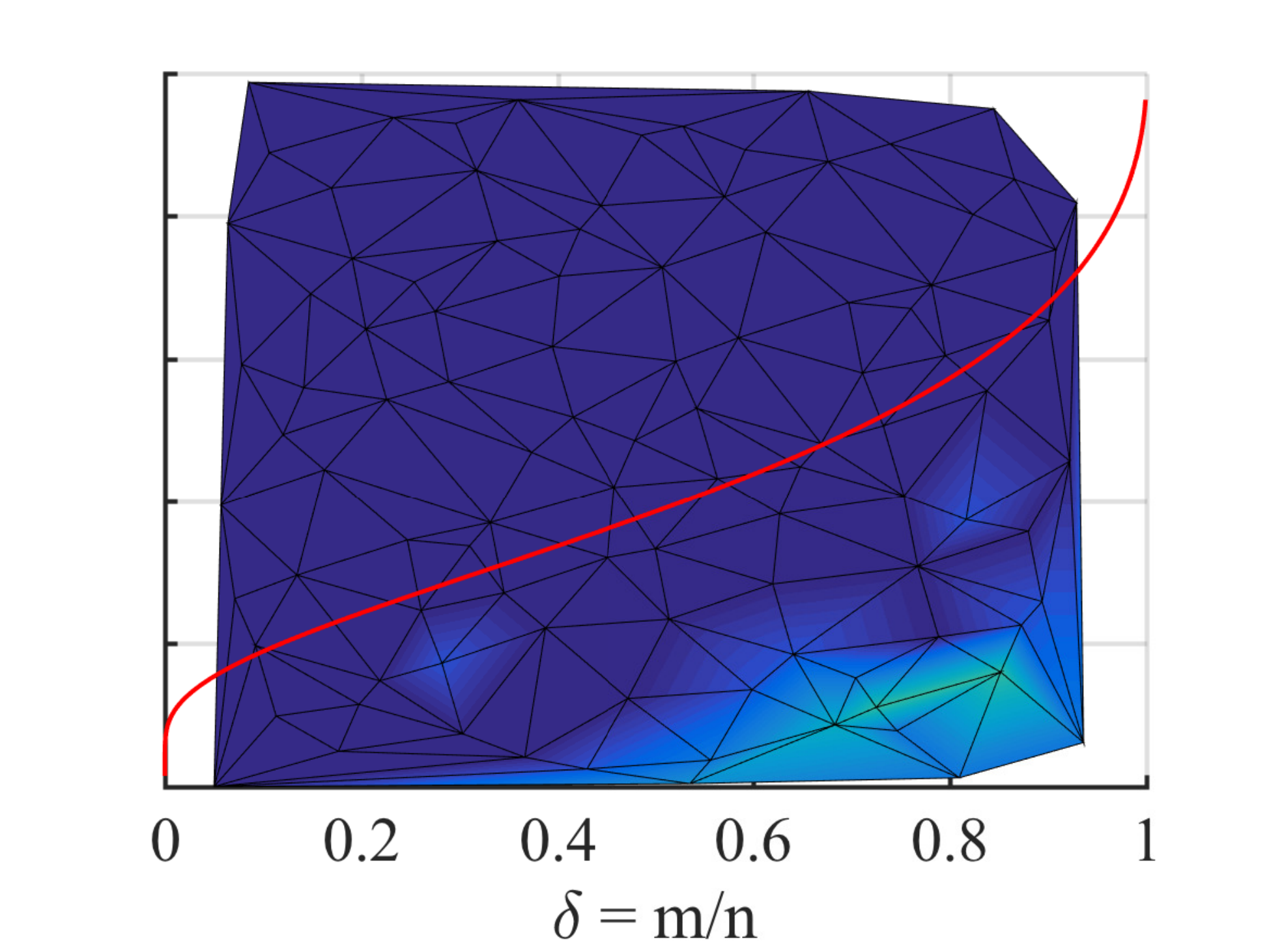}
  \includegraphics[width=0.32\textwidth, trim={1em 5.5em 4.5em 2em}, clip]{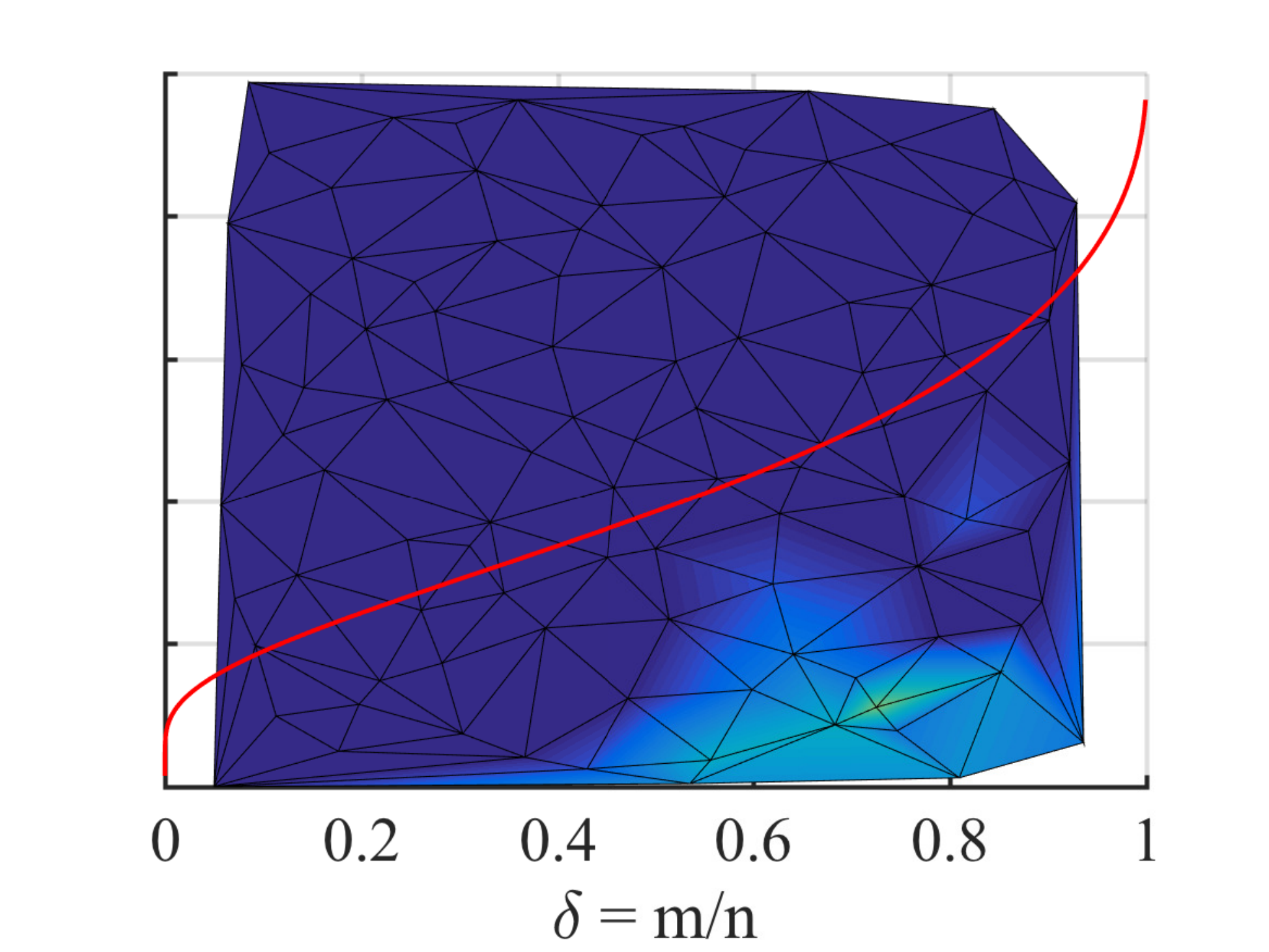}\\[0.5em]
  \includegraphics[width=0.32\textwidth, trim={1em 0 4.5em 2em}, clip]{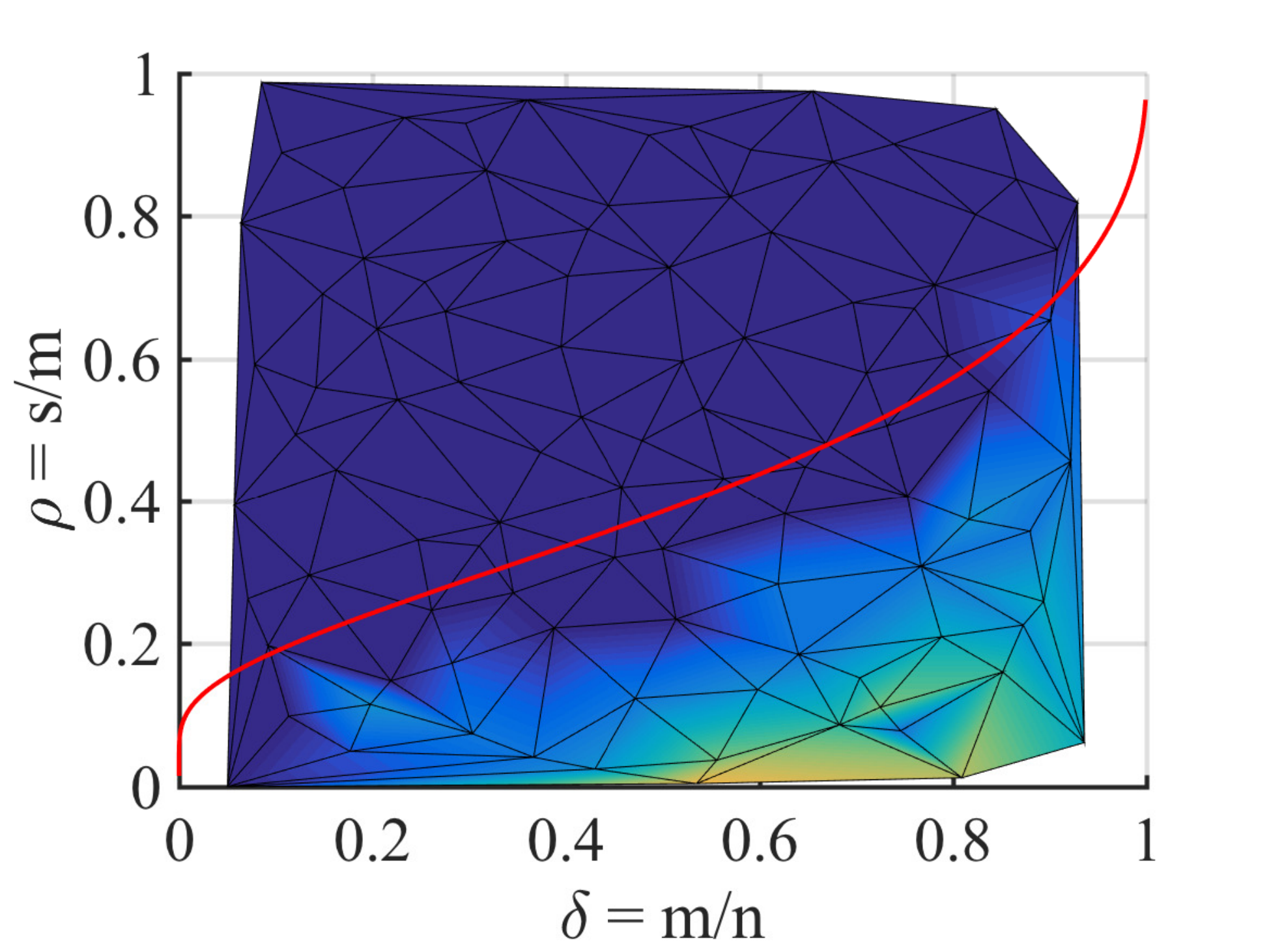}
  \includegraphics[width=0.32\textwidth, trim={1em 0 4.5em 2em}, clip]{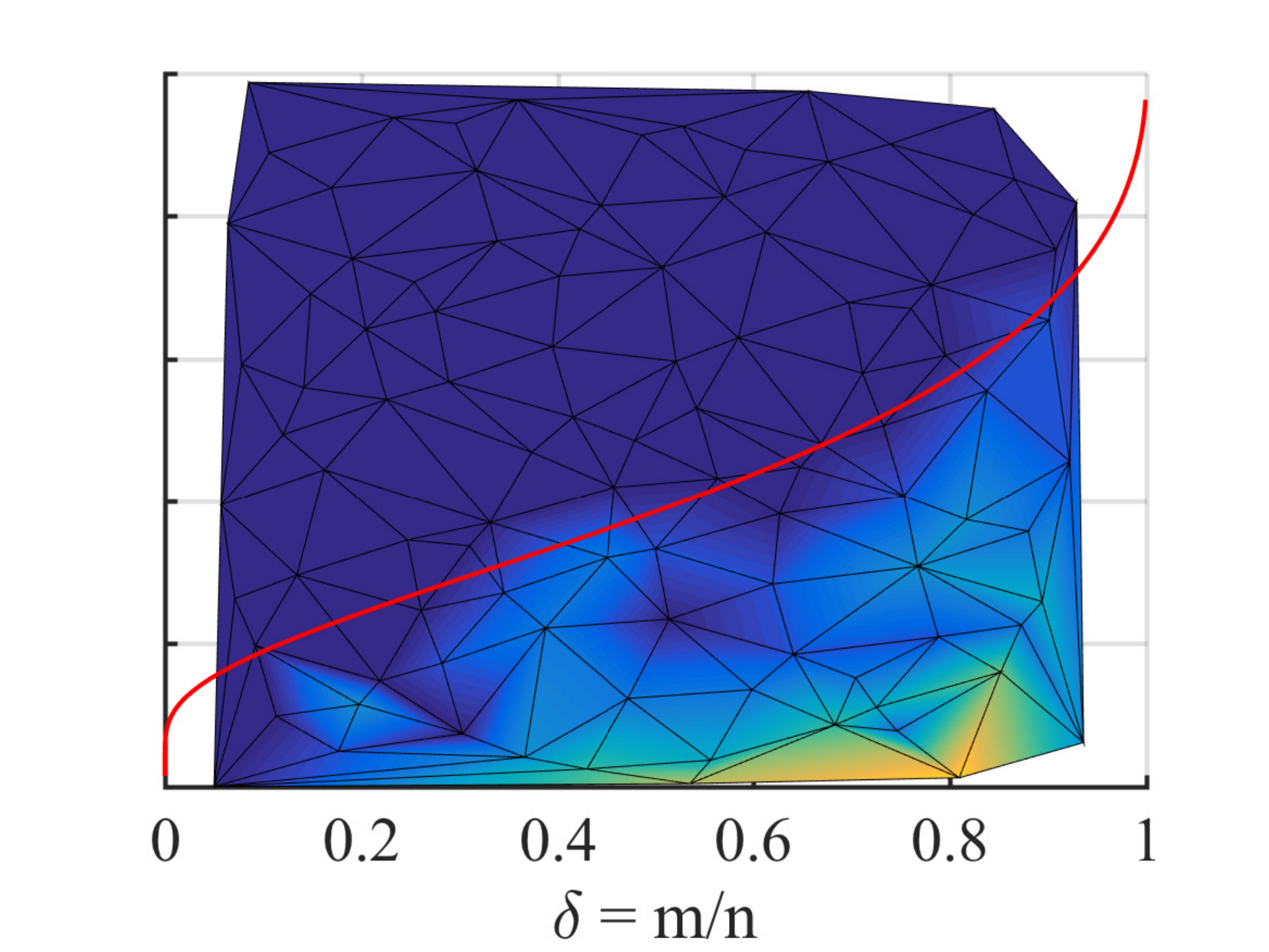}
  \includegraphics[width=0.32\textwidth, trim={1em 0 4.5em 2em}, clip]{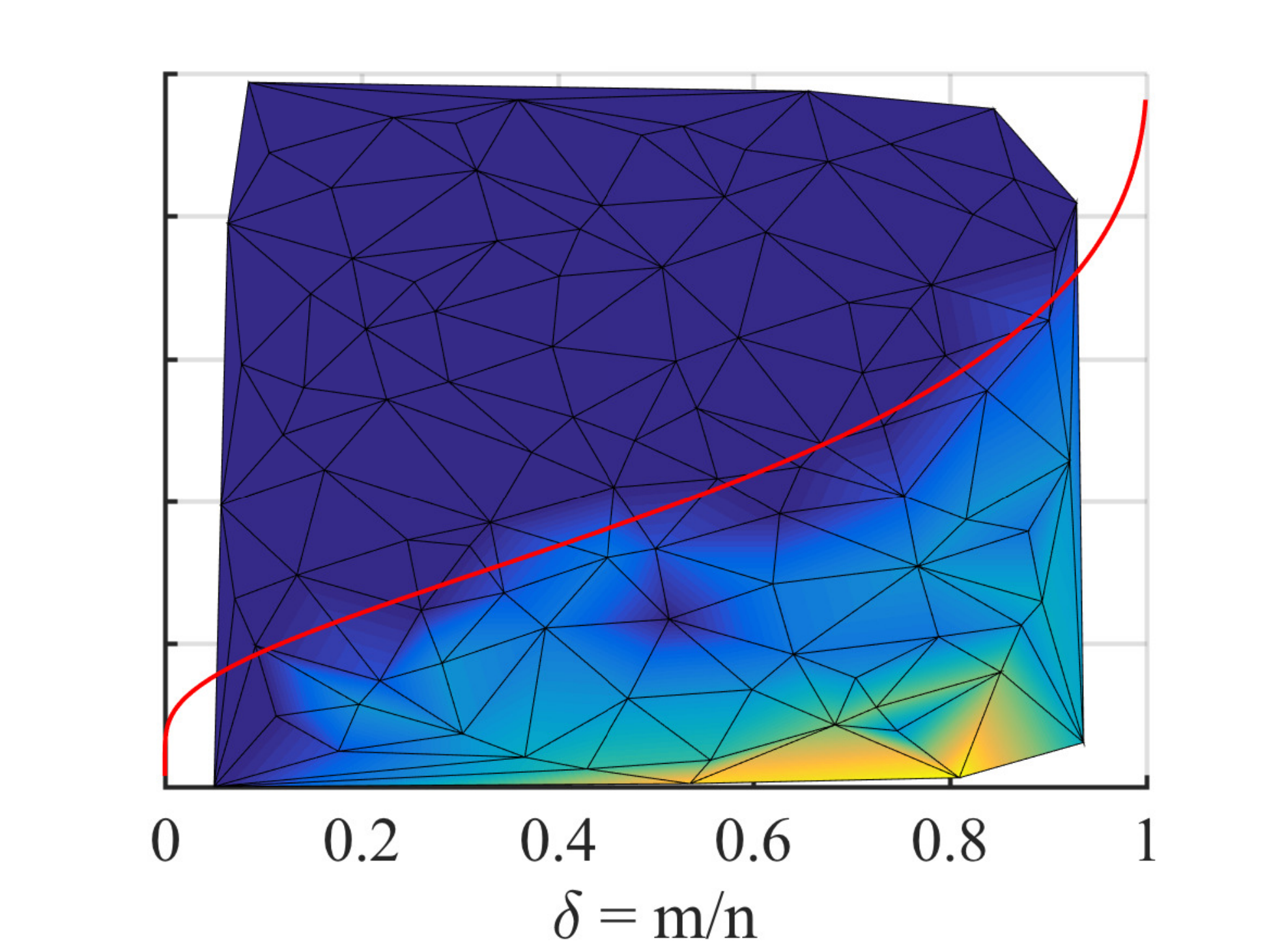}\\
    \includegraphics[width=0.95\textwidth, , trim={3em 0 0 43em}, clip]{figures/mycolorbar-eps-converted-to.pdf}
  \caption{Case 2 random PCE: phase-transition diagrams plotted from
    CV (left), validation (middle), and solution (right) errors where
    a success recovery is defined by when the respective error
    quantity is less than $0.1$, using \fpcas{} (top), and \admm{}
    (bottom). Plots using \lls{}, \sparsa{}, and \cgist{} are very
    similar to the \admm{} results, and thus omitted to avoid
    repetition.}
              \label{f:case2_PTDiagrams}
\end{figure}

\Cref{f:case2_PTDiagrams} shows the phase-transition diagrams using
\fpcas{} (top) and \admm{} (bottom), and plotted based on assessing CV
(left), validation (middle), and solution (right) errors.  Again, we
define a successful reconstruction if the error quantity of interest
from a run is less than $0.1$, and the diagram plots the empirical
success rate based on $b=10$ repeated trials at each node.  The
correlation structure of the $A$ matrix induced by the PCE basis leads
to a more difficult problem for sparse reconstruction, and the success
rates are much lower compared to those in \cref{f:case1_PTDiagrams}.
Donoho and Tanner hypothesized that the transition curve from Gaussian
random matrix systems exhibits universality properties for many other
``well-behaved'' distributions as well, particularly with large
$n$~\cite{Donoho2009}. We overlay this theoretical curve in
\cref{f:case2_PTDiagrams}, and the empirical transition curves
(0.5-probability contours) appear to be much worse in comparison; this
phenomenon is also consistent with observations in other papers
(e.g.,~\cite{Hampton2015, Jakeman2017}). This PCE-induced distribution
of $A$ matrix exhibits a complex correlation structure that likely
pushes it outside the ``well-behaved'' regime, and results in an
example that does not comply with the universality hypothesis.  Plots
using \lls{}, \sparsa{}, and \cgist{} are similar to the \admm{}
results and thus omitted to avoid repetition, while \fpcas{} is
observed to produce slightly lower success rates than the other
solvers.

\begin{figure}[htb]
  \centering
  \includegraphics[width=0.32\textwidth]{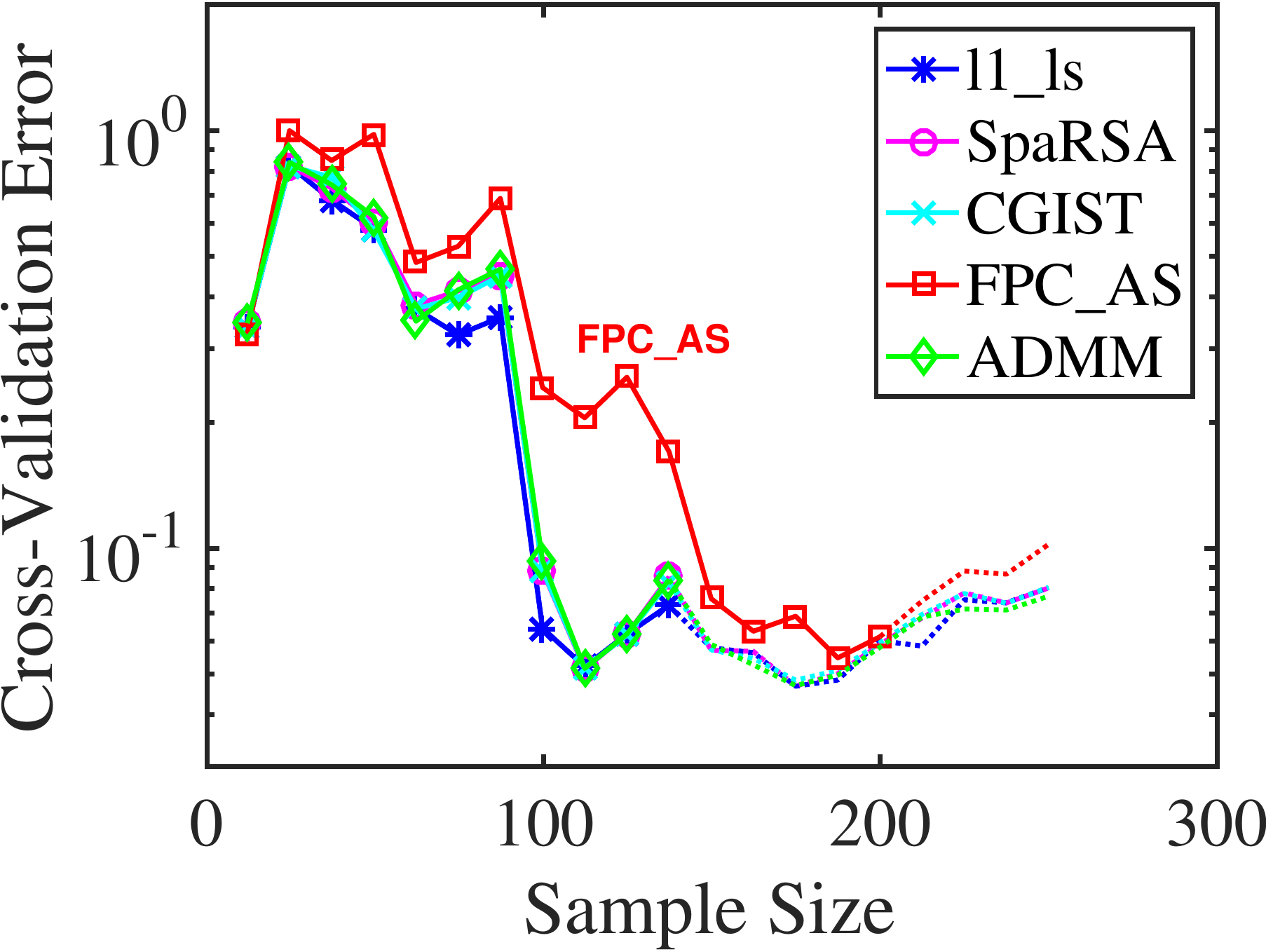}
  \includegraphics[width=0.32\textwidth]{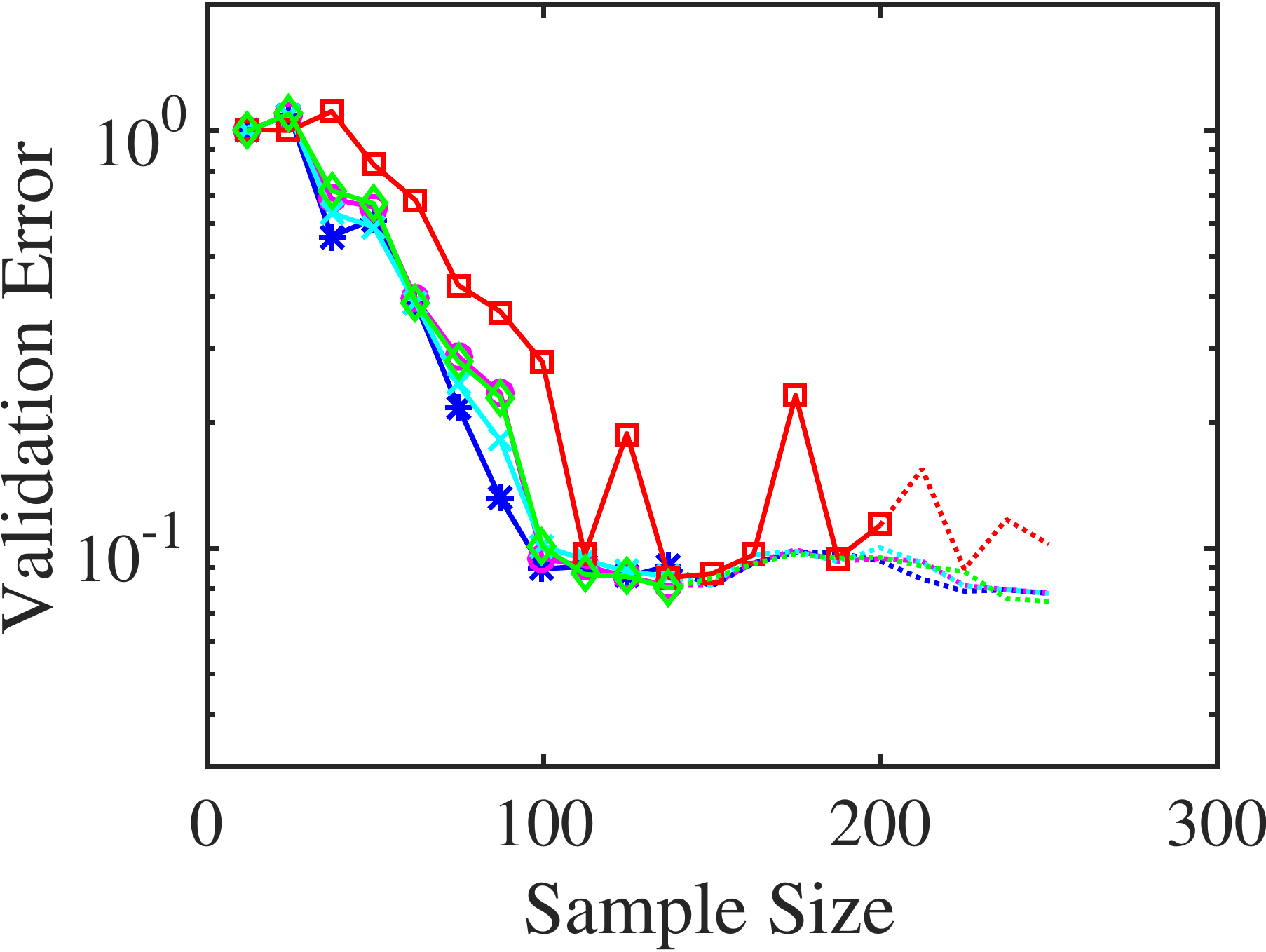}
  \includegraphics[width=0.32\textwidth]{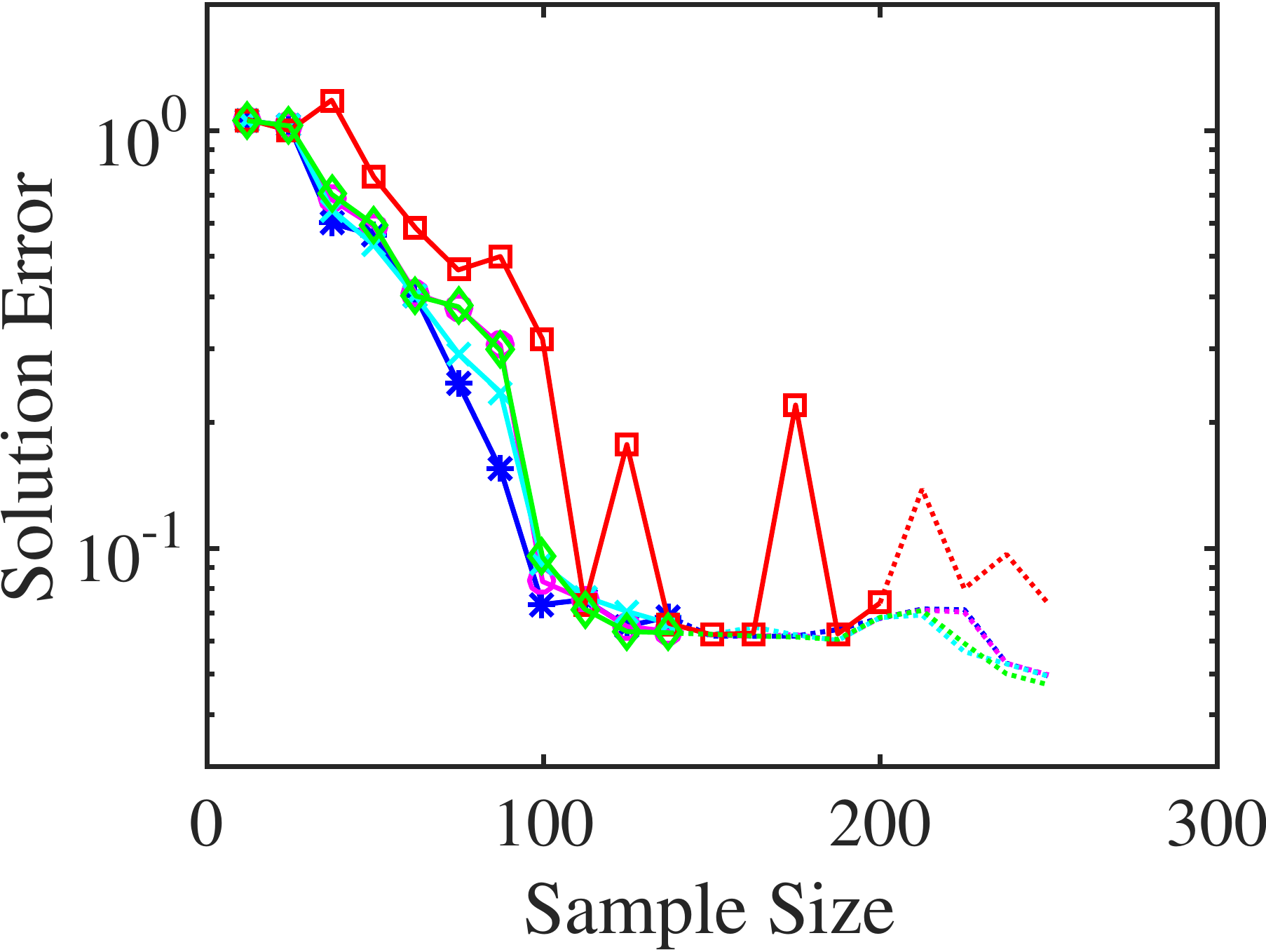}
  \caption{Case 2 random PCE: CV (left), validation (middle), and
    solution (right) errors for fixed problem instance with $n_s=5$,
    $p=5$ ($n=251$), and $s=20$. Stop-sampling is activated when the
    plot line turns from solid (with symbols) to dotted (without
    symbols).}
  \label{f:case2_single}
\end{figure}

\Cref{f:case2_single} illustrates results for a fixed problem instance
where the true solution $x^{\ast}$ is constructed to have $s=20$
randomly selected nonzero elements with values drawn from
i.i.d. standard normal $\CN(0,1)$.  All solvers yield similar error
levels, and a sharp drop of error can be observed around $m=100$.
\fpcas{} is observed to exhibit large error oscillation after $m=100$,
which can potentially introduce difficulties for our stop-sampling
strategy. However, stop-sampling still does a good job at terminating
around the base of the sharp drop.  It is interesting to note that
while \fpcas{} has some undesirable behavior in this case, it yielded
the lowest errors in Case 1; the inconsistency in performance may be
undesirable overall.  The more difficult problem structure can also be
observed through the overall higher error values (asymptotes are just
below $10^{-1}$) compared to those in the random Gaussian matrix case
(achieving $10^{-3}$ or better).

\subsection{Case 3: Genz-exponential function}

We now explore a more realistic scenario where the solution is not
strictly sparse, but compressible (i.e., the sorted coefficient
magnitudes decay rapidly),
through the Genz-exponential function:
\begin{align}
  y(\xi) = \exp\({\sum_{j = 1}^{n_s} a_j \xi_j}\).\label{e:Genz_exp}
\end{align}
Let the stochastic dimension be $n_s=5$, and the exponential
coefficients endowed with a power rule decay of
$a_j=j^{-1}$. Gauss-Hermite PCEs of different polynomial orders are
constructed to approximate \cref{e:Genz_exp} via evaluations at
$\xi_j$ samples drawn from i.i.d. standard normal $\CN(0,1)$. This
problem is challenging on several fronts. First, there is truncation
error for any finite-order polynomial approximation to the exponential
form of \cref{e:Genz_exp}---i.e., modeling error is always
present. Second, the best polynomial approximation is not expected to
be strictly sparse as a result of both truncation and that $a_j$ is
not sparse even though it is decaying. However, it may be near-sparse,
and it would still be valuable to find a solution that balances
approximation error and sparsity.

We compare PCEs of total-order $p=3$, $5$, and $7$ in representing
\cref{e:Genz_exp}.  Intuitively, one expects higher order polynomials
to offer richer basis sets and thus smaller modeling error (i.e.,
lower achievable error). However, larger basis sets also mean larger
$n$, and produce systems that could be harder to solve, both in terms
of computational cost and numerical accuracy (i.e., higher numerical
error). Let us elaborate on these two points. Imagine we are comparing
two total-order polynomial basis sets with degrees $p_1$ and $p_2$
(assume $p_1 < p_2$) under a fixed data set (and thus also the same
sample size $m$). If the true data-generating model were less than
$p_1$ degree polynomial, then neither choice has modeling error. The
richer basis set form $p_2$ has no modeling benefit while the
undersampling ratio $\delta=m/n$ (recall $n$ is the number of columns
of $A$) would be lower, which generally translates to a lower
probability of successful recovery if everything else remains the same
(c.f. phase-transition diagrams in
\cref{f:case1_PTDiagrams,f:case2_PTDiagrams}). In the case where there
is modeling error, the coefficient of the new terms from $p_2$ may or
may not have non-negligible magnitudes, and so the overall sparsity
ratio $\rho=s/m$ (recall $s$ is the number of non-zero elements) can
change in either direction\footnote{Strictly speaking, $s$ (and thus
  $\rho$ for a fixed $m$) cannot decrease when considering a
  higher-order (nested) polynomial basis set, and so $p_2$ would have
  a higher $\rho$ which again translates to a lower probability of
  successful recovery (c.f. phase-transition diagrams in
  \cref{f:case1_PTDiagrams,f:case2_PTDiagrams}). However, for example,
  if a new coefficient with a dominating magnitude were introduced,
  the ``practical'' sparsity may be reduced since the other existing
  coefficients, although non-zero, could be dwarfed.}. Since the
polynomial spectrum of an exponential form \cref{e:Genz_exp} generally
decays with order, we expect $\rho$ to not decrease for this
case. Overall, the effects of achievable error and numerical error are
unclear when polynomial order is increased, and certainly are problem
dependent.

\begin{figure}[htb]
  \centering
  \includegraphics[width=0.365\textwidth, trim={0 4.1em -0.5em 0}, clip]{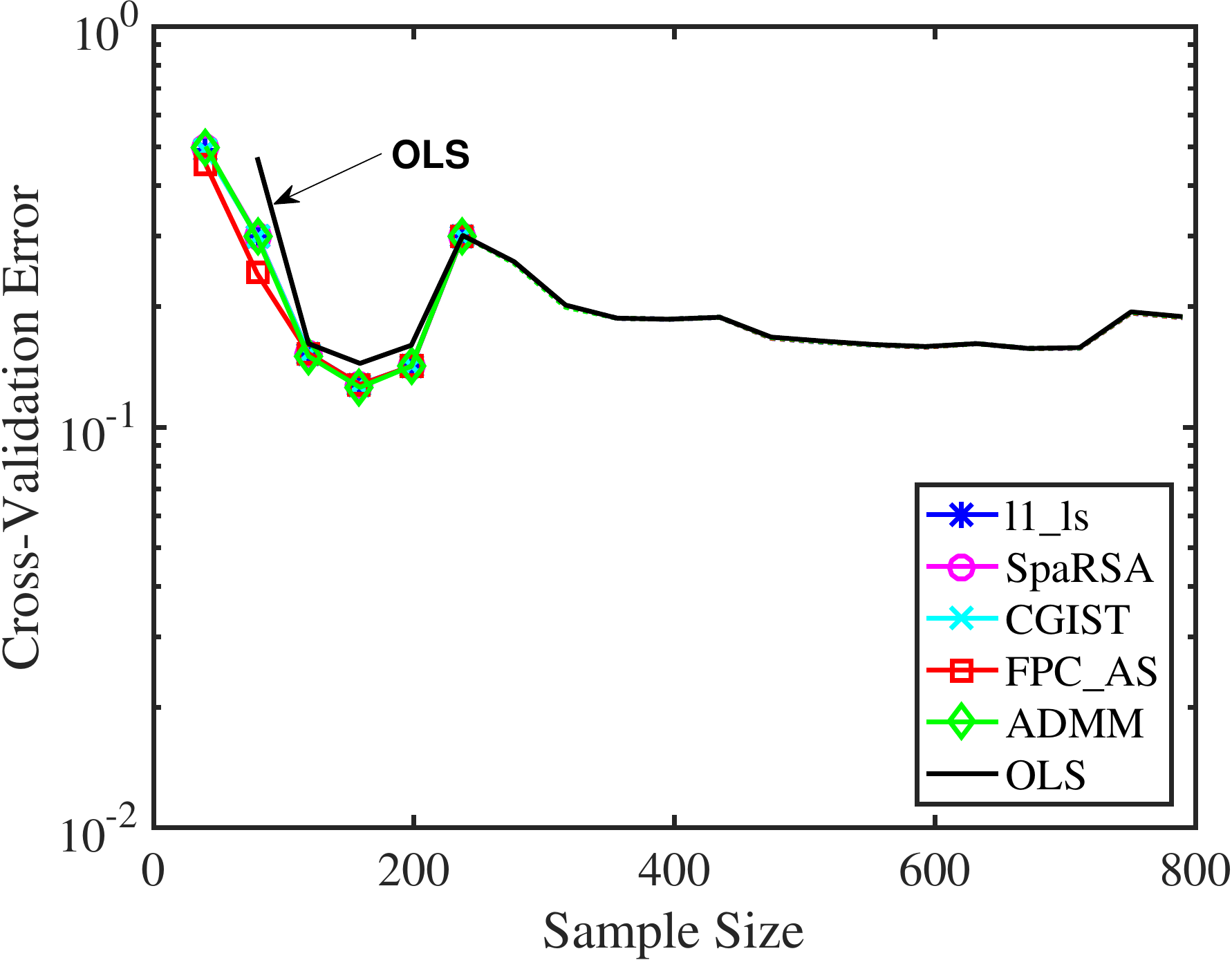}
  \includegraphics[width=0.365\textwidth, trim={0 4.1em -0.5em 0}, clip]{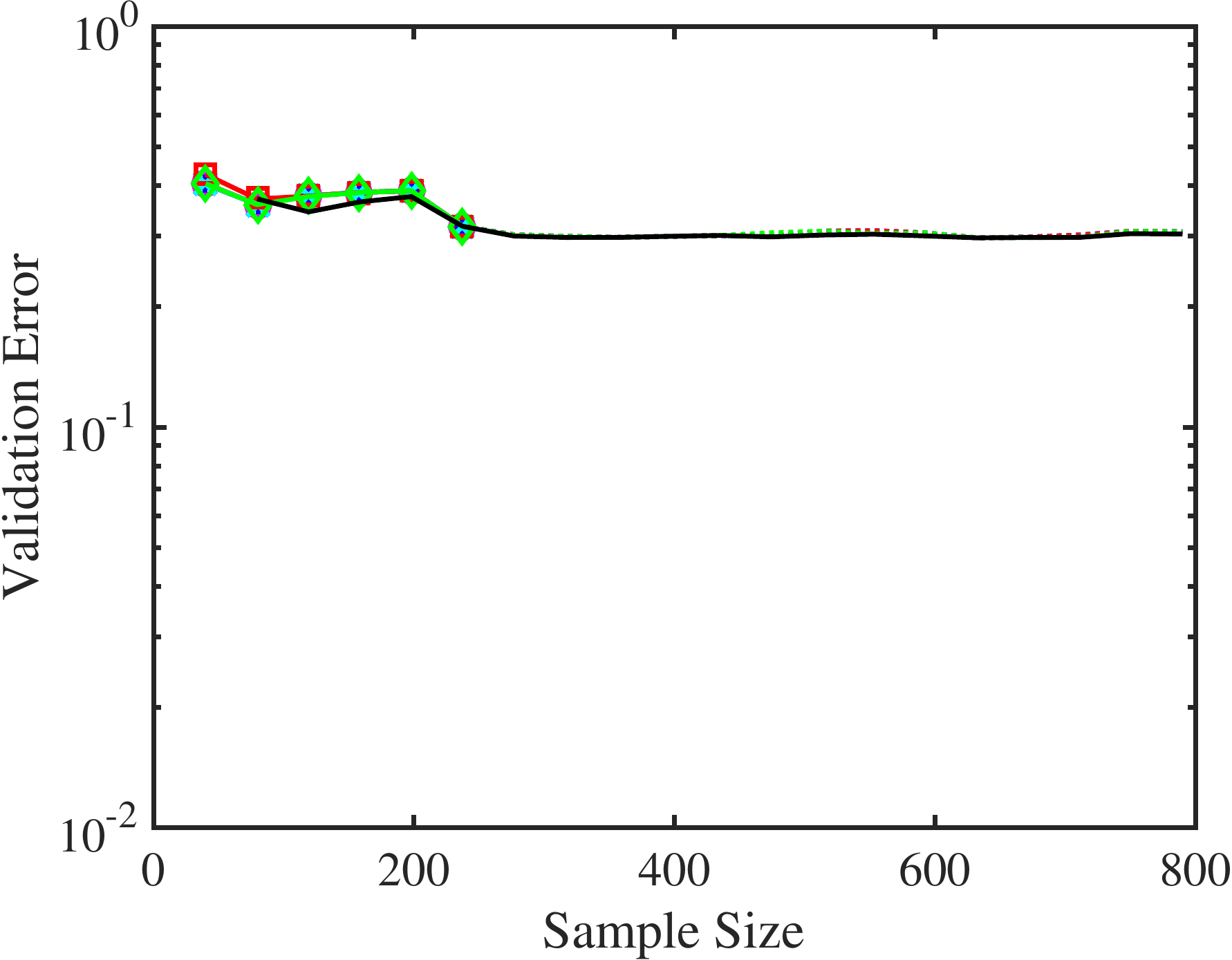}\\
  \includegraphics[width=0.365\textwidth, trim={0 4.1em -0.5em 0}, clip]{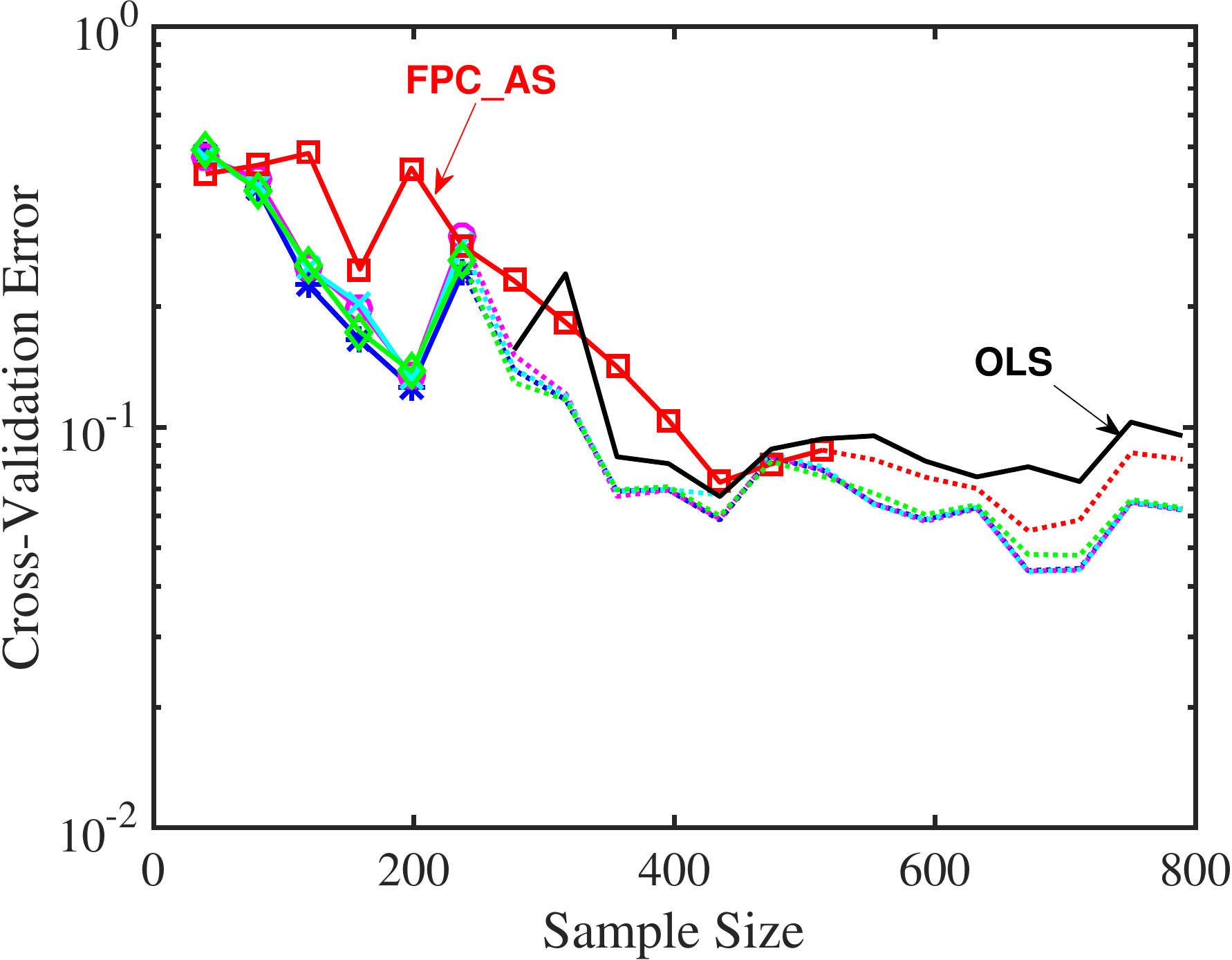}
  \includegraphics[width=0.365\textwidth, trim={0 4.1em -0.5em 0}, clip]{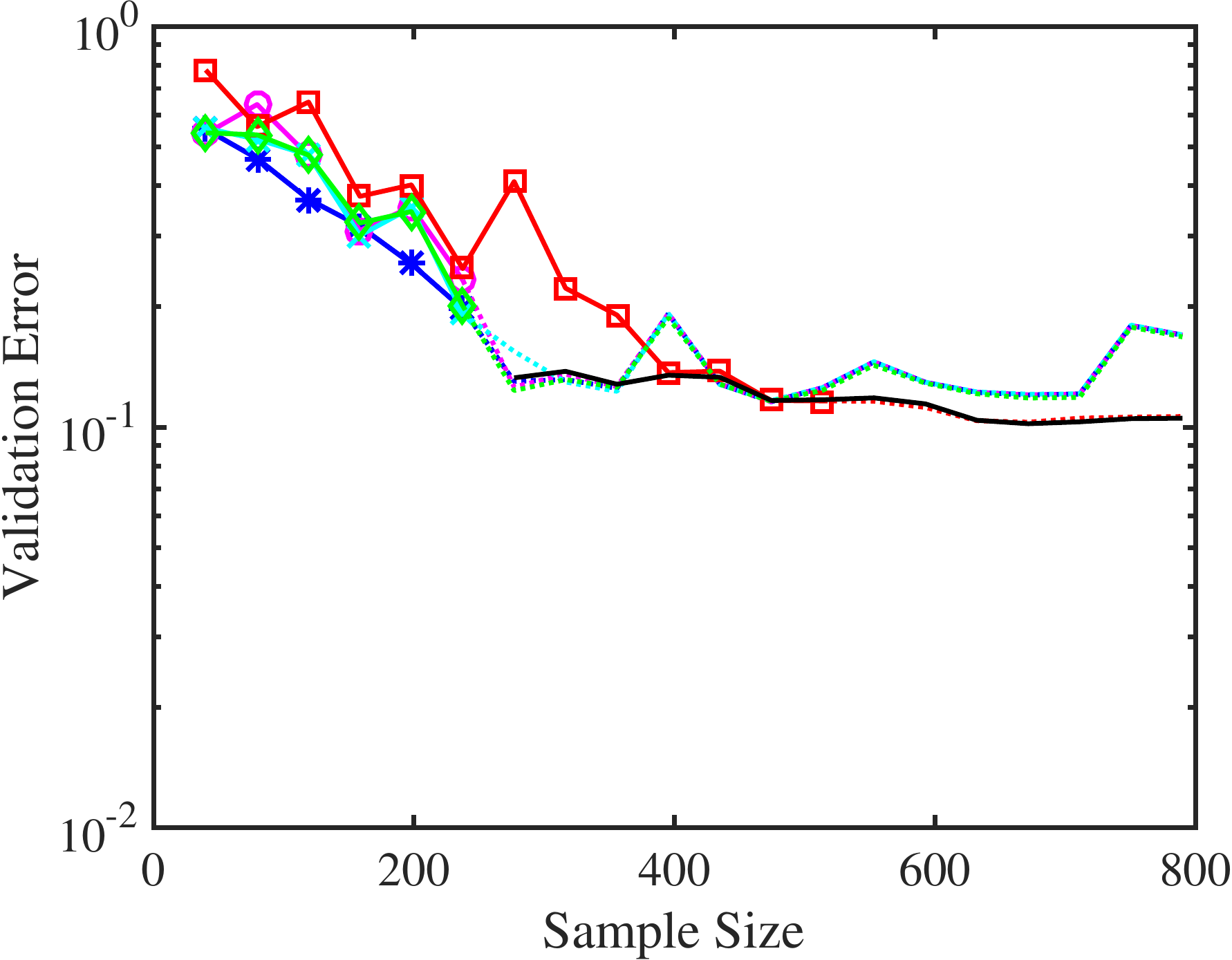}\\
  \includegraphics[width=0.365\textwidth]{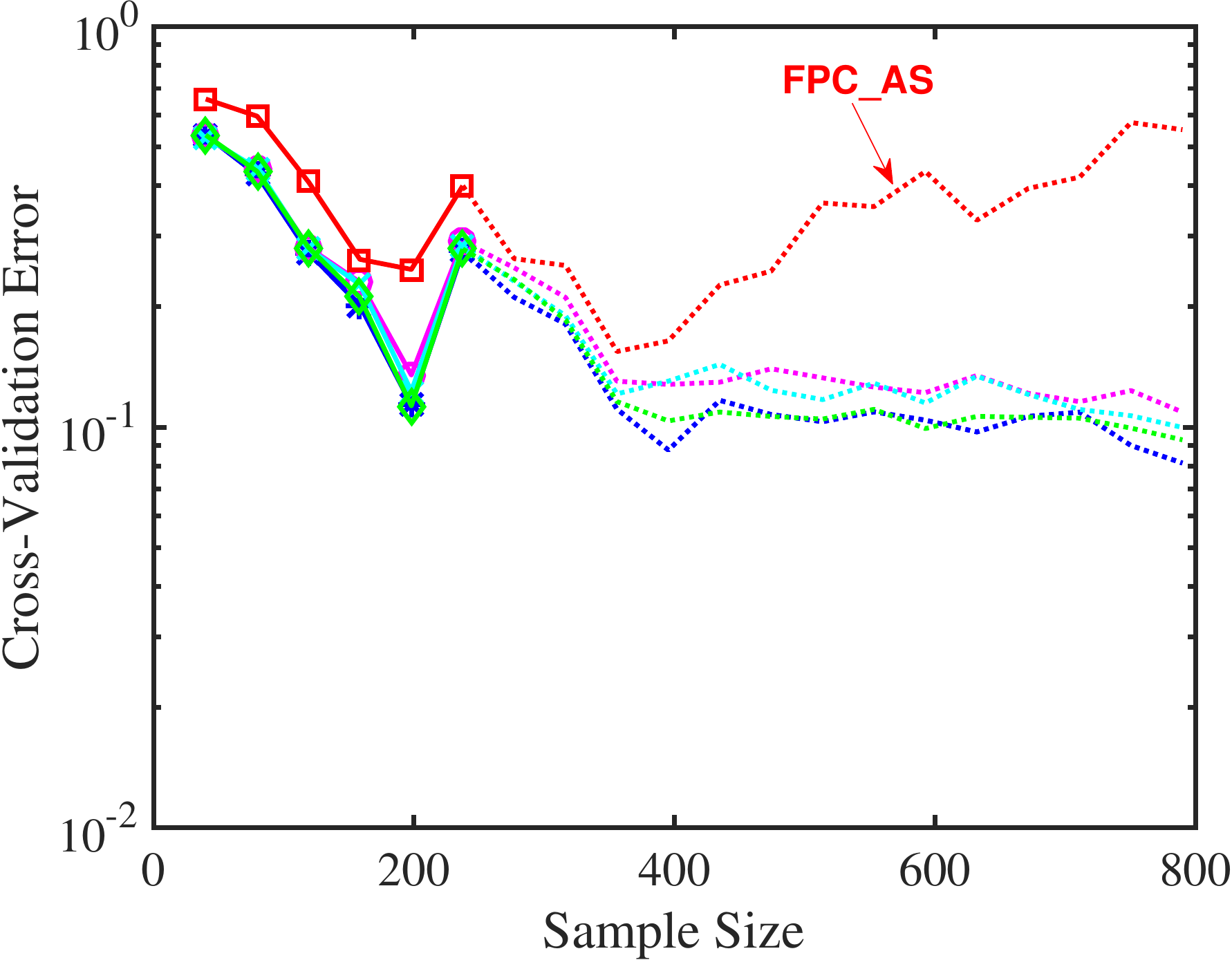}
  \includegraphics[width=0.365\textwidth]{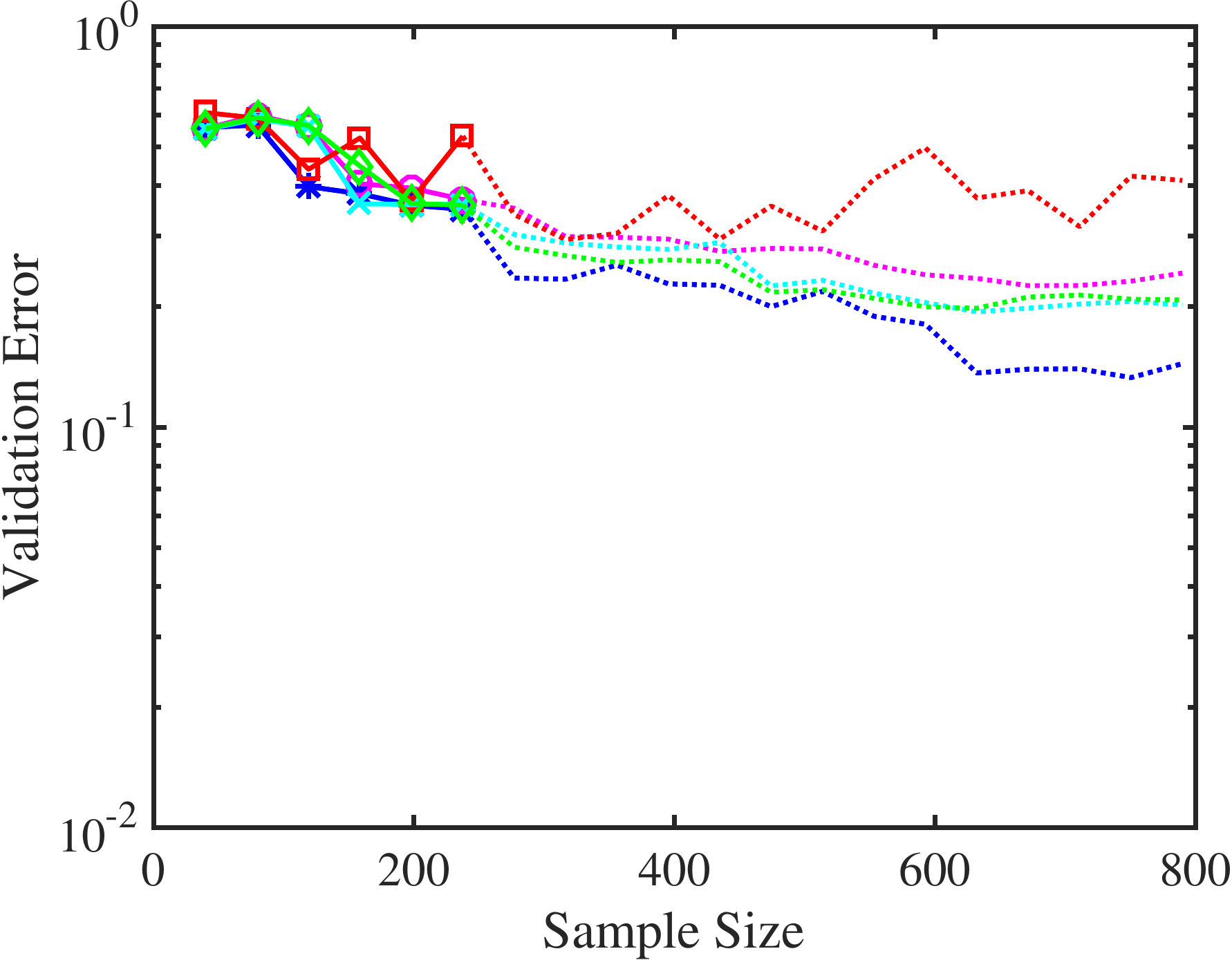}
  \caption{Case 3 Genz-exponential function: CV (left) and validation
    (right) errors for fixed problem instance with $n_s=5$, using
    $p=3$ ($n=55$) (top), $p=5$ ($n=251$) (middle), and $p=7$
    ($n=791$) (bottom). Stop-sampling is activated when the plot line
    turns from solid (with symbols) to dotted (without symbols). OLS
    results are also included for when $p=3$ and $p=5$ systems become
    overdetermined ($m>55$ and $m>251$, respectively), and are shown
    as solid black lines without symbols.}
  \label{f:case3_single_i1_p3_p5_p7_on_p7grid}
\end{figure}

\Cref{f:case3_single_i1_p3_p5_p7_on_p7grid} illustrates the CV (left)
and validation errors (right) for $p=3$ (top), $p=5$ (middle), and
$p=7$ (bottom), which correspond to $n=55$, $251$, and $791$,
respectively.  To ensure a fair comparison, a common $m$ range is
used, up to $m=790$, which corresponds to just before when $p=7$
systems are no longer underdetermined. Even though ordinary least
squares (OLS) becomes a possible option for overdetermined regions of
$p=3$ and $5$, we provide results using the same CS solvers for a
consistent assessment. Separate OLS results are also plotted with
solid black lines, and we can see they are generally very close to
results from the CS solvers. This is not surprising, since as $m$
increases, the contribution from the $\ell_2$-norm term increases
while the $\ell_1$-norm term remains fixed, and so the uLASSO
\cref{e:CS_uLASSO} tends toward OLS (although the OLS solution is much
cheaper to obtain).  From the new plots, the error levels are visibly
better when moving from $p=3$ to $5$, but only improves marginally (in
fact, slightly worse in certain regions) when switching to $p=7$. This
implies improvement to modeling error dominates up to $p=5$, but the
increased complexity starts to outweigh this benefit at $p=7$.

\begin{figure}[htb]
  \centering
  \includegraphics[width=0.365\textwidth, trim={0 4.1em -0.5em 0}, clip]{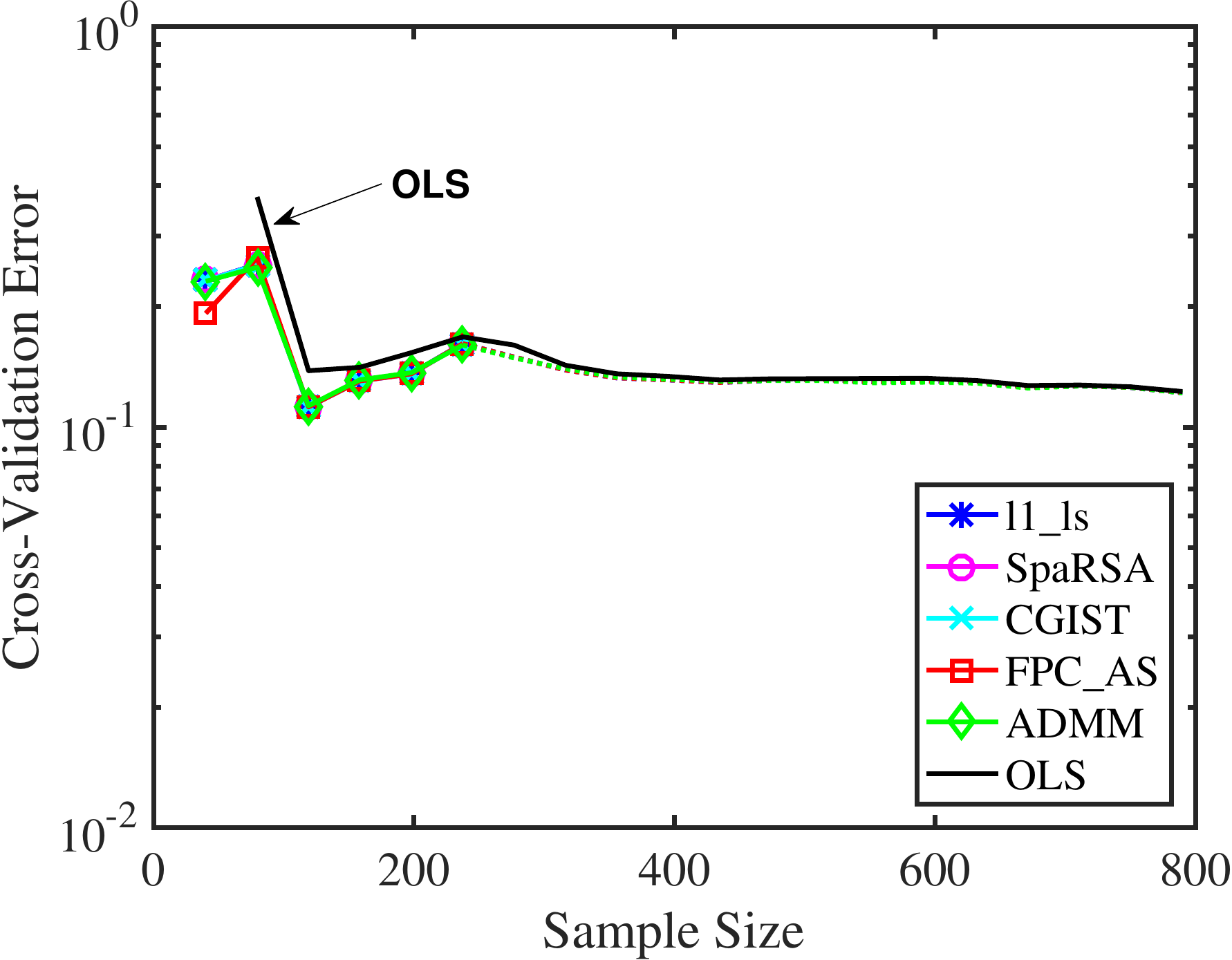}
  \includegraphics[width=0.365\textwidth, trim={0 4.1em -0.5em 0}, clip]{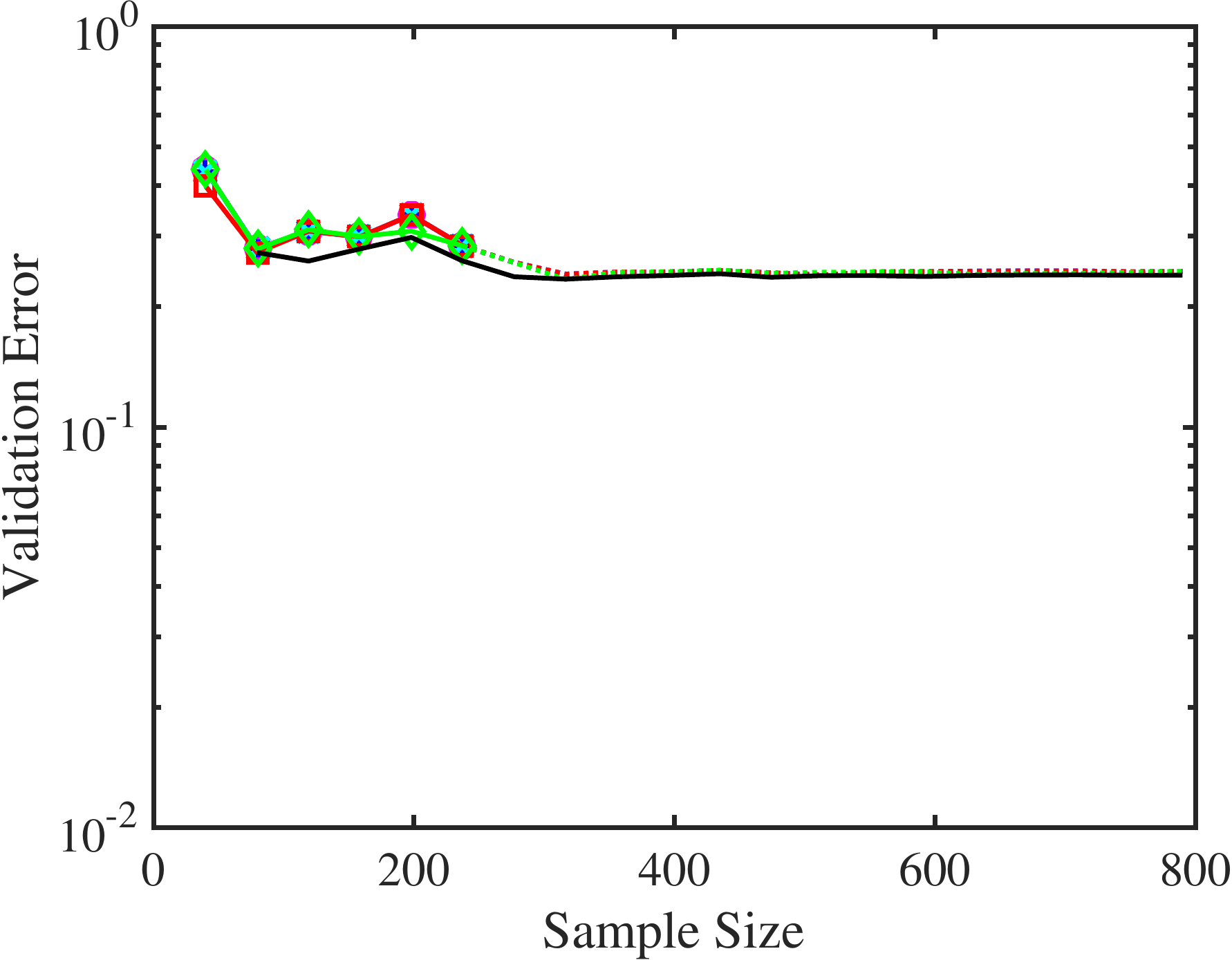}\\
  \includegraphics[width=0.365\textwidth, trim={0 4.1em -0.5em 0}, clip]{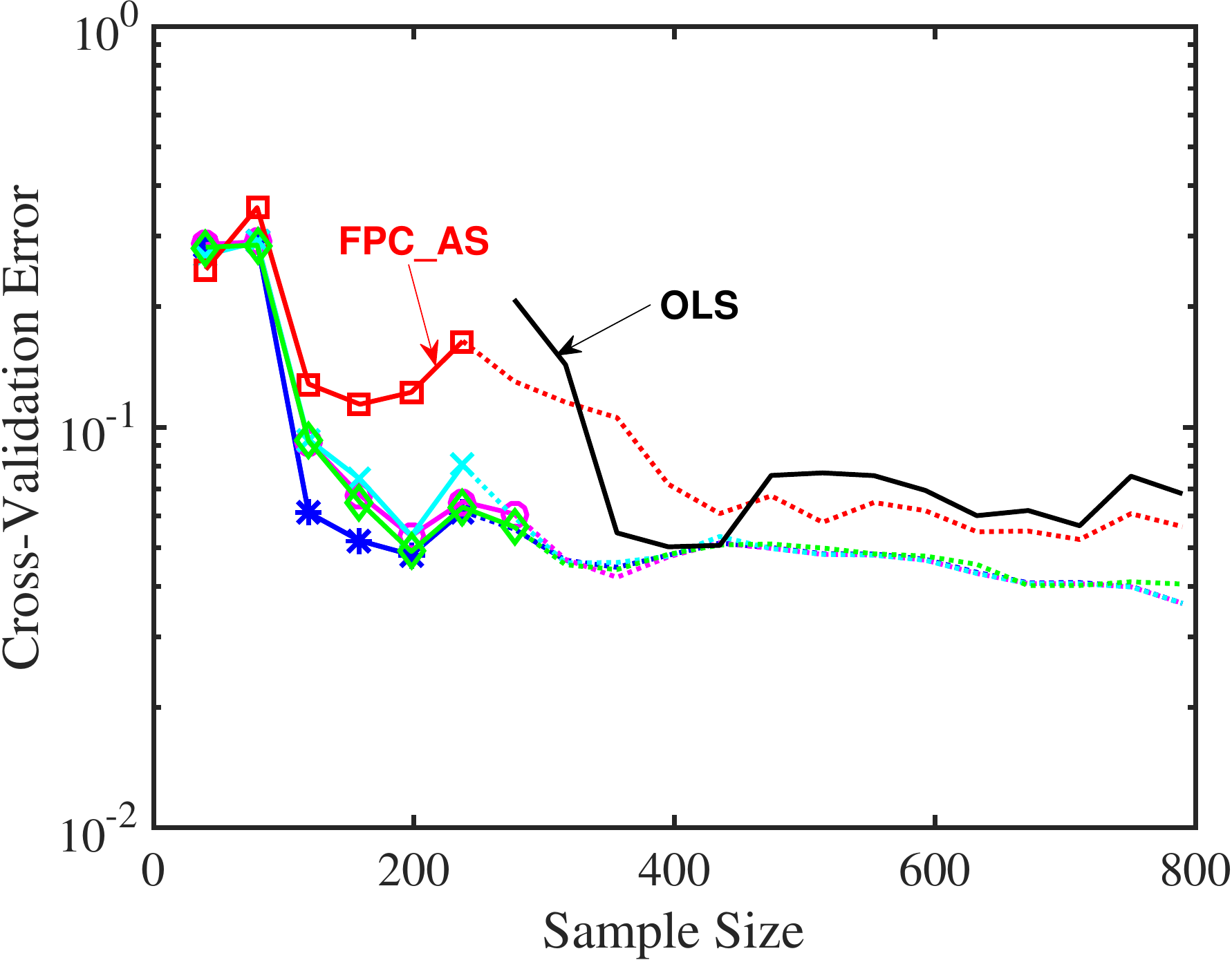}
  \includegraphics[width=0.365\textwidth, trim={0 4.1em -0.5em 0}, clip]{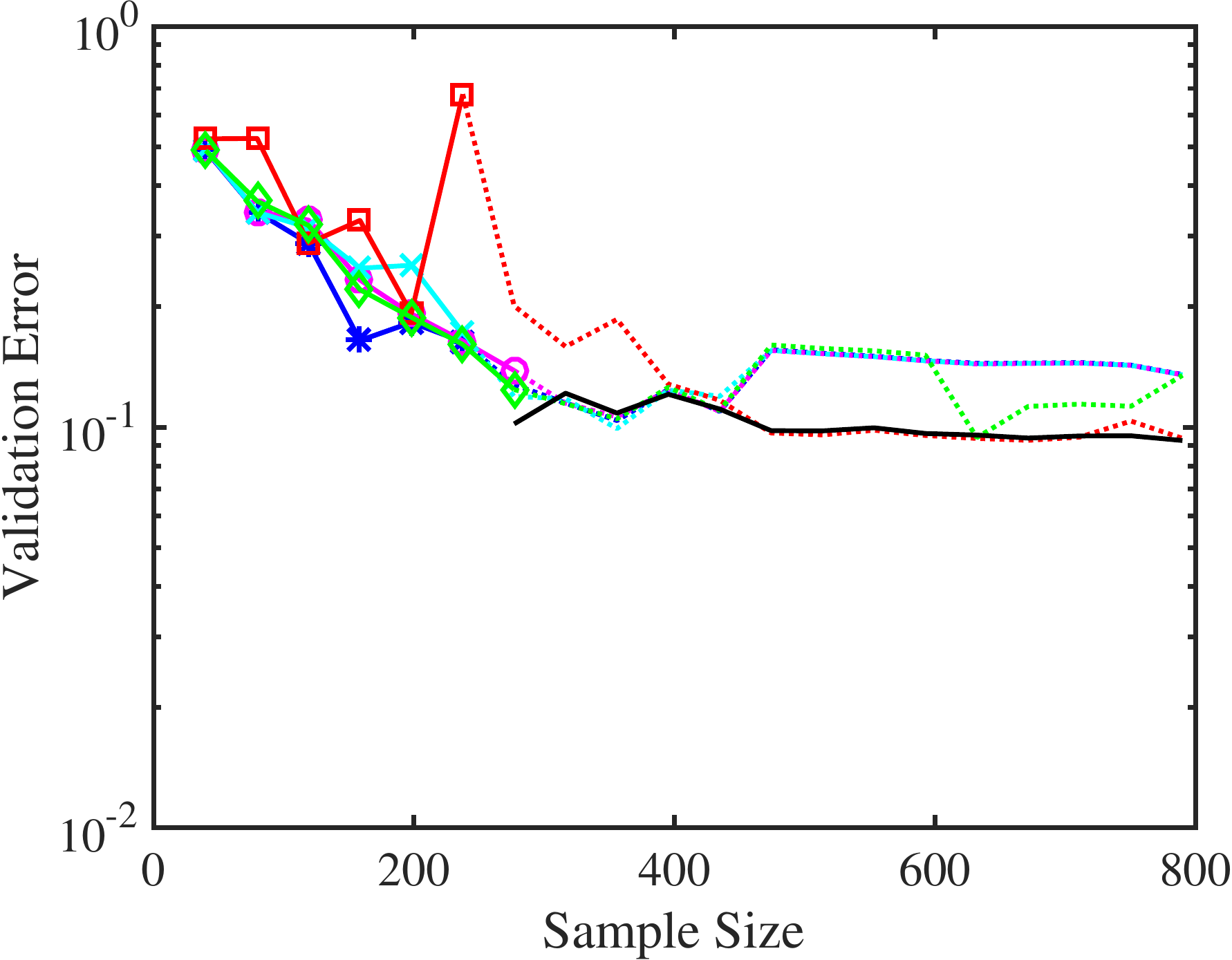}\\
  \includegraphics[width=0.365\textwidth]{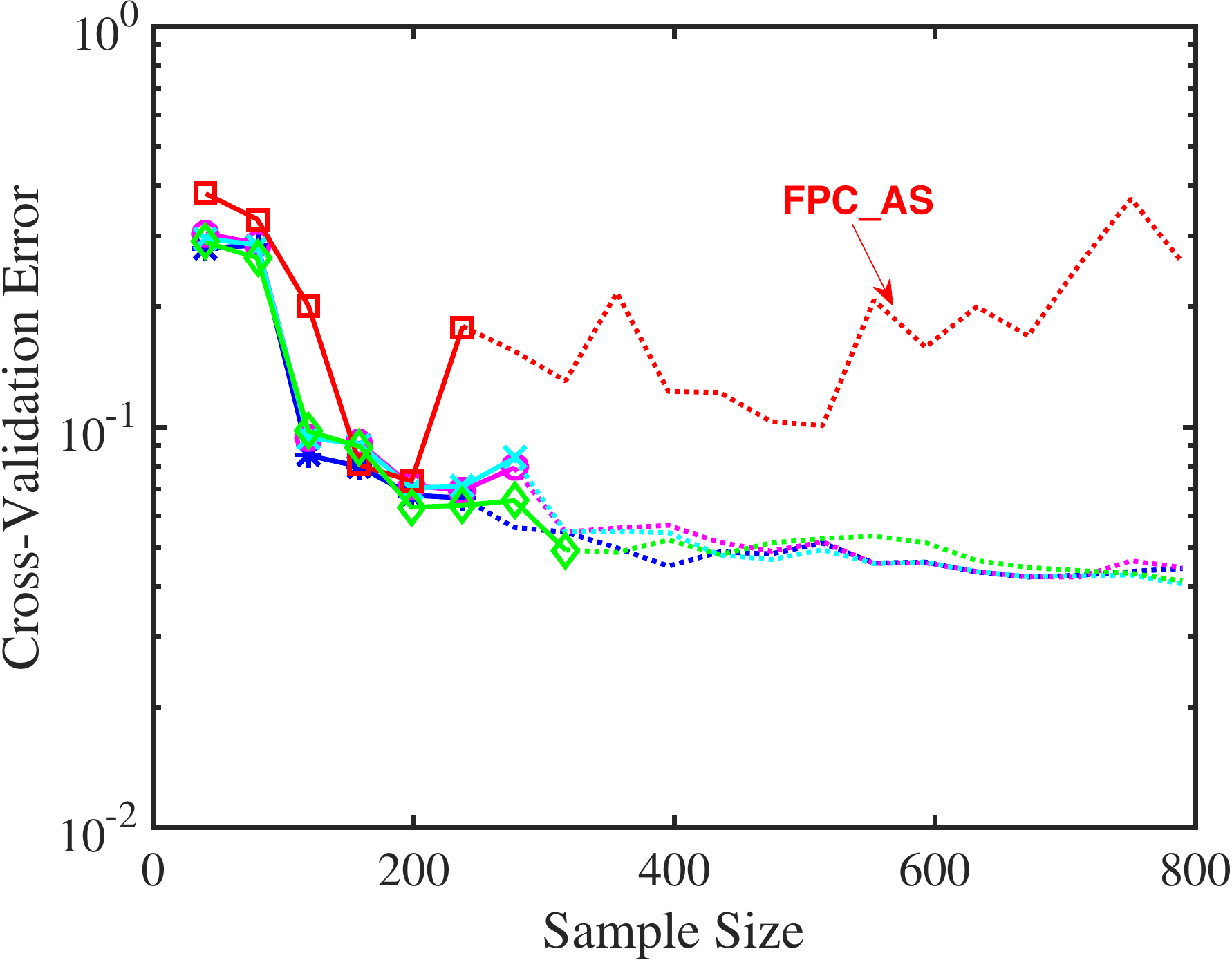}
  \includegraphics[width=0.365\textwidth]{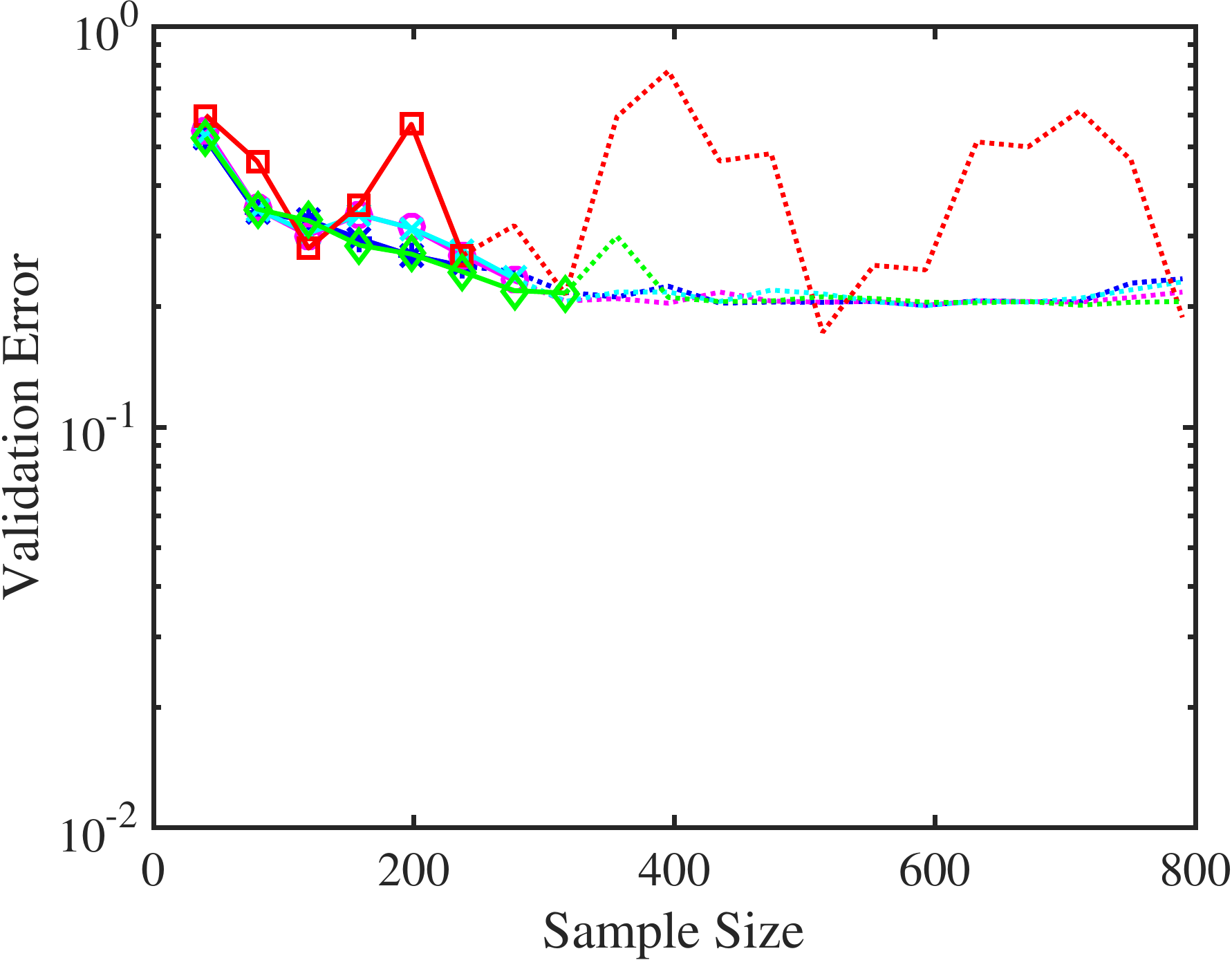}
  \caption{Case 3 Genz-exponential function with artificially imposed
    sparsity by setting $a_2=a_3=0$ in \cref{e:Genz_exp}: CV (left)
    and validation (right) errors for fixed problem instance with
    $n_s=5$, using $p=3$ ($n=55$) (top), $p=5$ ($n=251$) (middle), and
    $p=7$ ($n=791$) (bottom). Stop-sampling is activated when the plot
    line turns from solid (with symbols) to dotted (without
    symbols). OLS results are also included for when $p=3$ and $p=5$
    systems become overdetermined ($m>55$ and $m>251$, respectively),
    and are shown as solid black lines without symbols.}
  \label{f:case3_single_i1_sparse2_and_3_p3_p5_p7_on_p7grid}
\end{figure}

\begin{figure}[htb]
  \centering
  \includegraphics[width=0.47\textwidth]{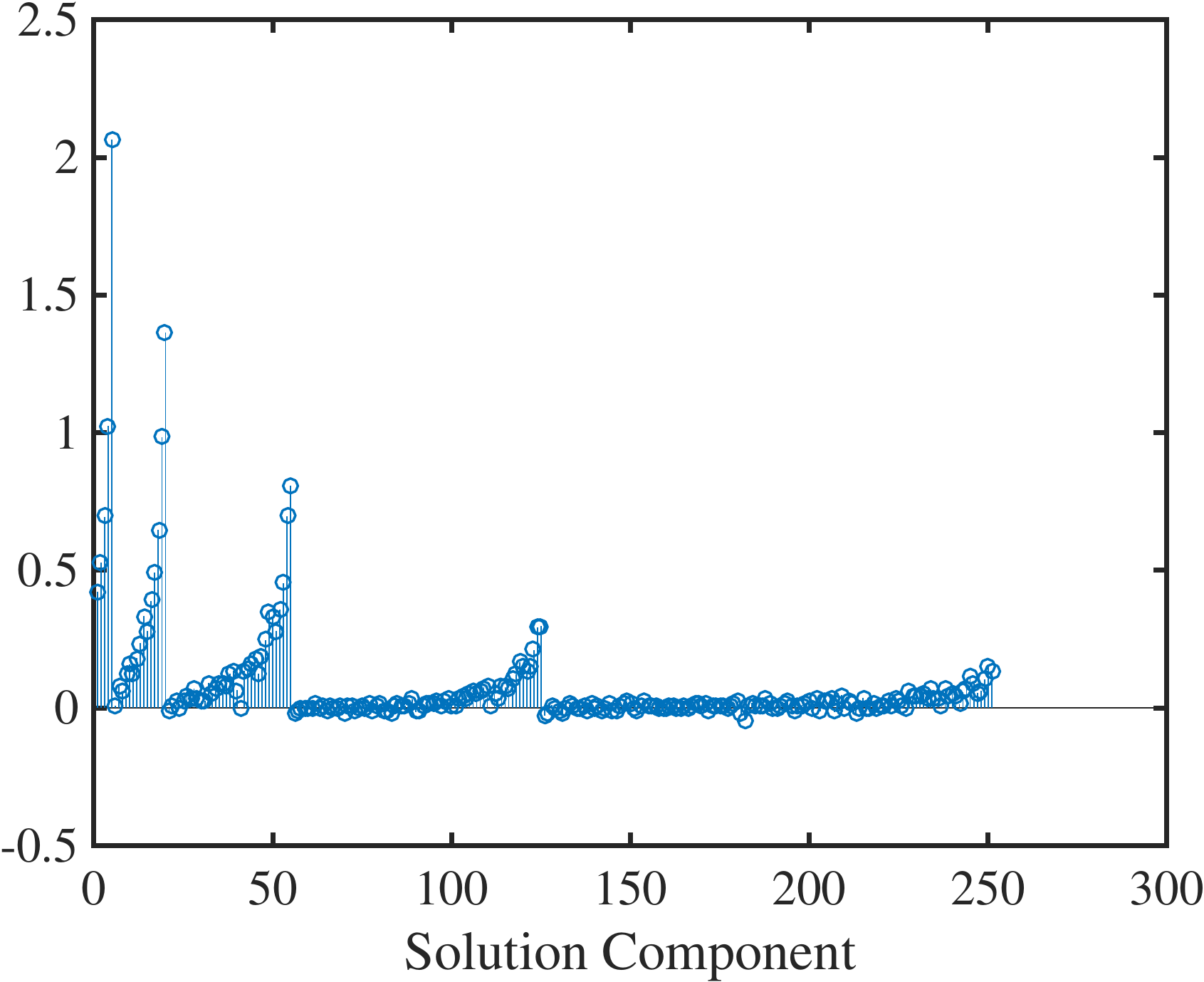}\hspace{2em}
  \includegraphics[width=0.47\textwidth]{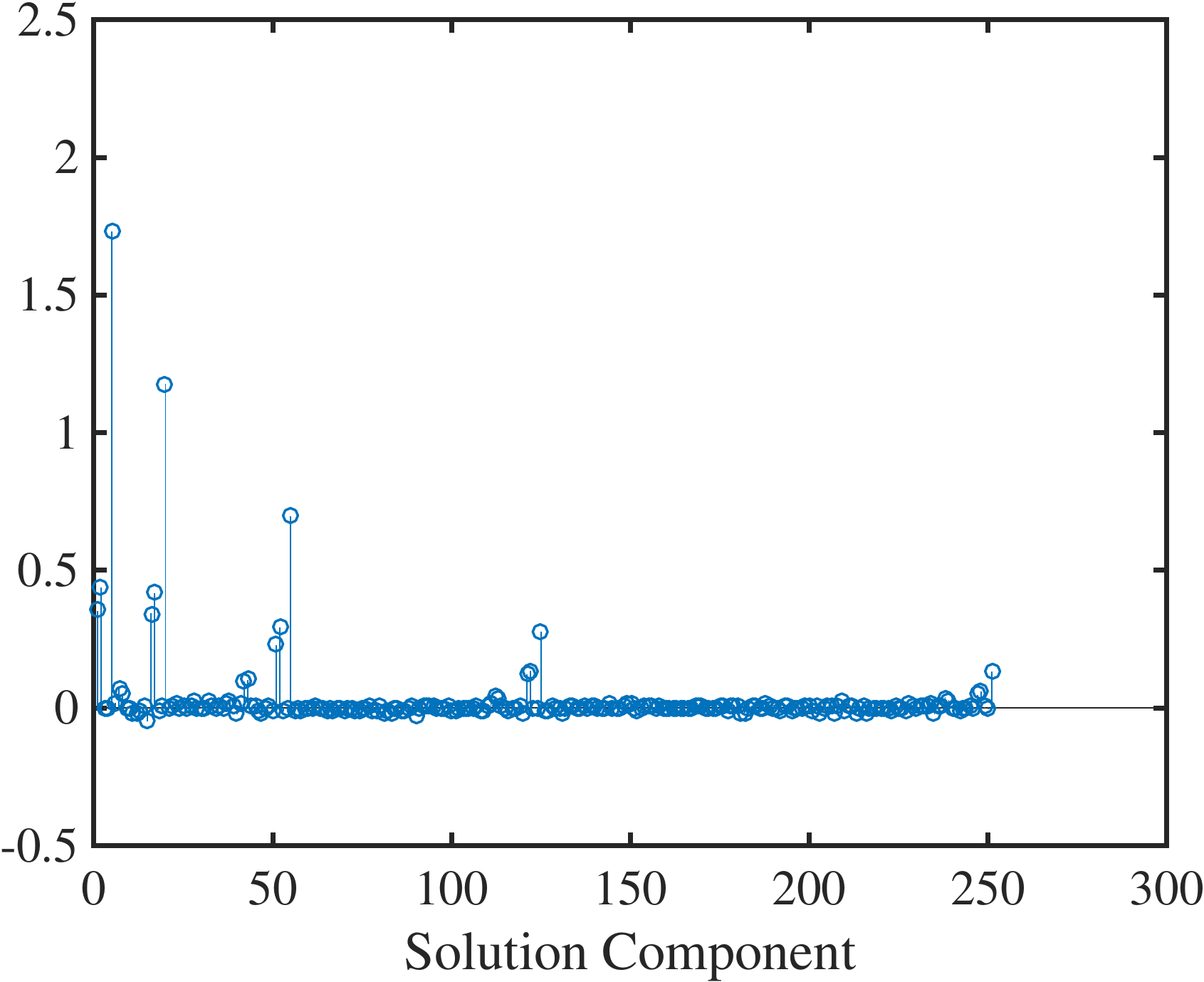}
  \caption{Case 3 Genz-exponential function: solution stem plot for
    $n_s=5$, $p=5$ ($n=251$), using OLS with $m=790$
    (\cref{f:case3_single_i1_p3_p5_p7_on_p7grid} middle row right-most
    end of the black line) (left), and its sparse version counterpart
    (\cref{f:case3_single_i1_sparse2_and_3_p3_p5_p7_on_p7grid} middle
    row right-most end of the black line) (right).}
  \label{f:case3_single_p5_on_p7_grid_select}
\end{figure}

\Cref{f:case3_single_i1_sparse2_and_3_p3_p5_p7_on_p7grid} shows
results for \cref{e:Genz_exp} with $a_2=a_3=0$ while all other
coefficients remain $a_j=j^{-1}$ as before, which induces a more
compressible solution. Compared to the compressible (decaying
coefficients) version in \cref{f:case3_single_i1_p3_p5_p7_on_p7grid},
the new system has similar trends but with sharper error decreases
that also occur at smaller $m$ values, and the error levels are
overall lower. Solution stem plots from $p=5$ using OLS with $m=790$
are shown in \cref{f:case3_single_p5_on_p7_grid_select} for both the
compressible and sparse versions; solutions from CV solvers are
similar and omitted.  All these observations match our intuitive
expectations, where a sparser problem would be easier to
solve. Furthermore, the difference in performance is small between the
two versions, indicating that the numerical methods would still work
well even when zero-coefficients are contaminated. 

All solvers yield similar error levels for this case, with the
exception of \fpcas{}, which produces higher error levels that can
also be quite oscillatory especially at higher polynomial orders.  The
stop-sampling heuristic performs well in detecting the lower end
regions of the largest error drops.

\subsection{Case 4: Jet-in-Crossflow (JXF) application}

We now describe a study involving simulations of the physical
phenomenon when a jet interacts in a crossflow.  JXF is frequently
encountered in many different engineering applications, such as
fuel injection and combustor cooling in gas turbine engines and plume dispertion.
A comprehensive review can be found in Mahesh~\cite{Mahesh2013}.
In this paper we focus on a setup that corresponds to the HiFiRE program~\cite{HiFiRE2010}.
This configuration is relevant for the design of supersonic
combusting ramjet (scramjet) engines~\cite{Huan2018a}. Having a
fundamental understanding of this physical behavior is thus extremely
valuable for producing accurate simulations and high-performing
designs.

\begin{figure}
  \centering
  \includegraphics[width=1\textwidth]{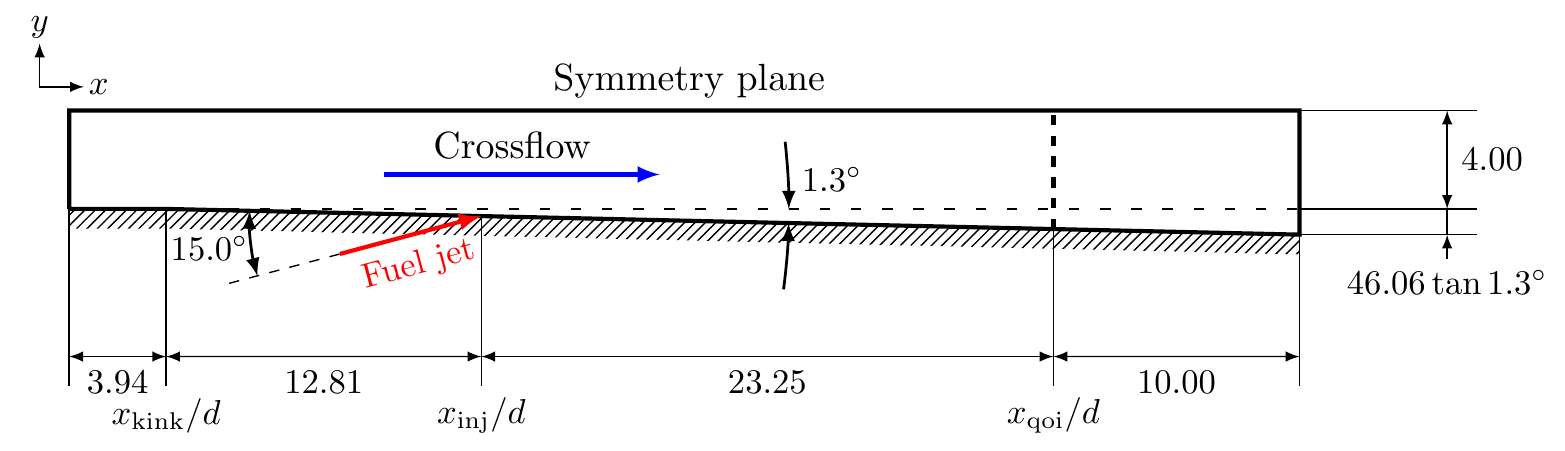}
  \caption{Case 4 JXF application: computational schematic of the
    injected fuel jet in a supersonic crossflow. All distance
    quantities shown are normalized by the injector diameter $d =
    3.175$ mm. For example, $x_{\textnormal{qoi}}=40 \,d = 12.7$ cm.}
  \label{f:LES_schematic}
\end{figure}

As an initial exploratory step of the overall design project, this
setup involves simulations of supersonic turbulent fuel jet and
crossflow in a simplified two-dimensional computational domain
presented in \cref{f:LES_schematic}. While 2D turbulence phenomenon is
not physically realistic, it is useful in providing physical insights
and for testing our CS methodology. The crossflow travels from left to
right in the $x$-direction (streamwise), and remains supersonic
throughout the entire domain. The geometry is symmetric about the top
in the $y$-direction (wall-normal), and is endowed with symmetry
boundary conditions.  Normalized spatial distance units $x/d$ and
$y/d$ are used in subsequent descriptions and results, where $d=3.175$
mm is the diameter of the fuel injector.  The bottom of the geometry
is a solid wall, with a downward slope of $1.3^{\circ}$ starting from
$x_{\textnormal{kink}}/d=3.94$. Fuel is introduced at
$x_{\textnormal{inj}}/d=16.75$ through an injector aligned an angle of
$15^{\circ}$ from the wall.  The fuel is the JP-7 surrogate, which
consists of 36\% methane and 64\% ethylene.  Combustion is turned off
for this study, allowing a targeted investigation of the interaction
between the fuel jet and the supersonic crossflow without the effects
of chemical heat release.

Large eddy simulation (LES) calculations are then carried out using
the RAPTOR code framework developed by Oefelein~\cite{Oefelein2006,
  Oefelein1997}.  The code performs compressible numerical simulation
and has been optimized to meet the strict algorithmic requirements
imposed by the LES formalism.  Specifically, it solves the
fully-coupled conservation equations of mass, momentum, and
total-energy, under high Reynolds number, high Mach number, and with
real-gas and liquid phases. It also accounts for detailed
thermodynamics and transport processes at the molecular level. A
relatively coarse grid is used, where each grid has size $d/8$. Each
simulation requires around 12 total CPU-hours to produce statistically
converged results.

\begin{figure}[htb]
  \centering
  \includegraphics[width=0.6\textwidth]{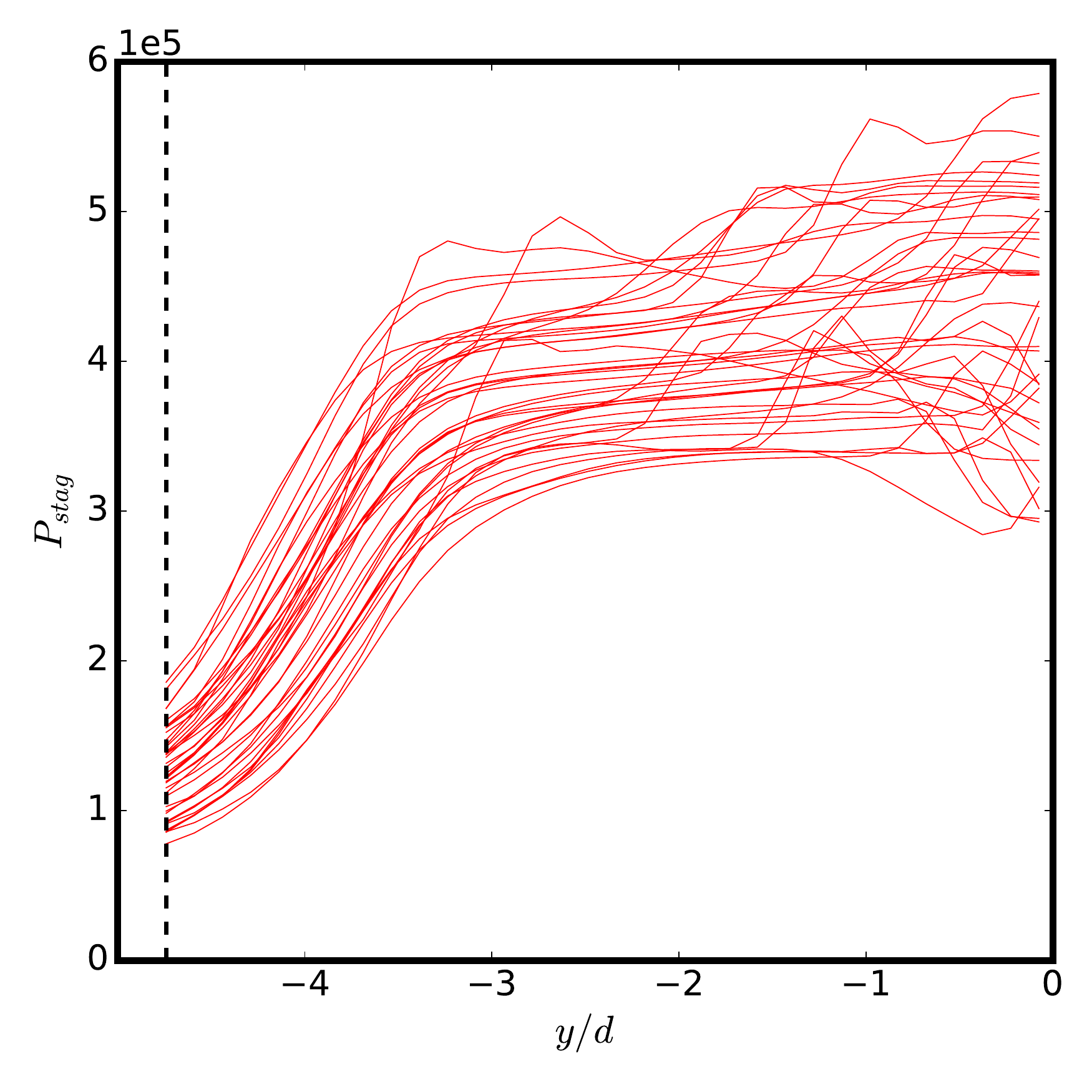}
  \caption{Case 4 JXF application: profiles of time-averaged
    (normalized) $P_{\textnormal{stag}}$ plotted over $y/d$, at fixed
    $x_{\textnormal{qoi}}/d=100$, for selected runs of different
    parameter settings. Each red line is the profile of an independent
    simulation drawn from the 24-dimensional parameter space according
    to the distributions described in \cref{t:scram_params}. The
    vertical dotted line on the left side is the bottom wall.}
  \label{f:qoi1_profile_examples}
\end{figure}

We let $n_s=24$ model parameters be uncertain with distributions
described in \cref{t:scram_params}. The wall temperature ($T_{w}$)
boundary condition is a function of $x/d$, and hence is a random field
(RF). $T_{w}$ is thus represented using the Karhunen-Lo\`{e}ve
expansion (KLE) (e.g.,~\cite{Ghanem1991}), which is built employing
the eigenstructure of the covariance function of the RF to achieve an
optimal representation. We employ a Gaussian RF with a square
exponential covariance structure along with a correlation length that
is similar to the largest turbulent eddies (i.e., the size of the
crossflow inlet). The mean temperature profile is constructed by
averaging temperature profile results from a small set of separate
adiabatic simulations. The correlation length employed leads to a
rapid decay in characteristic-mode amplitudes, allowing us to capture
about 90\% of the total variance of this RF with only a
ten-dimensional KLE. The wall temperature is further assumed to be
constant in time.  All other parameters are endowed with uniform
distributions. More details on the KLE model are presented in Huan et
al.~\cite{Huan2018a}. The mixture of Gaussian and uniform
distributions is accommodated by a hybrid Gauss-Hermite and
Legendre-Uniform PCE, with the appropriate type used for each
dimension.  The model output quantity of interest (QoI) is selected to
be the time-averaged stagnation pressure $P_{\textnormal{stag}}$
located at $x_{\textnormal{qoi}}/d=100$ (near the outlet) and
spatially averaged over $y/d$. $P_{\textnormal{stag}}$ is an important
output in engine design, since the stagnation pressure differential
between inlet and outlet is a relevant component in computing the
thrust.  Examples of $P_{\textnormal{stag}}$ plotted over $y/d$,
i.e. before spatial-averaging, are shown in
\cref{f:qoi1_profile_examples}; the wall boundary effect of reducing
the $P_{\textnormal{stag}}$ (left side of the plot) is evident.

\begin{table}[htb]
\caption{Case 4 JXF application: uncertain model input parameters. The
  uncertain distributions are uniform across the ranges shown, with
  the exception for the bottom wall temperature which is expressed as
  a KLE involving 10 standard Gaussian random variables.}
\label{t:scram_params}
\begin{center}
\begin{tabular}{cll}
\hline
Parameter &  Range  & Description \\
\hline
\multicolumn{3}{c}{Inlet boundary conditions}\\
\hline
$p_{0}$ & $[1.406, 1.554]\times 10^{6}$ Pa & Stagnation pressure   \\
$T_{0}$ & $[1472.5, 1627.5]$ K & Stagnation temperature  \\
$M_0$ & $[2.259, 2.761]$ & Mach number  \\
$\delta_{i}$ & $[2, 6] \times 10^{-3}$ m & Boundary layer thickness \\
$I_{i}$ & $[0, 0.05]$ & Turbulence intensity magnitude  \\
$L_{i}$ & $[0, 8]\times 10^{-3}$ m & Turbulence length scale  \\
\\
\hline
\multicolumn{3}{c}{Fuel inflow boundary conditions}\\
\hline
$\dot{m}_f$ & $[6.633, 8.107] \times 10^{-3}$ kg/s & Mass flux   \\
$T_{f}$ & $[285, 315]$ K & Static temperature \\
$M_f$ & $[0.95, 1.05]$ & Mach number \\
$I_{f}$ & $[0, 0.05]$ & Turbulence intensity magnitude \\
$L_{f}$ & $[0, 1]\times 10^{-3}$ m & Turbulence length scale  \\
\\
\hline
\multicolumn{3}{c}{Turbulence model parameters} \\ \hline
$C_R$ & $[0.01, 0.06]$  & Modified Smagorinsky constant  \\
$Pr_t$ & $[0.5, 1.7]$ & Turbulent Prandtl number \\
$Sc_t$ & $[0.5, 1.7]$ & Turbulent Schmidt number \\
\\
\hline
\multicolumn{3}{c}{ Wall boundary conditions}    \\  \hline
$T_{w}$  & Expansion in 10 params & Wall temperature represented via \\
& of i.i.d. $\CN(0,1)$ & Karhunen-Lo\`{e}ve expansion (KLE) \\
\hline
\end{tabular}
\end{center}
\end{table}

\begin{figure}[htb]
  \centering
  \includegraphics[width=0.49\textwidth]{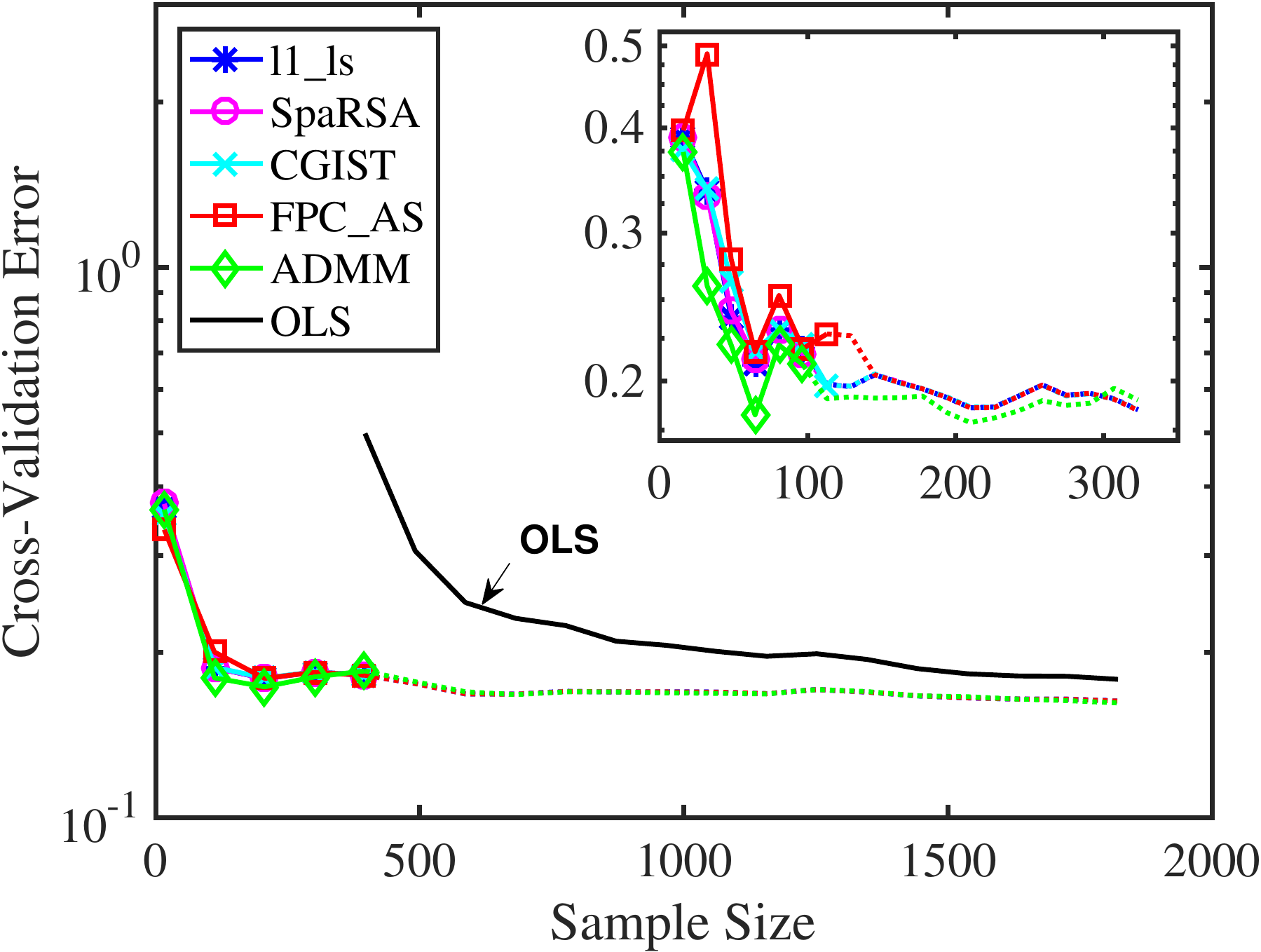}
  \includegraphics[width=0.49\textwidth]{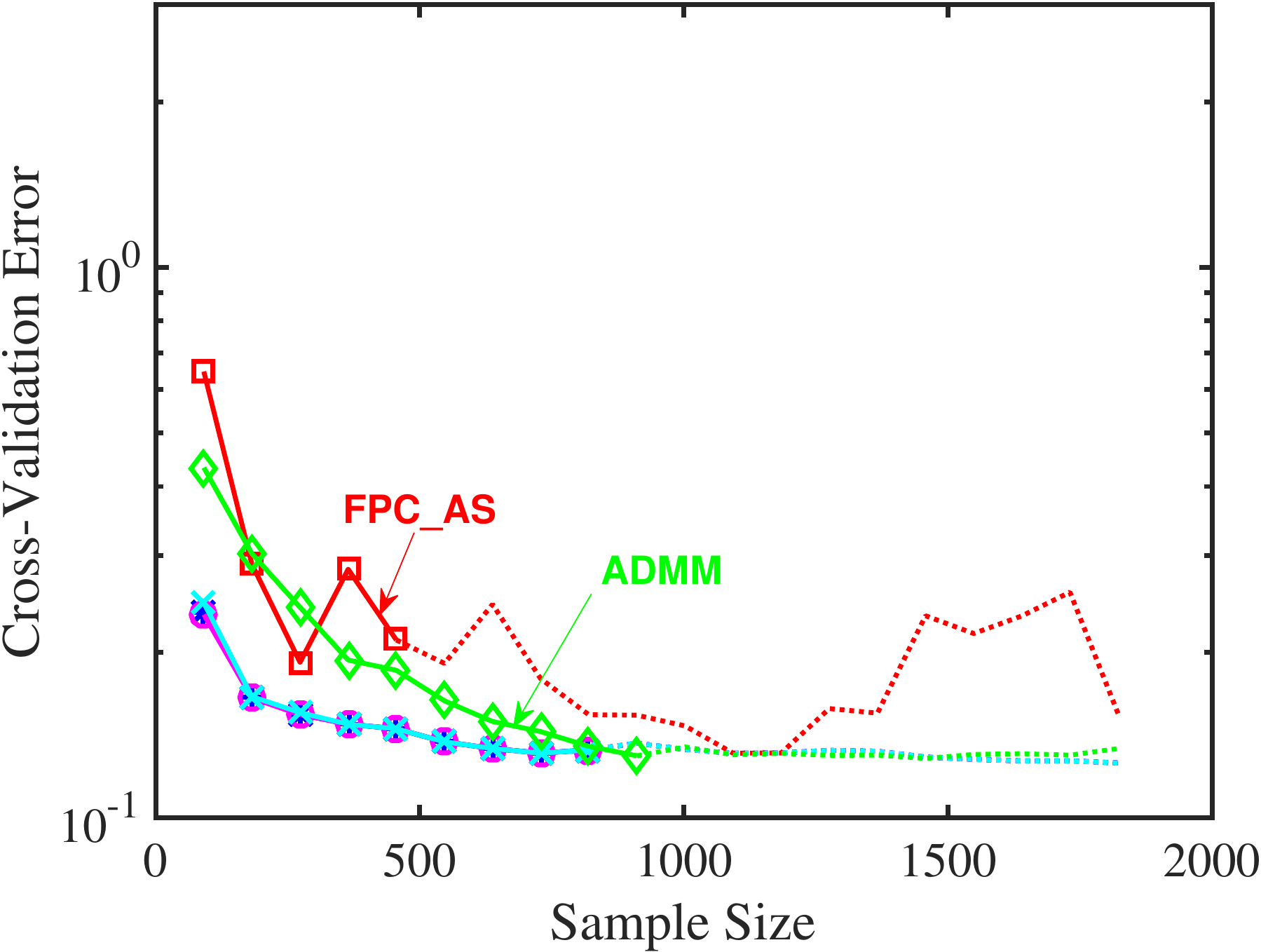}
  \caption{Case 4 JXF application: CV error for $n_s=24$, using $p=2$
    ($n=324$) (left) and $p=3$ ($n=2924$) (right). The inset on the
    left panel shows results using a finer $m$ discretization up to
    $m=323$, which corresponds to just before when $p=2$ systems
    become underdetermined. Stop-sampling is activated when the plot
    line turns from solid (with symbols) to dotted (without
    symbols). OLS results are also included for when $p=2$ systems
    become overdetermined ($m>324$), and are shown as solid black
    lines without symbols.}
  \label{f:case4_single_p3_qoi1_CV}
\end{figure}

PCEs with total-order polynomial degrees $p=2$ ($n=324$) and $p=3$
($n=2924$) are constructed using \cref{a:CS_alg_overall}, and their CV
errors are shown in \cref{f:case4_single_p3_qoi1_CV}. The left and
right plots correspond to results from $p=2$ and $3$ for $m$ sizes up
to 1822, and the inset on the left panel shows results using a finer
$m$ discretization up to $m=324$, which corresponds to just before
when $p=2$ systems are no longer underdetermined. OLS results are also
included when $p=2$ systems become overdetermined, and the difference
between CS and OLS results decreases as $m$ increases, as expected.
The error levels are slightly lower for $p=3$ than $2$, implying that
the overall advantage from the enriched $p=3$ basis is present, but
not significant. (Additional testing using $p=4$, not shown in this
paper, also produced similar solutions, thus supporting that $p=3$ is
appropriate for this application.)  The trends appear smooth and
well-behaved, with a rather rapid drop at around $m=100$ for $p=2$
($n=324$), and $m=300$ for $p=3$ ($n=2924$), hinting at a sparse or
compressible solution. Stop-sampling is able to capture the drop well
for the $p=3$ case ($p=2$ errors do not drop sufficiently).  All
solvers perform similarly for $p=2$ but move apart for $p=3$
especially at low $m$, with \admm{} and \fpcas{} yielding higher
errors than others. \fpcas{} also experiences large error
oscillations.

\begin{figure}[htb]
  \centering
  \includegraphics[width=0.47\textwidth]{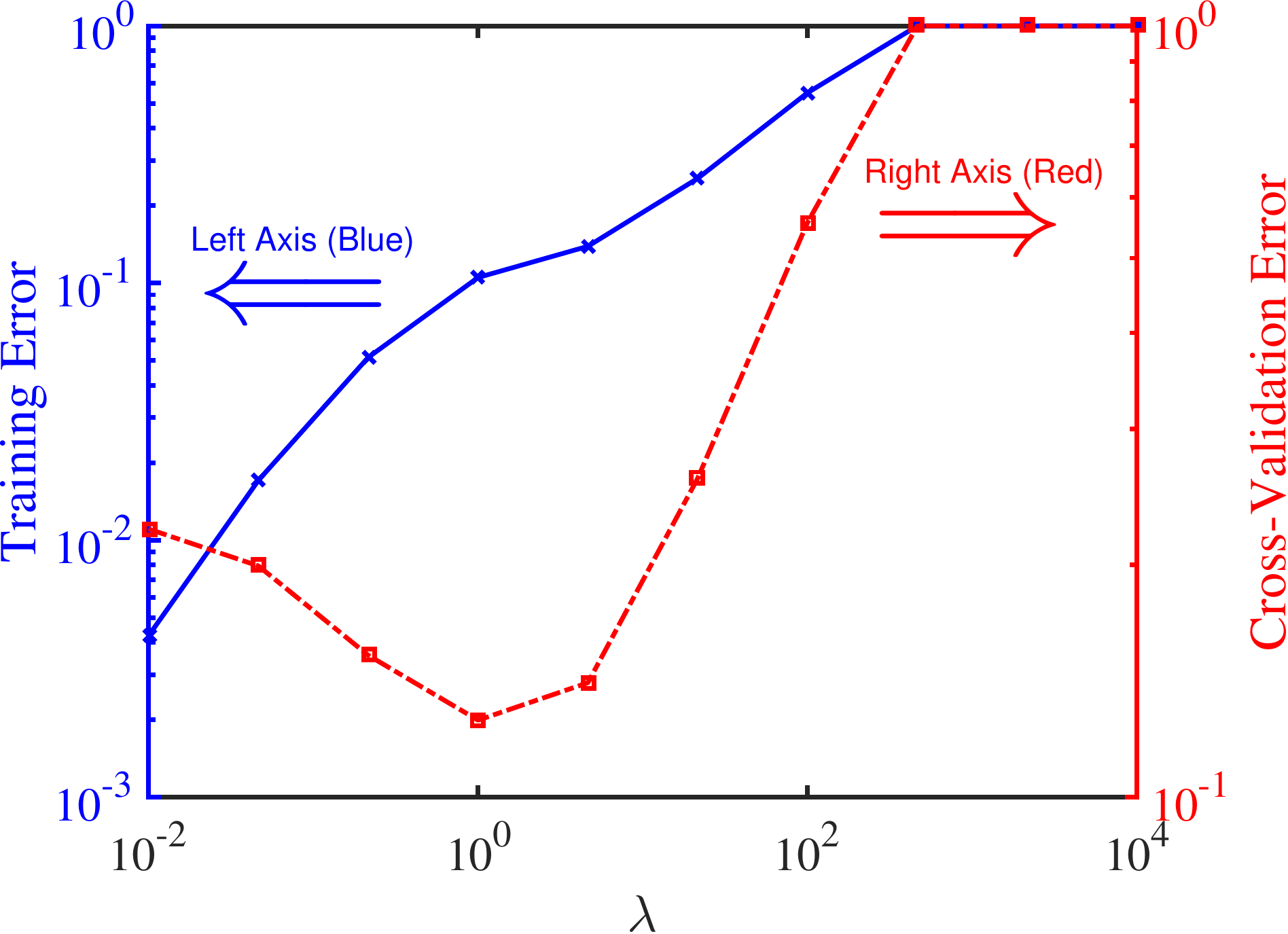}\hspace{2em}
  \includegraphics[width=0.425\textwidth]{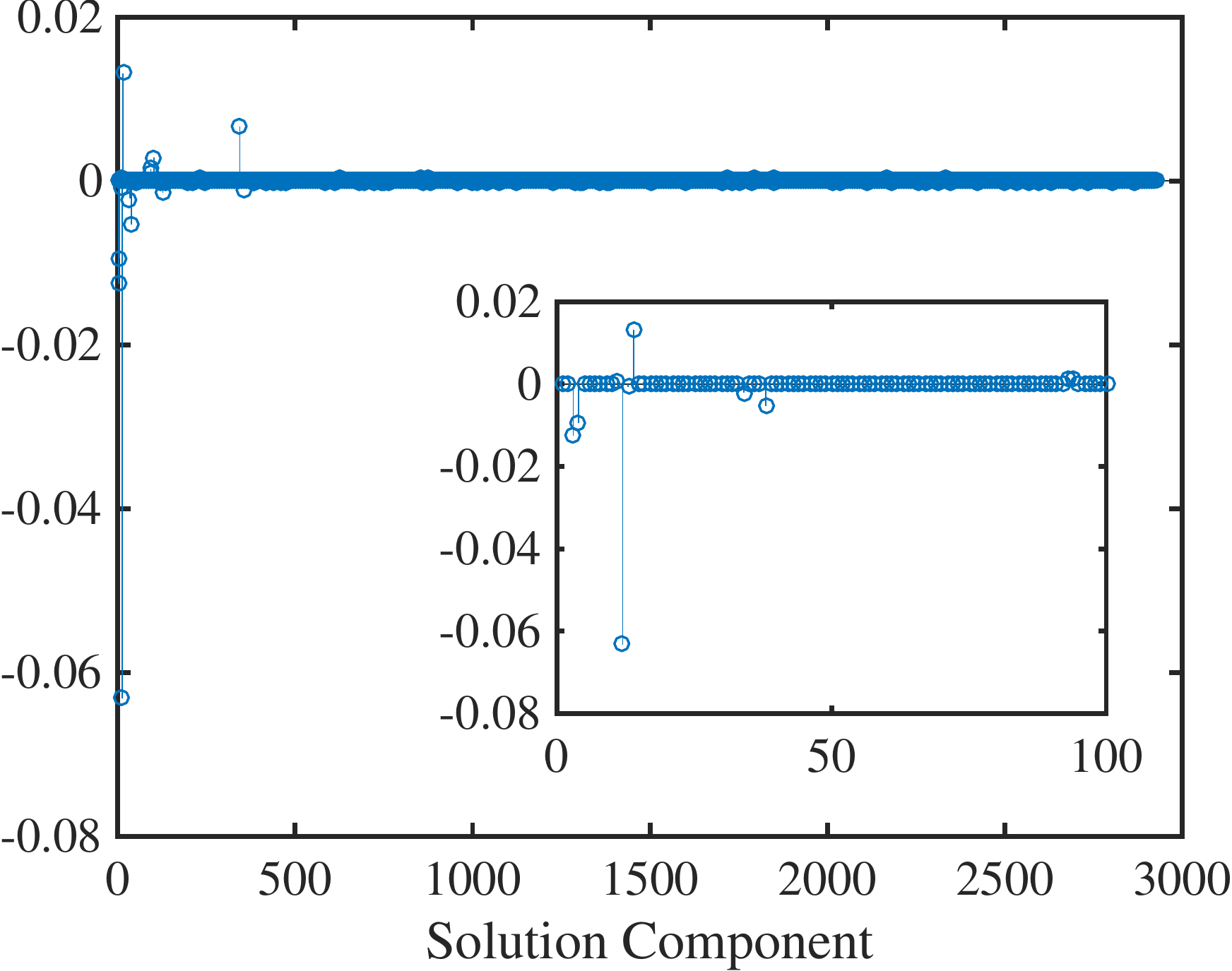}
  \caption{Case 4 JXF application: training and CV errors (left;
    training error in solid blue using left $y$-axis, and CV error in
    dotted red using right $y$-axis) and corresponding lowest-CV-error
    solution stem plot (right) for $n_s=24$, $p=3$ ($n=2924$), using
    \sparsa{} with $m=1822$ (\cref{f:case4_single_p3_qoi1_CV} right
    panel right-most end of the magenta line). The inset on the right
    panel shows the stem plot zoomed in for the first 100 components.
    This is the best solution obtained for case 4, and serves as a
    reference for subsequent comparisons.}
  \label{f:case4_single_p3_qoi1_best}
\end{figure}

Since the true solution is unknown, we study the best numerical
solution encountered in our computations, which is produced by $p=3$
using \sparsa{} with $m=1822$ (\cref{f:case4_single_p3_qoi1_CV} right
panel, right-most end of the magenta line). The training and CV errors
as a function of $\lambda$ are shown in the left panel in
\cref{f:case4_single_p3_qoi1_best}, and the solution stem plot for the
lowest-CV-error solution is shown on the right. The solution indeed
appears sparse, confirming our earlier hypothesis. The most dominating
coefficients, in decreasing magnitude, correspond to the linear terms
in $M_0$, $p_0$, $\delta_i$, and $L_i$. This is consistent with the
global sensitivity analysis performed in~\cite{Huan2018a}, where the
leading total sensitivity Sobol indices for $P_{\textnormal{stag}}$
were $\{M_0: 0.904\}$, $\{p_0: 0.033\}$, $\{\delta_i: 0.032\}$, and
$\{L_i: 0.031\}$.  While the Sobol indices are not exactly equal to
the linear coefficients (Sobol indices also include sum of
coefficients-squared from higher order orthnormal basis terms), they
are dominated by the linear coefficients in this case. This also
explains the rather small improvement of $p=3$ compared to $p=2$
results in \cref{f:case4_single_p3_qoi1_CV}, as most of the
third-order terms have negligible coefficients. These observations are
consistent with physics-based intuition: since the current setup does
not involve combustion, one would expect the crossflow conditions to
dictate the impact on the QoI behavior.

\begin{figure}[htb]
  \centering
  \includegraphics[width=0.47\textwidth]{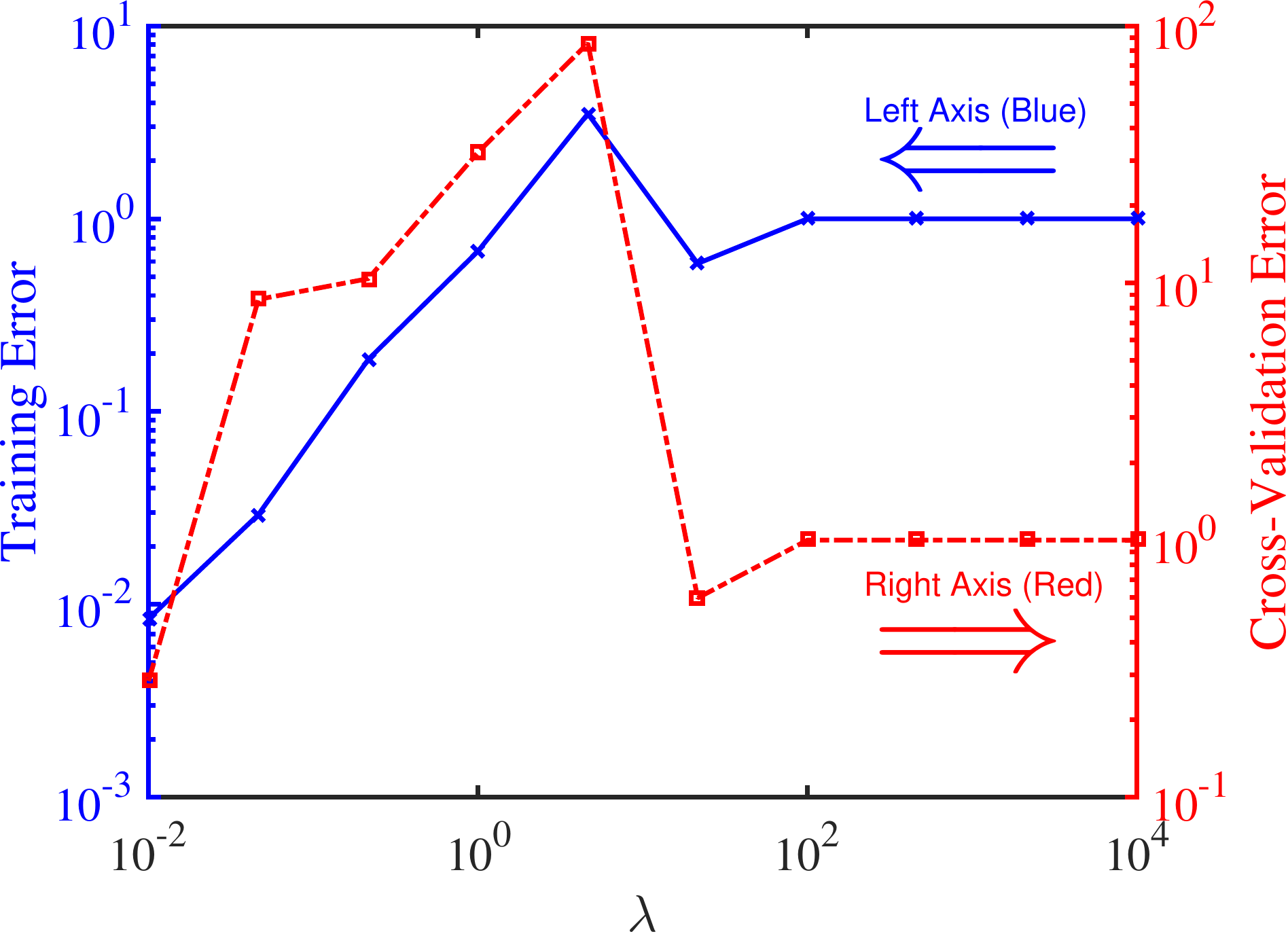}\hspace{2em}
  \includegraphics[width=0.425\textwidth]{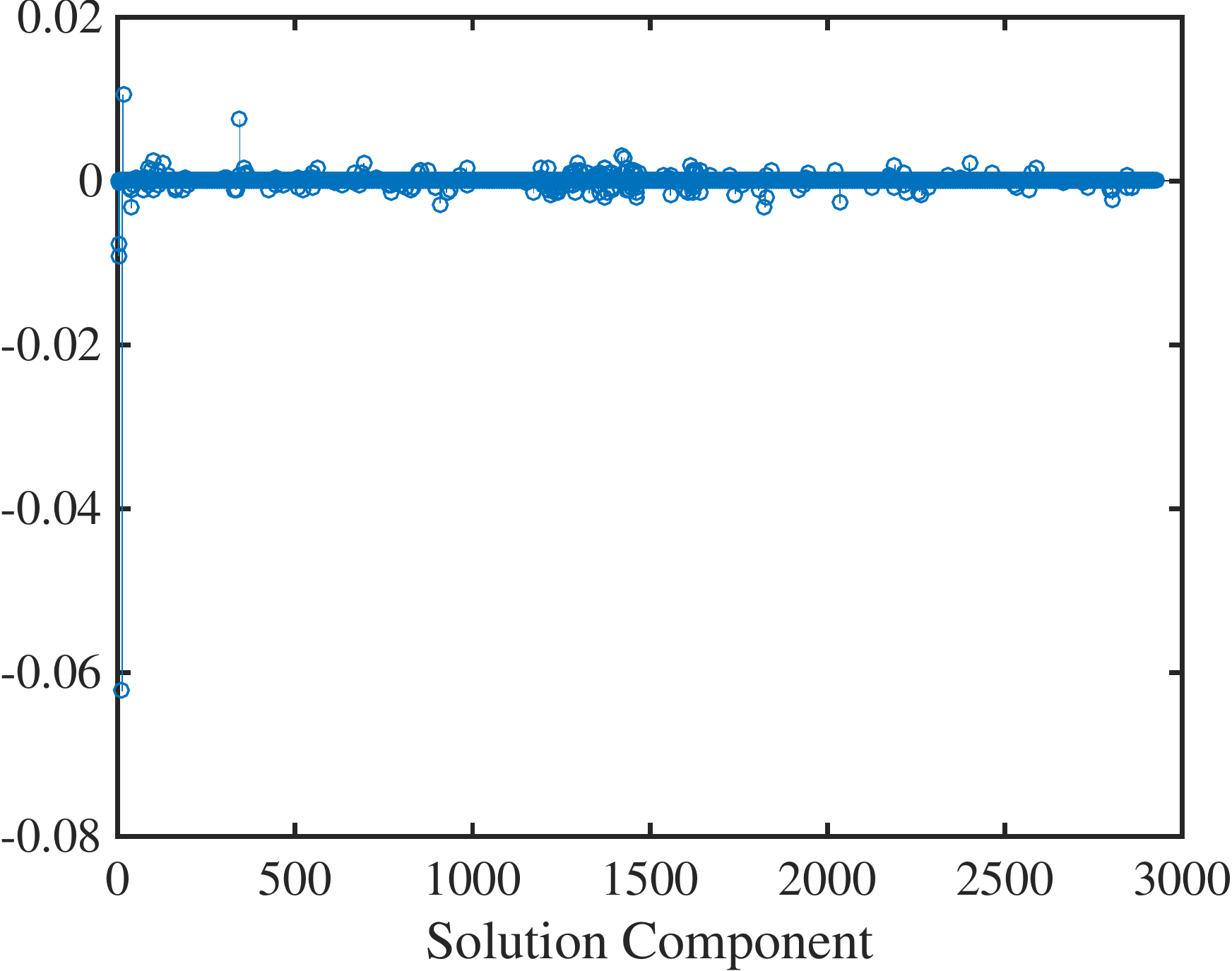}
  \caption{Case 4 JXF application: training and CV errors (left;
    training error in solid blue using left $y$-axis, and CV error in
    dotted red using right $y$-axis) and corresponding lowest-CV-error
    solution stem plot (right) for $n_s=24$, $p=3$ ($n=2924$), using
    \fpcas{} with $m=364$ (\cref{f:case4_single_p3_qoi1_CV} right
    panel fourth point of red line). This is an example of bad
    solution compared to the reference in
    \cref{f:case4_single_p3_qoi1_best}.}
  \label{f:case4_single_p3_qoi1_select}
\end{figure}

In contrast to \cref{f:case4_single_p3_qoi1_best}, an example of a bad
solution is shown in \cref{f:case4_single_p3_qoi1_select}, produced by
$p=3$ using \fpcas{} with $m=364$ (\cref{f:case4_single_p3_qoi1_CV}
right panel fourth point of red line). Similar to the example in
\cref{f:case1_single_select}, the CV error has a ``bump'' instead of a
``dip''. In this case, the lowest-CV-error solution is at the smallest
$\lambda$ in the range investigated. The solution stem plot is shown
on the right panel, and the small $\lambda$ induces a dense vector
with many non-zero components clearly visible. The highest magnitude
coefficients, however, are still correctly identified.

\subsection{Timing discussions}

\begin{figure}[htb]
  \centering
  \includegraphics[width=0.45\textwidth]{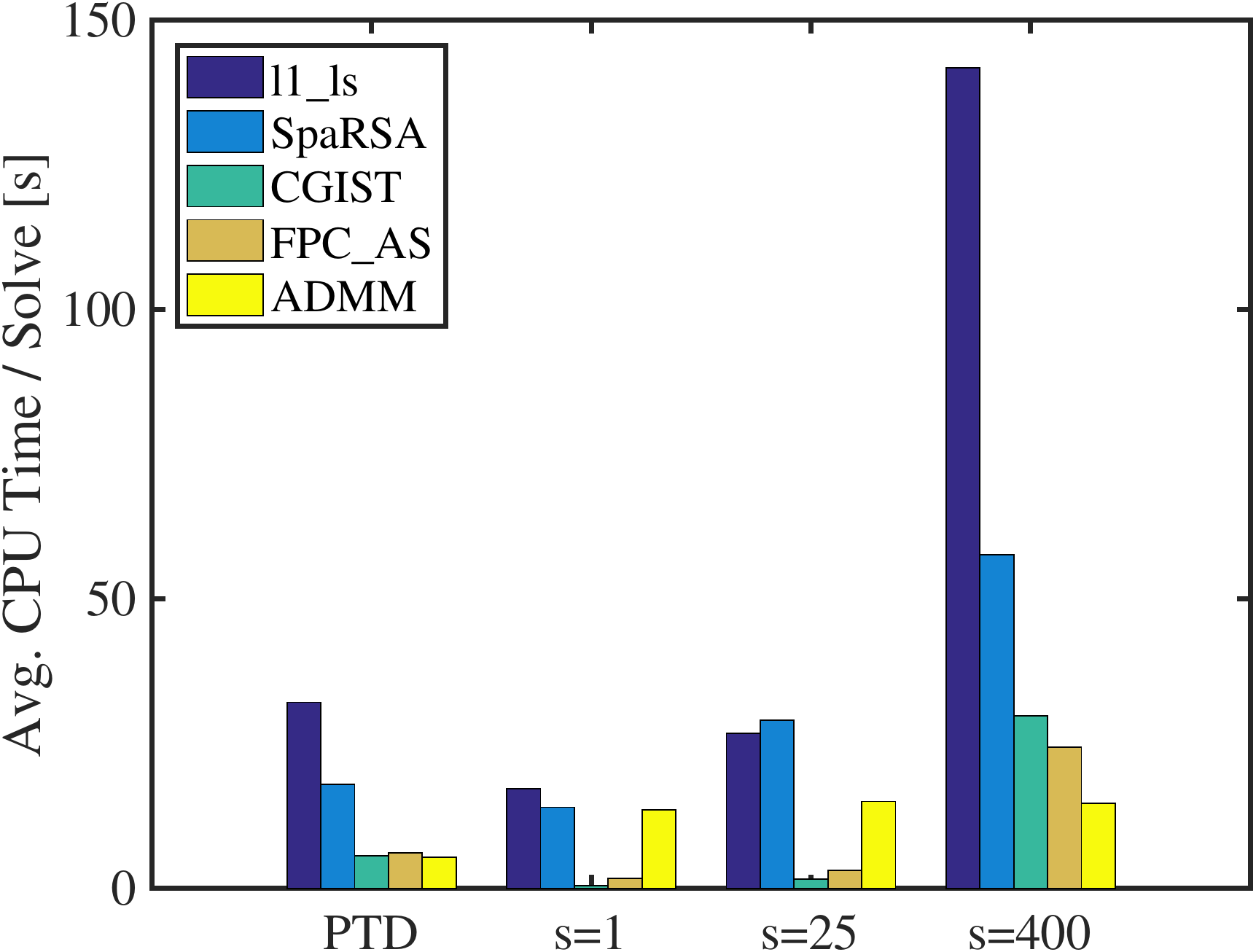}\hspace{1.2em}
  \includegraphics[width=0.45\textwidth]{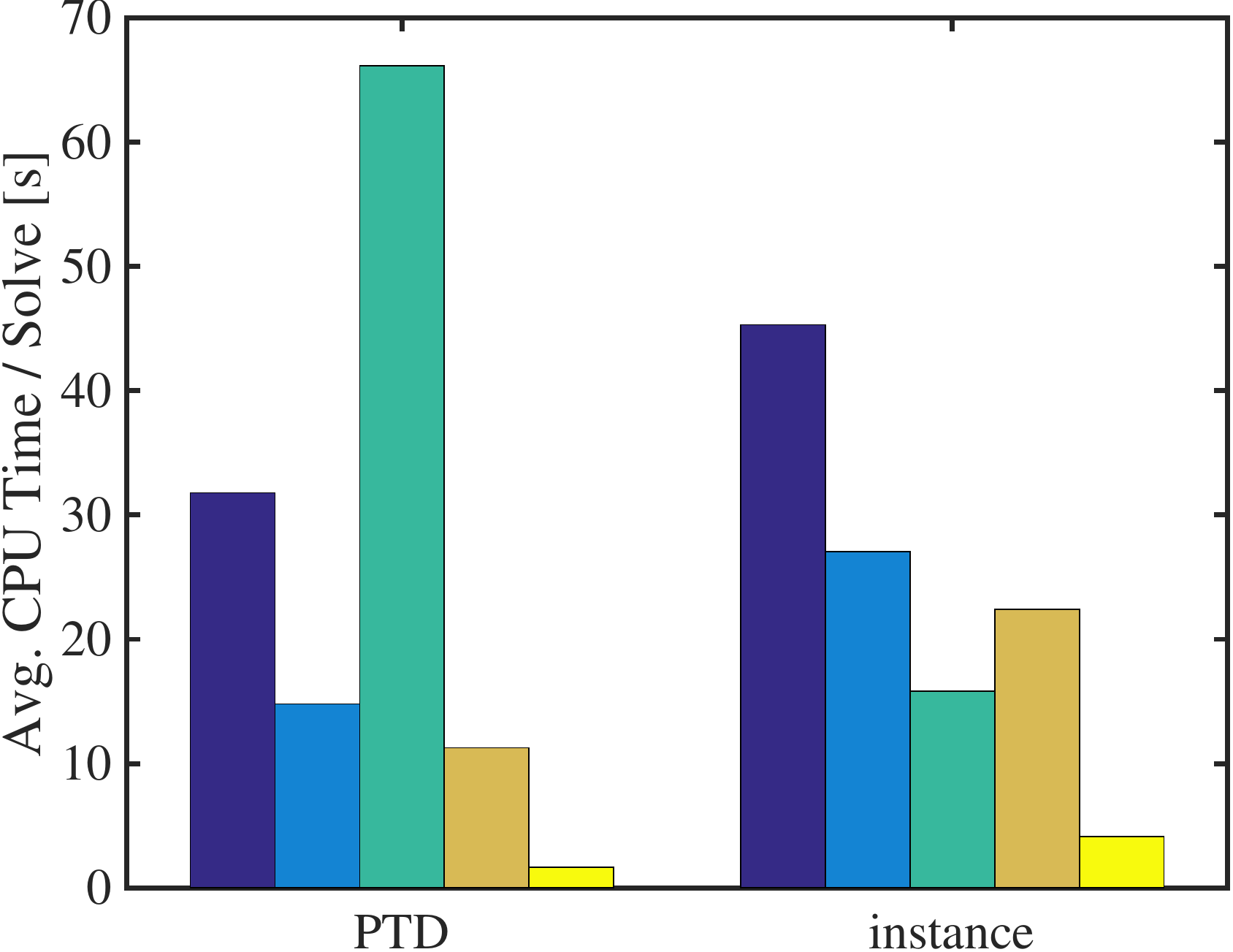}\\[1.5em]
  \includegraphics[width=0.45\textwidth]{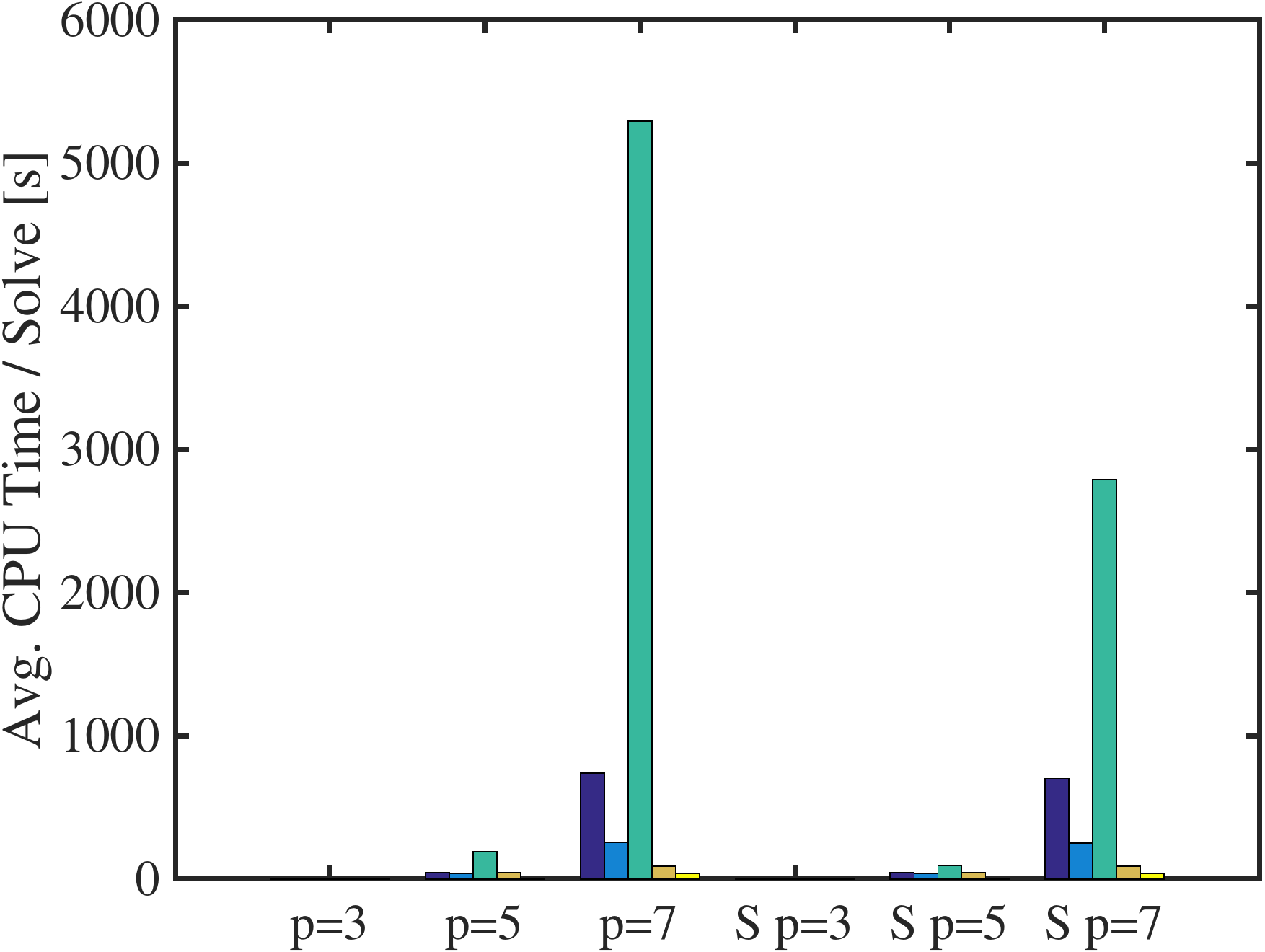}\hspace{1.2em}
  \includegraphics[width=0.45\textwidth]{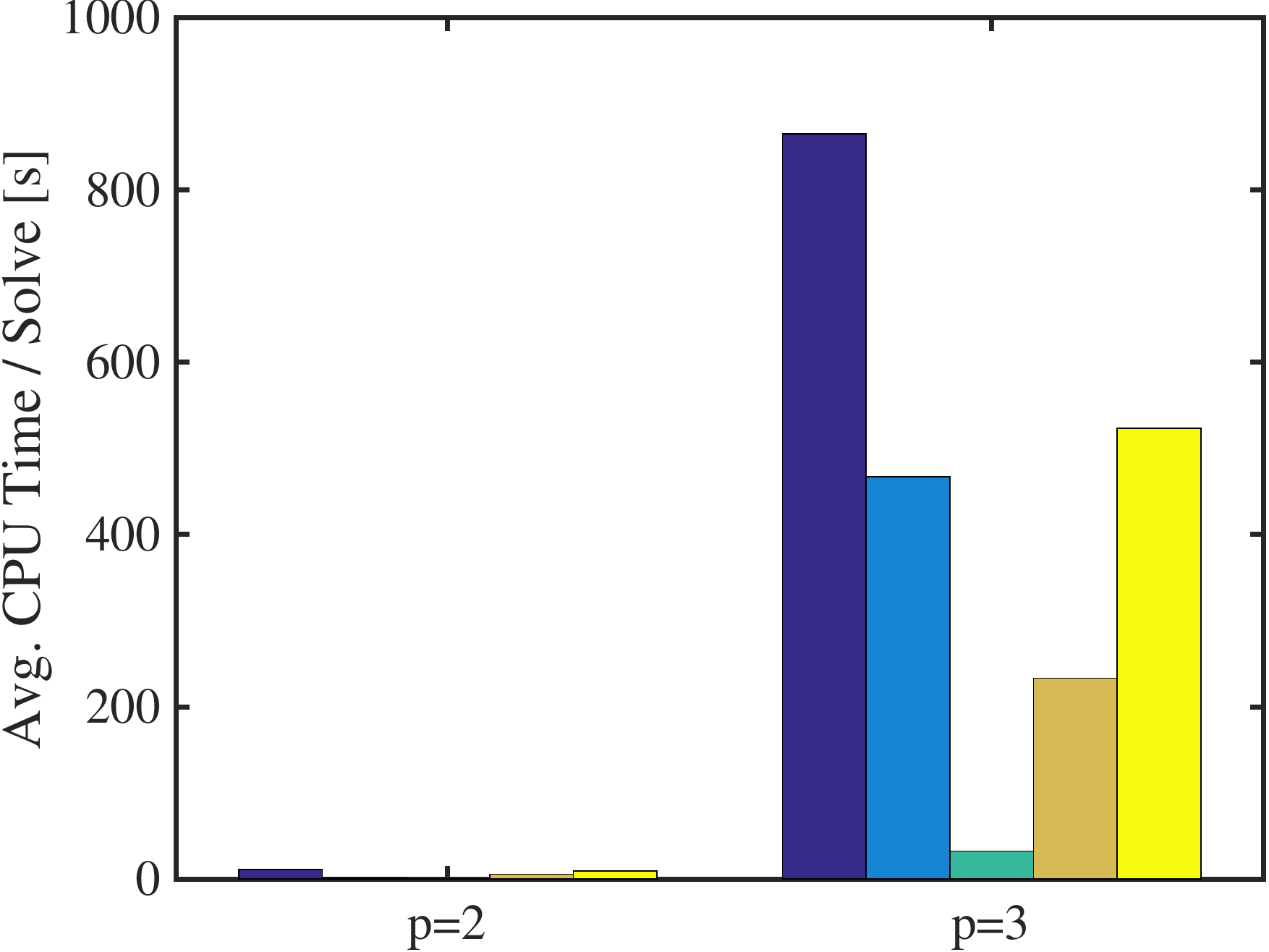}
  \caption{Average CPU time of \cref{a:CS_alg_overall} per system
    solve for case 1 (top left), case 2 (top right), case 3 (bottom
    left), case 4 (bottom right), and their subcases (each group
    across the horizontal axis). ``PTD'' refers to phase-transition
    diagram computations, ``instance'' refers to the main fixed
    problem instance, and ``S p=3'' corresponds to the
    Genz-exponential case with artificially imposed sparsity.}
  \label{f:timing_all_cases}
\end{figure}

All computations in \cref{a:CS_alg_overall} are performed on a 1400
MHz AMD processor with a very small memory requirement, and the
average CPU times per system solve (including corresponding CV solves)
are summarized in \cref{f:timing_all_cases}.  Overall, \fpcas{} and
\admm{} produce CPU times that are lower in the group, \lls{} and
\cgist{} have less consistent performance and can be very fast in some
situations while slow in others, and \sparsa{} is usually in the
middle. Since \fpcas{} is sometimes observed to have slightly higher
and oscillatory error levels in our test cases, we find \admm{} to be
a good practical solver to start with.

\begin{figure}[htb]
  \centering
  \includegraphics[width=0.42\textwidth]{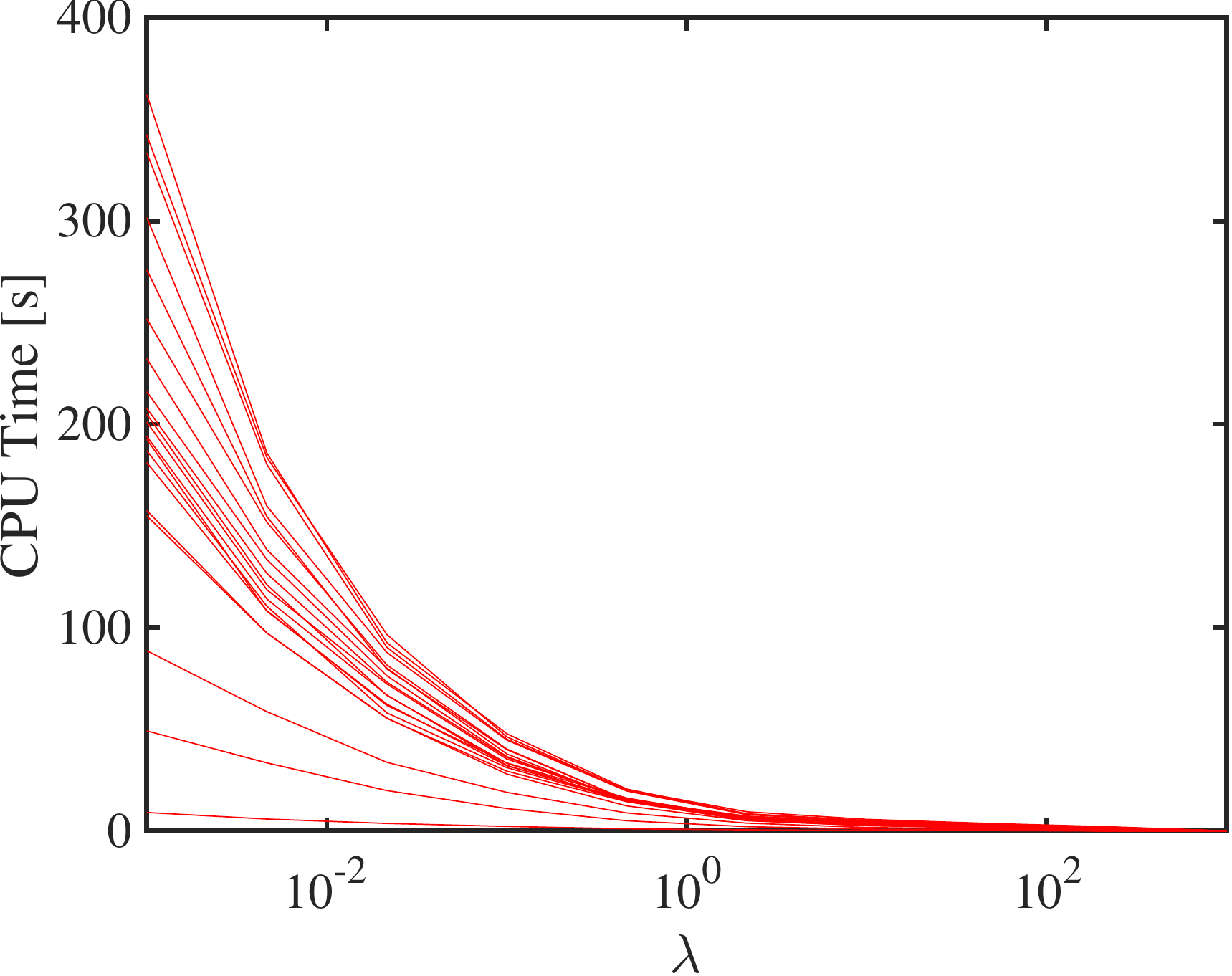}\hspace{1em}
  \includegraphics[width=0.42\textwidth]{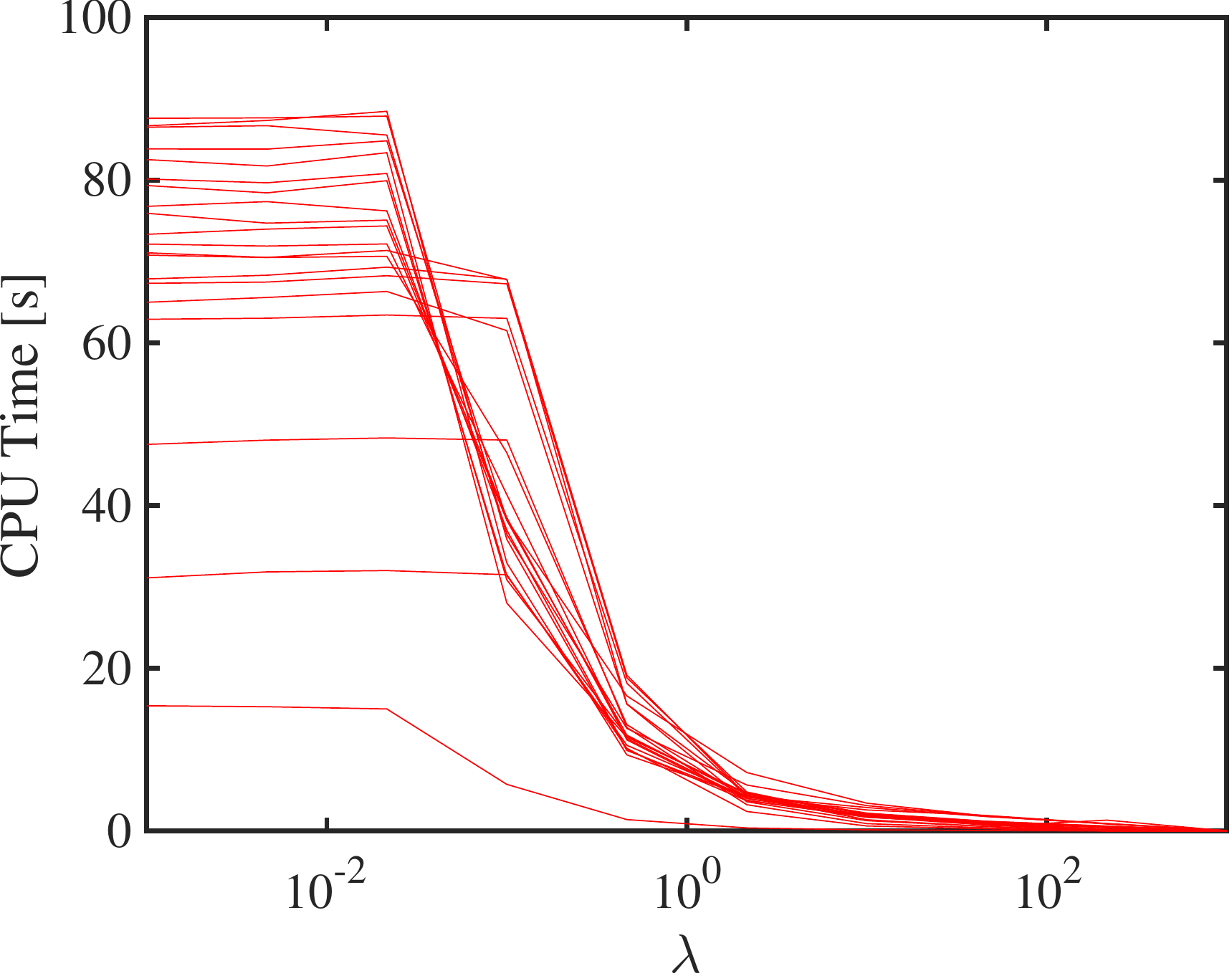}\\[0.5em]
  \includegraphics[width=0.42\textwidth]{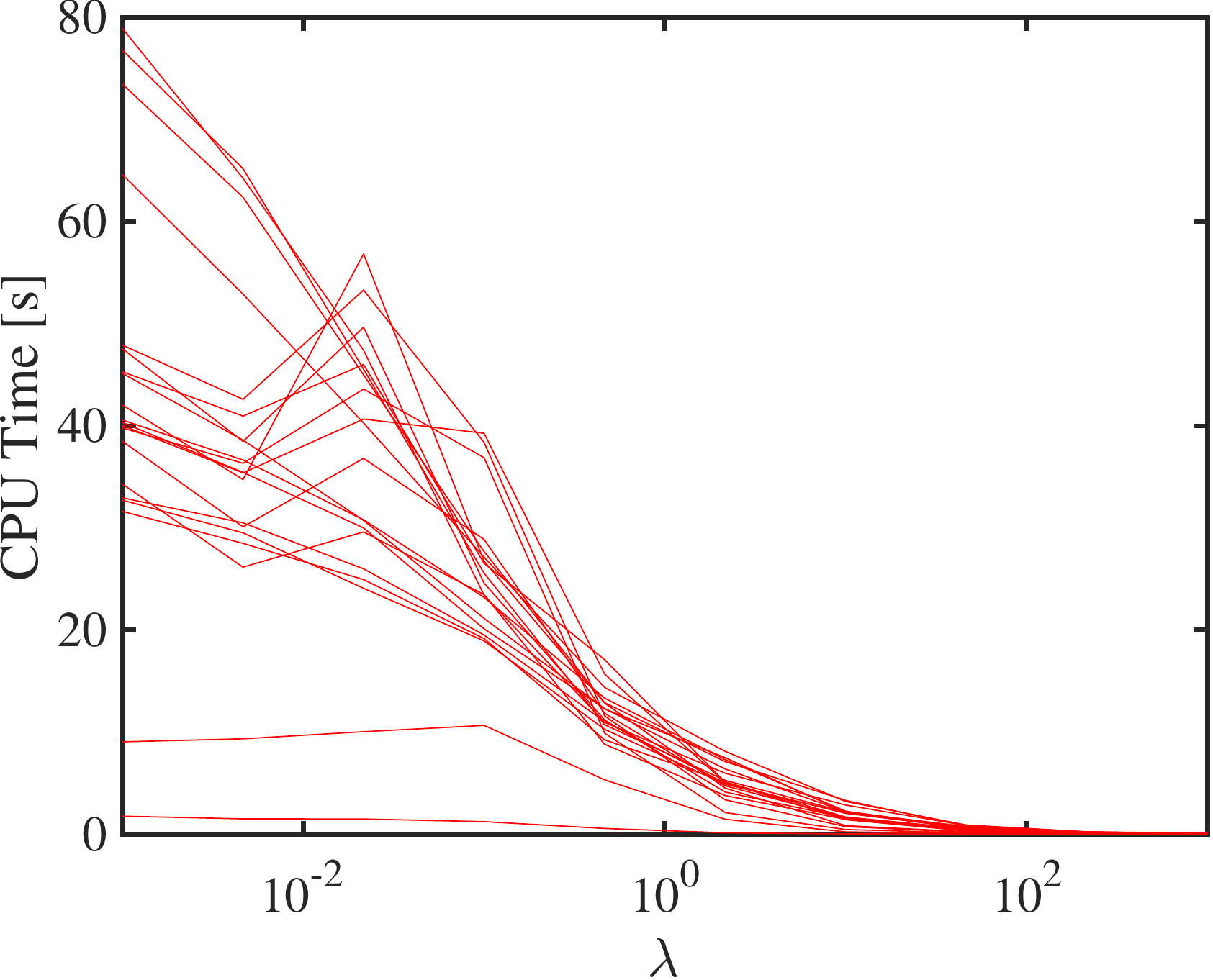}\hspace{1em}
  \includegraphics[width=0.42\textwidth]{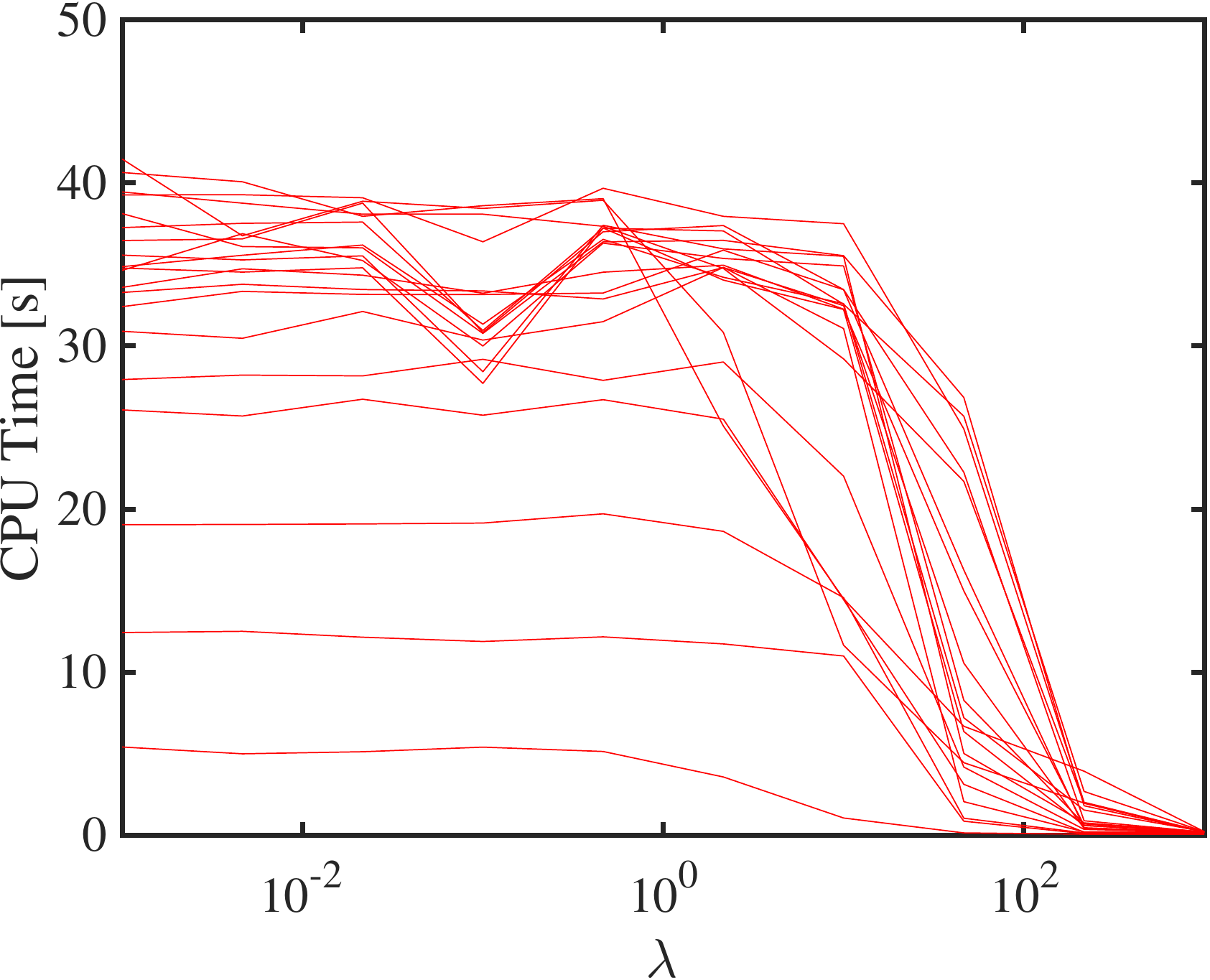}\\
  \includegraphics[width=0.42\textwidth]{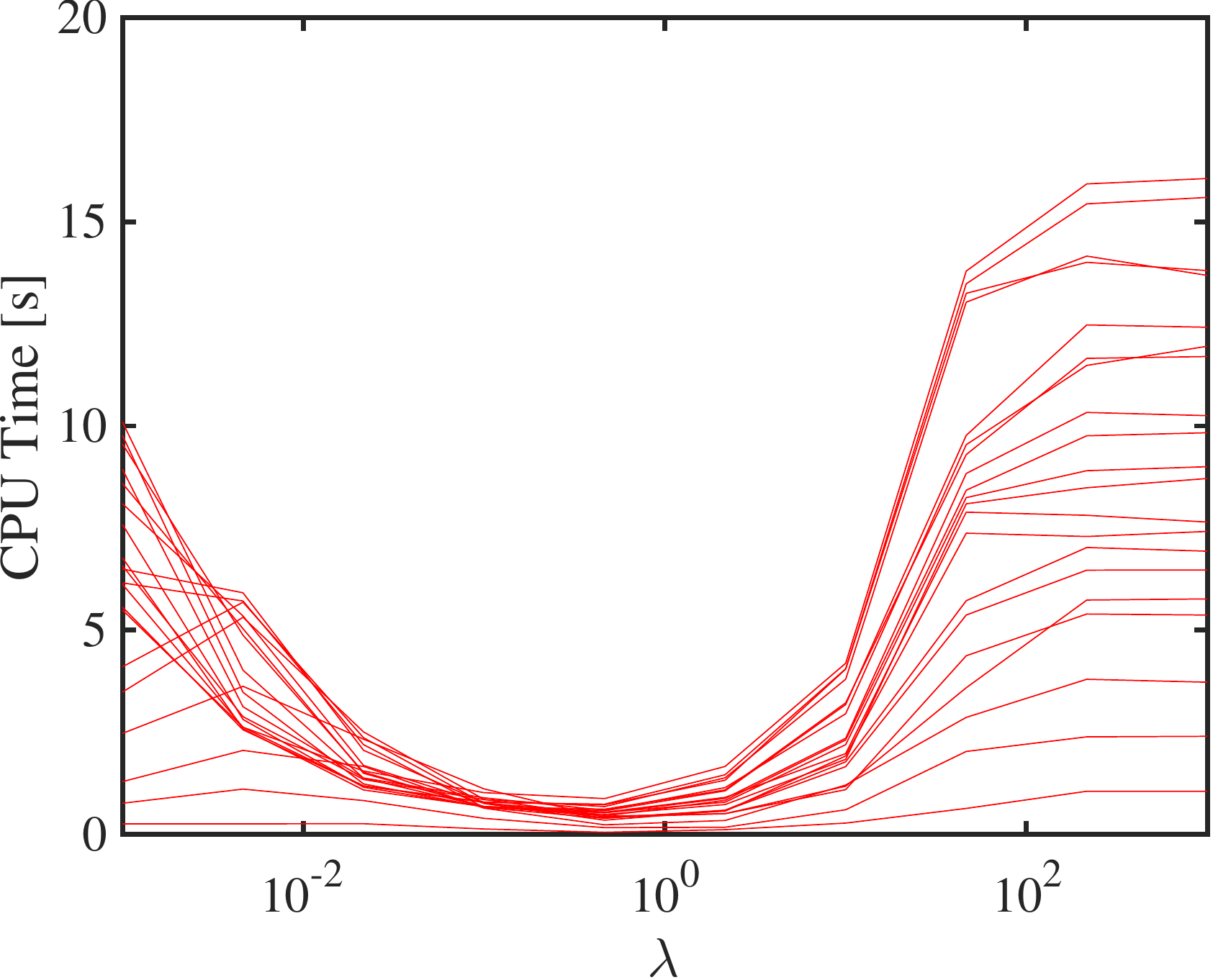}
  \caption{CPU time as a function of $\lambda$ for case 2 random PCE,
    fixed problem instance with $n_s=5$, $p=5$ ($n=251$), and $s=20$,
    using solver \lls{} and \sparsa{} (top left and right), \cgist{}
    and \fpcas{} (middle left and right), and \admm{} (bottom). Each
    red line corresponds to a system solve (including CV) at a
    different $m$.}
  \label{f:case2_single_timing_vs_lambda}
\end{figure}

\Cref{f:case2_single_timing_vs_lambda} further illustrates the CPU
time trend across $\lambda$ for case 2's fixed problem instance
(similar trends are observed in other cases). Each red line represents
the system solve (including CV) at a different $m$. Although not
labeled in the figure, the more computationally expensive plot lines
generally correspond to larger $m$ values. The computational costs
increase with smaller $\lambda$ for all solvers, and with the
exception for \admm{} they also decrease with larger $\lambda$ (and
the solution tending towards the zero vector). \admm{} has an
interesting ``dip'' shape, is overall computationally less expensive
in this case, but can become the more expensive choice when $\lambda$
is large enough. A solver other than \admm{} thus may be advisable for
large values of $\lambda$.

%% file: sections/conclusions.tex
\section{Conclusions}
\label{s:conclusions}

In this paper, we performed numerical investigations employing several
CS solvers that target the unconstrained LASSO formulation, with a
focus on linear systems that rise in PCE constructions. With core
solvers of \lls{}, \sparsa{}, \cgist{}, \fpcas{}, and \admm{}, we
implemented techniques to mitigate overfitting through an automated
selection of regularization constant $\lambda$ based on minimizing the
$K$-fold CV error, and a heuristic strategy to guide the stop-sampling
decision using trends on CV error decay. Practical recommendations on
parameter settings for these techniques were provided and discussed.

The overall method is then applied to a series of numerical examples
of increasing complexity: Gaussian random matrix, random PCE, PCE
approximations to a Genz-exponential model, and large eddy simulations
of supersonic turbulent jet-in-crossflow involving a 24-dimensional
input.  We produced phase-transition diagrams for the Gaussian random
matrix test that matched well with theoretical results, and also for
the random PCE example that experienced more difficult recovery. The
latter involved a complex correlation structure in the linear system
matrices induced by PCE, resulting in an example that does not comply
with the phase-transition universality hypothesis. PCEs of different
degrees were explored for the Genz-exponential study, illustrating the
effects of modeling error and tradeoff between accuracy and
computational costs.  The jet-in-crossflow case produced results that
are consistent with physical intuition and previous global sensitivity
analysis investigation. Furthermore, it demonstrated the
practicability of conducting CS for a realistic, high-dimensional
physical application. Overall, the accuracy and computational
performance for all CS solvers were similar, with \admm{} showing some
advantages with consistent low errors and computational times for
several of the test cases studied in this paper.

Interesting future directions of research include comparisons with BCS
methods, the incorporation of front-tracking (adaptive basis
enrichment) informed by CV error, and investigations on the effects of
overfitting when multiple models of different fidelity are available.

%% file: sections/acknowledgments.tex
\section*{Acknowledgments}

Support for this research was provided by the Defense Advanced
Research Projects Agency (DARPA) program on Enabling Quantification of
Uncertainty in Physical Systems (EQUiPS).
Sandia National Laboratories is a multimission laboratory managed and
operated by National Technology and Engineering Solutions of Sandia,
LLC, a wholly owned subsidiary of Honeywell International, Inc., for
the U.S. Department of Energy's National Nuclear Security
Administration under contract DE-NA-0003525.
The views expressed in the article do not necessarily represent the
views of the U.S. Department of Energy or the United States
Government.

%% file: sections/appendix.tex
\appendix
\section{Default Parameters for CS Solvers}
\label{app:default_params}

\Cref{t:l1_ls_params,t:sparsa_params,t:cgist_params,t:fpcas_params,t:admm_params}
show the default parameters for CS solvers adopted in this study.

\begin{table}[htb]
\caption{Default parameters for \lls{}.}
\label{t:l1_ls_params}
\begin{center}
\begin{tabular}{cc}
\hline
Parameter &  Value \\
\hline
\verb|tar_gap| & $10^{3}$  \\
\verb|eta| & $10^{3}$  \\
\verb|pcgmaxi| & $5000$  \\
\hline
\end{tabular}
\end{center}
\end{table}

\begin{table}[htb]
\caption{Default parameters for \sparsa{}.}
\label{t:sparsa_params}
\begin{center}
\begin{tabular}{cc|cc|cc}
\hline
Parameter &  Value & Parameter &  Value & Parameter &  Value \\
\hline
\verb|StopCriterion| & $2$ & \verb|Initialization| & $0$ & \verb|Eta| & $2$ \\
\verb|ToleranceA| & $10^{-2}$ & \verb|BB_variant| & $1$ & \verb|Continuation| & $0$  \\
\verb|Debias| & $0$ & \verb|BB_cycle| & $1$  & \verb|AlphaMin| & $10^{-30}$\\
\verb|MaxiterA| & $1000$  & \verb|Monotone| & $0$ & \verb|AlphaMax| & $10^{30}$ \\
\verb|MiniterA| & $5$  & \verb|Safeguard| & $0$  \\
\hline
\end{tabular}
\end{center}
\end{table}

\begin{table}[htb]
\caption{Default parameters for \cgist{}.}
\label{t:cgist_params}
\begin{center}
\begin{tabular}{cc}
\hline
Parameter &  Value \\
\hline
\verb|guess| & $0$  \\
\verb|continuation_if_needed| & true  \\
\verb|tol| & $10^{-4}$  \\
\verb|max_iter| & $25000$  \\
\hline
\end{tabular}
\end{center}
\end{table}

\begin{table}[htb]
\caption{Default parameters for \fpcas{}. To enable the solver for
  overdetermined systems, replace all ``$m$'' terms by
  ``$\min(m,n)$''.}
\label{t:fpcas_params}
\begin{center}
\begin{tabular}{cc|cc|cc}
\hline
Parameter &  Value & Parameter &  Value & Parameter &  Value \\
\hline
\verb|x0| & $0$ & \verb|tau_min| & $10^{-4}$ & \verb|eta| & $0.1$   \\
\verb|init| & $2$ & \verb|tau_max| & $10^3$ & \verb|sub_mxitr| & $50$  \\
\verb|tol_eig| & $10^{-4}$  & \verb|mxitr| & $1000$ & \verb|lbfgs_m| & $5$  \\
\verb|scale_A| & $0$  & \verb|gtol| & $10^{-6}$ & \verb|ls_meth| & Hybridls  \\
\verb|eps| & $10^{-16}$  & \verb|gtol_scale_x| & $10^{-12}$ &
\verb|sub_opt_meth| & pcg \\
\verb|zero| & $10^{-10}$ & \verb|f_rel_tol| & $10^{-20}$ & \verb|kappa_g_d| & $10$ \\
\verb|dynamic_zero| & $1$ & \verb|f_value_tol| & $0$ &  \verb|kappa_rho| & $10$ \\
\verb|minK| & $\lfloor m/2 \rfloor$  & \verb|ls_mxitr| & $5$  &
\verb|tol_start_sub| & $10^{-6}$  \\
\verb|maxK| & $m$ & \verb|gamma| & $0.85$ & \verb|min_itr_shrink| & $3$ \\
\verb|hard_truncate| & $1$ & \verb|c| & $10^{-3}$ & \verb|max_itr_shrink| & $20$ \\
\verb|tauD| & $\begin{aligned} &\min( 1.999,\\&-1.665\frac{m}{n} + 2.665)\end{aligned}$ & \verb|beta| & $0.5$  \\
\hline
\end{tabular}
\end{center}
\end{table}

\begin{table}[htb]
\caption{Default parameters for \admm{}.}
\label{t:admm_params}
\begin{center}
\begin{tabular}{cc}
\hline
Parameter &  Value \\
\hline
\verb|rho| & $1.0$  \\
\verb|alpha| & $1.0$  \\
\hline
\end{tabular}
\end{center}
\end{table}

%% file: main.bbl
\begin{thebibliography}{10}

\bibitem{Akaike1974}
{\sc H.~Akaike}, {\em {A New Look at the Statistical Model Identification}},
  IEEE Transactions on Automatic Control, 19 (1974), pp.~716--723,
  \url{https://doi.org/10.1109/TAC.1974.1100705}.

\bibitem{Babacan2009}
{\sc S.~Babacan, R.~Molina, and A.~Katsaggelos}, {\em {Bayesian Compressive
  Sensing Using Laplace Priors}}, IEEE Transactions on Image Processing, 19
  (2010), pp.~53--63, \url{https://doi.org/10.1109/TIP.2009.2032894}.

\bibitem{Barthelmann2000}
{\sc V.~Barthelmann, E.~Novak, and K.~Ritter}, {\em {High dimensional
  polynomial interpolation on sparse grids}}, Advances in Computational
  Mathematics, 12 (2000), pp.~273--288,
  \url{https://doi.org/10.1023/A:1018977404843}.

\bibitem{Blatman:2011}
{\sc G.~Blatman and B.~Sudret}, {\em Adaptive sparse polynomial chaos expansion
  based on least angle regression}, Journal of Computational Physics, 230
  (2011), pp.~2345--2367, \url{https://doi.org/10.1016/j.jcp.2010.12.021}.

\bibitem{Boyd2010}
{\sc S.~Boyd, N.~Parikh, E.~Chu, B.~Peleato, and J.~Eckstein}, {\em
  {Distributed Optimization and Statistical Learning via the Alternating
  Direction Method of Multipliers}}, Foundations and Trends in Machine
  Learning, 3 (2010), pp.~1--122, \url{https://doi.org/10.1561/2200000016}.

\bibitem{Boyd2011a}
{\sc S.~Boyd, N.~Parikh, E.~Chu, B.~Peleato, and J.~Eckstein}, {\em {Solve
  LASSO via ADMM}}, 2011,
  \url{https://web.stanford.edu/~boyd/papers/admm/lasso/lasso.html} (accessed
  2017-07-02).

\bibitem{Candes2006a}
{\sc E.~J. Cand{\`{e}}s, J.~Romberg, and T.~Tao}, {\em {Robust Uncertainty
  Principles: Exact Signal Reconstruction From Highly Incomplete Frequency
  Information}}, IEEE Transactions on Information Theory, 52 (2006),
  pp.~489--509, \url{https://doi.org/10.1109/TIT.2005.862083}.

\bibitem{Chen2001a}
{\sc S.~S. Chen, D.~L. Donoho, and M.~A. Saunders}, {\em {Atomic Decomposition
  by Basis Pursuit}}, SIAM Review, 43 (2001), pp.~129--159,
  \url{https://doi.org/10.1137/S003614450037906X}.

\bibitem{Davis1997}
{\sc G.~Davis, S.~Mallat, and M.~Avellaneda}, {\em {Adaptive greedy
  approximations}}, Constructive Approximation, 13 (1997), pp.~57--98,
  \url{https://doi.org/10.1007/BF02678430}.

\bibitem{Donoho2006a}
{\sc D.~L. Donoho}, {\em {Compressed sensing}}, IEEE Transactions on
  Information Theory, 52 (2006), pp.~1289--1306,
  \url{https://doi.org/10.1109/Tit.2006.871582}.

\bibitem{Donoho2006}
{\sc D.~L. Donoho}, {\em {For Most Large Underdetermined Systems of Linear
  Equations the Minimal $\ell_1$-norm Solution Is Also the Sparsest Solution}},
  Communications on Pure and Applied Mathematics, 59 (2006), pp.~797--829,
  \url{https://doi.org/10.1002/cpa.20132}.

\bibitem{Donoho2009}
{\sc D.~L. Donoho and J.~Tanner}, {\em {Observed universality of phase
  transitions in high-dimensional geometry, with implications for modern data
  analysis and signal processing}}, Philosophical Transactions of the Royal
  Society A: Mathematical, Physical and Engineering Sciences, 367 (2009),
  pp.~4273--4293, \url{https://doi.org/10.1098/rsta.2009.0152}.

\bibitem{Donoho2010}
{\sc D.~L. Donoho and J.~Tanner}, {\em {Precise Undersampling Theorems}},
  Proceedings of the IEEE, 98 (2010), pp.~913--924,
  \url{https://doi.org/10.1109/JPROC.2010.2045630}.

\bibitem{Efron2004}
{\sc B.~Efron, T.~Hastie, I.~Johnstone, and R.~Tibshirani}, {\em {Least angle
  regression}}, The Annals of Statistics, 32 (2004), pp.~407--499,
  \url{https://doi.org/10.1214/009053604000000067}.

\bibitem{Eldred2015}
{\sc M.~S. Eldred, L.~W.~T. Ng, M.~F. Barone, and S.~P. Domino}, {\em
  {Multifidelity Uncertainty Quantification Using Spectral Stochastic
  Discrepancy Models}}, in Handbook of Uncertainty Quantification, Springer
  International Publishing, Cham, 2015, pp.~1--45,
  \url{https://doi.org/10.1007/978-3-319-11259-6_25-1}.

\bibitem{Ernst2012}
{\sc O.~G. Ernst, A.~Mugler, H.-J. Starkloff, and E.~Ullmann}, {\em {On the
  convergence of generalized polynomial chaos expansions}}, ESAIM: Mathematical
  Modelling and Numerical Analysis, 46 (2012), pp.~317--339,
  \url{https://doi.org/10.1051/m2an/2011045}.

\bibitem{Fajraoui2017}
{\sc N.~Fajraoui, S.~Marelli, and B.~Sudret}, {\em {On optimal experimental
  designs for sparse polynomial chaos expansions}}, 2017,
  \url{https://arxiv.org/abs/1703.05312}.

\bibitem{Gerstner1998}
{\sc T.~Gerstner and M.~Griebel}, {\em {Numerical integration using sparse
  grids}}, Numerical Algorithms, 18 (1998), pp.~209--232,
  \url{https://doi.org/10.1023/A:1019129717644}.

\bibitem{Gerstner2003}
{\sc T.~Gerstner and M.~Griebel}, {\em {Dimension-Adaptive Tensor-Product
  Quadrature}}, Computing, 71 (2003), pp.~65--87,
  \url{https://doi.org/10.1007/s00607-003-0015-5}.

\bibitem{Ghanem1991}
{\sc R.~G. Ghanem and P.~D. Spanos}, {\em {Stochastic Finite Elements: A
  Spectral Approach}}, Springer New York, New York, NY, 1st~ed., 1991.

\bibitem{Goldstein2010}
{\sc T.~Goldstein and S.~Setzer}, {\em {High-order methods for basis pursuit}},
  Tech. Report CAM Report 10-41, University of California Los Angeles, 2010.

\bibitem{Goldstein2011}
{\sc T.~Goldstein and S.~Setzer}, {\em {CGIST: A Fast and Exact Solver for L1
  Minimization}}, 2011, \url{http://tag7.web.rice.edu/CGIST.html} (accessed
  2017-07-02).

\bibitem{Halton1960}
{\sc J.~H. Halton}, {\em {On the efficiency of certain quasi-random sequences
  of points in evaluating multi-dimensional integrals}}, Numerische Mathematik,
  2 (1960), pp.~84--90, \url{https://doi.org/10.1007/BF01386213}.

\bibitem{Hampton2015}
{\sc J.~Hampton and A.~Doostan}, {\em {Compressive sampling of polynomial chaos
  expansions: Convergence analysis and sampling strategies}}, Journal of
  Computational Physics, 280 (2015), pp.~363--386,
  \url{https://doi.org/10.1016/j.jcp.2014.09.019}.

\bibitem{HiFiRE2010}
{\sc N.~E. Hass, K.~F. Cabell, and A.~M. Storch}, {\em {HIFiRE Direct-Connect
  Rig (HDCR) Phase I Ground Test Results from the NASA Langley Arc-Heated
  Scramjet Test Facility}}, Tech. Report CR-2010-002215, NASA, 2010.

\bibitem{Hastie2009}
{\sc T.~Hastie, R.~Tibshirani, and J.~Friedman}, {\em {The Elements of
  Statistical Learning}}, Springer, New York, NY, 2nd~ed., 2009.

\bibitem{Huan2018a}
{\sc X.~Huan, C.~Safta, K.~Sargsyan, G.~Geraci, M.~S. Eldred, Z.~P. Vane,
  G.~Lacaze, J.~C. Oefelein, and H.~N. Najm}, {\em {Global Sensitivity Analysis
  and Estimation of Model Error, Toward Uncertainty Quantification in Scramjet
  Computations}}, AIAA Journal, 56 (2018), pp.~1170--1184,
  \url{https://doi.org/10.2514/1.J056278}.

\bibitem{Jakeman2015}
{\sc J.~D. Jakeman, M.~S. Eldred, and K.~Sargsyan}, {\em {Enhancing
  $\ell_1$-minimization estimates of polynomial chaos expansions using basis
  selection}}, Journal of Computational Physics, 289 (2015), pp.~18--34,
  \url{https://doi.org/10.1016/j.jcp.2015.02.025}.

\bibitem{Jakeman2017}
{\sc J.~D. Jakeman, A.~Narayan, and T.~Zhou}, {\em {A Generalized Sampling and
  Preconditioning Scheme for Sparse Approximation of Polynomial Chaos
  Expansions}}, SIAM Journal on Scientific Computing, 39 (2017),
  pp.~A1114--A1144, \url{https://doi.org/10.1137/16M1063885}.

\bibitem{Ji2008}
{\sc S.~Ji, Y.~Xue, and L.~Carin}, {\em {Bayesian compressive sensing}}, IEEE
  Transactions on Signal Processing, 56 (2008), pp.~2346--2356,
  \url{https://doi.org/10.1109/TSP.2007.914345}.

\bibitem{Kass1995}
{\sc R.~E. Kass and A.~E. Raftery}, {\em {Bayes Factor}}, Journal of American
  Statistical Association, 90 (1995), pp.~773--795,
  \url{https://doi.org/10.2307/2291091}.

\bibitem{Kim2007}
{\sc S.-J. Kim, K.~Koh, M.~Lustig, S.~Boyd, and D.~Gorinevsky}, {\em {An
  Interior Point Method for Large-Scale $\ell_1$-Regularized Least Squares}},
  IEEE Journal of Selected Topics in Signal Processing, 1 (2007), pp.~606--617,
  \url{https://doi.org/10.1109/JSTSP.2007.910971}.

\bibitem{Koh2008}
{\sc K.~Koh, S.-J. Kim, and S.~Boyd}, {\em {Simple Matlab Solver for
  l1-regularized Least Squares Problems, beta version}}, 2008,
  \url{https://stanford.edu/~boyd/l1_ls/} (accessed 2017-07-02).

\bibitem{LeMaitre2010}
{\sc O.~P. {Le Ma{\^{i}}tre} and O.~M. Knio}, {\em {Spectral Methods for
  Uncertainty Quantification: with Applications to Computational Fluid
  Dynamics}}, Springer Netherlands, Houten, Netherlands, 2010.

\bibitem{Mahesh2013}
{\sc K.~Mahesh}, {\em {The Interaction of Jets with Crossflow}}, Annual Review
  of Fluid Mechanics, 45 (2013), pp.~379--407,
  \url{https://doi.org/10.1146/annurev-fluid-120710-101115}.

\bibitem{Malioutov2010}
{\sc D.~M. Malioutov, S.~R. Sanghavi, and A.~S. Willsky}, {\em {Sequential
  Compressed Sensing}}, IEEE Journal of Selected Topics in Signal Processing, 4
  (2010), pp.~435--444, \url{https://doi.org/10.1109/JSTSP.2009.2038211}.

\bibitem{Najm2009}
{\sc H.~N. Najm}, {\em {Uncertainty Quantification and Polynomial Chaos
  Techniques in Computational Fluid Dynamics}}, Annual Review of Fluid
  Mechanics, 41 (2009), pp.~35--52,
  \url{https://doi.org/10.1146/annurev.fluid.010908.165248}.

\bibitem{Oefelein1997}
{\sc J.~C. Oefelein}, {\em {Simulation and Analysis of Turbulent Multiphase
  Combustion Processes at High Pressures}}, PhD thesis, The Pennsylvania State
  University, 1997.

\bibitem{Oefelein2006}
{\sc J.~C. Oefelein}, {\em {Large eddy simulation of turbulent combustion
  processes in propulsion and power systems}}, Progress in Aerospace Sciences,
  42 (2006), pp.~2--37, \url{https://doi.org/10.1016/j.paerosci.2006.02.001}.

\bibitem{Peng2014}
{\sc J.~Peng, J.~Hampton, and A.~Doostan}, {\em {A weighted
  $\ell_1$-minimization approach for sparse polynomial chaos expansions}},
  Journal of Computational Physics, 267 (2014), pp.~92--111,
  \url{https://doi.org/10.1016/j.jcp.2014.02.024}.

\bibitem{Rauhut2012}
{\sc H.~Rauhut and R.~Ward}, {\em {Sparse Legendre expansions via
  $\ell_1$-minimization}}, Journal of Approximation Theory, 164 (2012),
  pp.~517--533, \url{https://doi.org/10.1016/j.jat.2012.01.008}.

\bibitem{Sargsyan2014}
{\sc K.~Sargsyan, C.~Safta, H.~N. Najm, B.~J. Debusschere, D.~Ricciuto, and
  P.~Thornton}, {\em {Dimensionality Reduction for Complex Models Via Bayesian
  Compressive Sensing}}, International Journal for Uncertainty Quantification,
  4 (2014), pp.~63--93,
  \url{https://doi.org/10.1615/Int.J.UncertaintyQuantification.2013006821}.

\bibitem{Schwarz1978}
{\sc G.~Schwarz}, {\em {Estimating the Dimension of a Model}}, The Annals of
  Statistics, 6 (1978), pp.~461--464,
  \url{https://doi.org/10.1214/aos/1176344136}.

\bibitem{Tanner2012}
{\sc J.~Tanner}, {\em {Phase Transitions of the Regular Polytopes and Cone}},
  2012, \url{https://people.maths.ox.ac.uk/tanner/polytopes.shtml} (accessed
  2017-07-03).

\bibitem{VandenBerg2009a}
{\sc E.~van~den Berg and M.~P. Friedlander}, {\em {Probing the Pareto Frontier
  for Basis Pursuit Solutions}}, SIAM Journal on Scientific Computing, 31
  (2009), pp.~890--912, \url{https://doi.org/10.1137/080714488}.

\bibitem{Wasserman2000}
{\sc L.~Wasserman}, {\em {Bayesian Model Selection and Model Averaging}},
  Journal of Mathematical Psychology, 44 (2000), pp.~92--107,
  \url{https://doi.org/10.1006/jmps.1999.1278}.

\bibitem{Wen2010}
{\sc Z.~Wen, W.~Yin, D.~Goldfarb, and Y.~Zhang}, {\em {A Fast Algorithm for
  Sparse Reconstruction Based on Shrinkage, Subspace Optimization, and
  Continuation}}, SIAM Journal on Scientific Computing, 32 (2010),
  pp.~1832--1857, \url{https://doi.org/10.1137/090747695}.

\bibitem{Wright2009a}
{\sc S.~J. Wright, R.~D. Nowak, and M.~A.~T. Figueiredo}, {\em {SpaRSA: Sparse
  Reconstruction by Separable Approximation, version 2.0}}, 2009,
  \url{http://www.lx.it.pt/~mtf/SpaRSA/} (accessed 2017-07-02).

\bibitem{Wright2009}
{\sc S.~J. Wright, R.~D. Nowak, and M.~A.~T. Figueiredo}, {\em {Sparse
  Reconstruction by Separable Approximation}}, IEEE Transactions on Signal
  Processing, 57 (2009), pp.~2479--2493,
  \url{https://doi.org/10.1109/TSP.2009.2016892}.

\bibitem{Xiu2009}
{\sc D.~Xiu}, {\em {Fast Numerical Methods for Stochastic Computations: A
  Review}}, Communications in Computational Physics, 5 (2009), pp.~242--272.

\bibitem{Xiu2002}
{\sc D.~Xiu and G.~E. Karniadakis}, {\em {The Wiener-Askey Polynomial Chaos for
  Stochastic Differential Equations}}, SIAM Journal on Scientific Computing, 24
  (2002), pp.~619--644, \url{https://doi.org/10.1137/S1064827501387826}.

\bibitem{Yin2010}
{\sc W.~Yin and Z.~Wen}, {\em {FPC{\_}AS: A MATLAB Solver for L1-Regularization
  Problems, version 1.21}}, 2010,
  \url{http://www.caam.rice.edu/~optimization/L1/FPC_AS/} (accessed
  2017-07-02).

\end{thebibliography}
